\documentclass[aps,prd,12pt,citeautoscript,superscriptaddress,showpacs,scrartcl,amsmath,amssymb]{revtex4-1}
\usepackage[utf8x]{inputenc}

\usepackage{xcolor}
\usepackage[colorlinks=true, pdfstartview=FitV,linkcolor=blue, citecolor=blue, urlcolor=blue]{hyperref}

\usepackage[bottom]{footmisc}

\usepackage{graphicx}

\makeatletter
\newcommand*\bigcdot{\mathpalette\bigcdot@{.5}}
\newcommand*\bigcdot@[2]{\mathbin{\vcenter{\hbox{\scalebox{#2}{$\m@th#1\bullet$}}}}}
\makeatother

\newcommand{\be}{\begin{eqnarray}}
\newcommand{\ee}{\end{eqnarray}}
\newcommand{\bea}{\begin{eqnarray}}
\newcommand{\eea}{\end{eqnarray}}

\newcommand{\nn}{\nonumber}

\newcommand{\thetaw}{\theta_{\mbox{\tiny W}}}
\newcommand{\mz}{m_{\mbox{\tiny Z}}}
\newcommand{\mh}{m_{\mbox{\tiny H}}}
\newcommand{\mw}{m_{\mbox{\tiny W}}}

\newcommand{\F}{{B}}
\newcommand{\WW}{{\rm W}}

\usepackage{natbib}
\usepackage{graphicx}
\usepackage{amsmath}
\usepackage{amssymb}
\usepackage{mathtools}
\usepackage{tensor}
\usepackage{accents}
\usepackage{cancel}
\usepackage{titlesec}
\usepackage{subfigure}
\usepackage[a4paper]{geometry}
\usepackage{psfrag}
\usepackage{latexsym,array,theorem,mathrsfs,subfigure,bm,float}
\usepackage{amsfonts}
\usepackage{amssymb}

\renewcommand{\theequation}{\arabic{section}.\arabic{equation}}

\newcommand{\A}{{B}}

\newcommand{\T}{{\rm T}}

\newcommand{\Om}{{\Omega}}

\DeclareMathOperator{\e}{e}

\newcommand\numberthis{\addtocounter{equation}{1}\tag{\theequation}}
\titleformat{\paragraph}
{\normalfont\normalsize\bfseries}{\theparagraph}{1em}{}
\titlespacing*{\paragraph}
{0pt}{3.25ex plus 1ex minus .2ex}{1.5ex plus .2ex}

\begin{document}

\title{Electroweak monopoles and their stability}

\author{Romain Gervalle}
\email{romain.gervalle@univ-tours.fr}
\affiliation{
Institut Denis Poisson, UMR - CNRS 7013, \\ 
Universit\'{e} de Tours, Parc de Grandmont, 37200 Tours, France}

\author{Mikhail~S.~Volkov}
\email{volkov@lmpt.univ-tours.fr}
\affiliation{
Institut Denis Poisson, UMR - CNRS 7013, \\ 
Universit\'{e} de Tours, Parc de Grandmont, 37200 Tours, France}


\ 

\vspace{1 cm}

\begin{abstract}

\vspace{0.5 cm}

We apply a generalized field ansatz to describe the spherically symmetric sector of classical solutions of the electroweak theory. 
This sector  contains Abelian magnetic monopoles labeled by their magnetic charge $n=\pm 1,\pm 2,\ldots$, 
the  non-Abelian monopole  for $n=\pm 2$ found previously by 
Cho and Maison (CM), and also the electric oscillating solutions. 
All magnetic monopoles  have infinite energy. 
We  analyze their perturbative stability and 
use the method of complex spacetime tetrad 
to separate  variables 
and reduce the perturbation equations
 to multi-channel Schroedinger-type
eigenvalue  problems. The spectra of perturbations around the CM monopole do not contain negative 
modes hence this solution is stable. 
The $n=\pm 1$ Abelian monopole is also stable, but all monopoles with $|n|\geq 2$ are unstable with respect to perturbations 
with angular momentum $j=|n|/2-1$.
The Abelian $|n|=2$ monopole is unstable only within the $j=0$ sector whereas  the  CM monopole  also has $|n|=2$ and belongs to the same sector, 
hence it may  be viewed as a stable remnant of the decay of the Abelian monopole. 
One may similarly conjecture that stable remnants exist also for monopoles with $|n|>2$, hence the CM monopole may  be just the 
first member of a sequence of non-Abelian monopoles with higher magnetic charges. 
 Only the CM monopole is spherically symmetric while all non-Abelian monopoles with $|n|>2$ are not rotationally invariant.

\end{abstract}


\maketitle


\section{INTRODUCTION}

Let us remind basic facts about magnetic monopoles. The magnetic monopole in the U(1) electrodynamics is described
by the Colombian magnetic field  $\vec{{\bm B}}=-\bm{P}\vec{\bm r}/\bm{r}^3$ (the minus sign here is for further convenience 
and we denote all dimensionfull quantities  boldfaced). One cannot find a globally regular vector potential $\bm A$
such that $\vec{{\bm B}}={\rm rot} \vec{{\bm A}}$, however, as was noticed by Dirac \cite{Dirac:1931kp} and later rectified 
by Wu and Yang \cite{Wu:1976ge}, one can use for this two locally regular potentials expressed in spherical coordinates by 
$
{\bm A}_\pm ={\bm P}\,(\cos\vartheta\mp 1) d\varphi\,.
$
The potential ${\bm A}_{+}$ is regular in the upper hemisphere but
singular on the negative $z$-axis (the Dirac string singularity), as can be seen by passing to Cartesian coordinates. 
Similarly, ${\bm A}_{-}$ is regular in
the lower hemisphere. Therefore,  using either ${\bm A}_{+}$ or ${\bm A}_{-}$ depending on the 
hemisphere yields a  regular description on the entire sphere. The two potentials are related in the equatorial region 
by the gauge transformation, ${\bm A}_{-}-{\bm A}_{+}=d(2{\bm P}\varphi)$, while the wavefunction in the Schroedinger equation 
acquires then the phase factor $\exp\left(2i{\bm e}{\bm P}\varphi /\bm{ \hbar c}\right)$, and since this should be a periodic 
function of $\varphi$, it follows that the magnetic charge fulfills the Dirac quantization condition 
\be
\bm{P}=\frac{\bm{\hbar c}}{2\bm e}\times n,~~~~n=\pm 1,\pm 2,\ldots
\ee
This construction with two locally regular gauges and the quantized magnetic charge is called the Dirac monopole. 

It was later noticed by Wu and Yang \cite{Wu} that the Dirac monopole can be embedded into the 
non-Abelian Yang-Mills theory by simply multiplying 
the U(1) vector potential by a constant matrix from the Lie algebra of the gauge group. Using the non-Abelian 
gauge transformations one can then globally remove the Dirac string instead of using two locally regular gauges, but 
the field potential still remains singular at the origin where the magnetic field diverges. 
However, as was discovered  independently by t'Hooft  \cite{tHooft:1974kcl} and by Polyakov \cite{Polyakov:1974ek}, the singularity 
at the origin can be removed by adding a Higgs field in the adjoint representation of the gauge group, choosing the latter 
to be SU(2) in the simplest case. This yields a completely regular configuration of soliton type which has a finite energy, 
contains massive fields in the central region, while at large distances only the massless U(1) gauge field 
survives and approaches that for the Dirac monopole. 

The discovery of t'Hooft and Polyakov triggered a large number of theoretical studies  and nowadays monopoles
find applications in various  branches of theoretical physics 
(see \cite{Goddard:1977da,Coleman:1982cx,Konishi:2007dn,Manton:2004tk,Shnir:2005vvi} for reviews
and, e.g., \cite{Chamseddine:1997nm,Forgacs:2003yh} for particular aspects of monopoles). 
However, the experimental 
search for magnetic monopoles has always been giving negative results (see   \cite{Rajantie:2016paj,Mitsou:2019mrs} 
for recent reviews). A possible explanation for this is that the  t'Hooft-Polyakov monopole is not described by the Standard Model, 
because the latter contains in the electroweak sector the Higgs field in the fundamental and not adjoint representation. 
As a result, the standard topological arguments \cite{Manton:2004tk} insuring  the existence and stability  of monopoles do not apply. 

Since there no topological arguments for their existence, 
one may wonder if there are any monopoles in the electroweak theory at all ? 
The answer should  of course be positive because, as the U(1) electrodynamics is a part of the electroweak theory,  
the Dirac monopoles should be solutions of the theory.  However, properties of such embedded monopoles may be not the same
as in the electrodynamics. 
For example, Dirac monopoles are stable within the U(1) theory, but this does not mean that they should be stable in the 
SU(2)$\times$U(1) electroweak theory as well. One can also wonder if there exist in addition some other electroweak monopoles,
maybe some extended solutions similar to the  t'Hooft-Polyakov  monopole ?

The best known classical solution in the electroweak theory is the sphaleron \cite{Klinkhamer:1984di},  
but it is unstable and neutral, hence it is  not at all similar to monopoles. 
Moreover, the electroweak sphaleron  is not even spherically symmetric \cite{Kleihaus:1991ks}, unless for  vanishing 
mixing angle $\thetaw$ when the U(1) hypercharge field decouples \cite{Dashen:1974ck}, \cite{Yaffe:1989ms}.
This may suggest that the electroweak theory admits spherically symmetric solutions only in the $\thetaw\to 0$  limit 
 \cite{Ratra:1987dp,Farhi:2005rz}, 
 while for a finite mixing angle the spherical symmetry should be broken \cite{Graham:2006vy,Graham:2007ds}. 
 
 However, an  essentially non-Abelian and spherically symmetric monopole solution in the full electroweak theory was found by  
 Cho and Maison (CM) \cite{Cho:1996qd}. 
 This solution is somewhat  similar to  the t'Hooft-Polyakov monopole but there is one important difference: 
in addition to the regular SU(2) gauge field it  contains a Colombian U(1) field which diverges  
at the origin thus rendering  the energy infinite. This feature of the solution is  not very appealing and 
there have been attempts  to regularize  the monopole energy  in some way, 
 but they require to modify the Lagrangian of the theory \cite{Cho:2013vba,Pak:2013jaa,Blaschke:2017pym,Ellis:2020bpy,Hung:2020vuo}. 
At the same time, since the Standard Model describes the real world extremely well, it seems to be more logical to take and study 
the CM monopole as it is, with infinite energy.  In any case, its energy 
certainly becomes finite  when gravity is taken into account \cite{Bai:2020ezy}. 
 
A remarkable feature of the CM solution  is the fact that  it is spherically symmetric 
and yet exists for nonzero values of the Weinberg angle $\thetaw$. 
The ansatz constructed by Cho and Maison uses 
exactly the same SU(2) gauge field as for the spherically symmetric for $\thetaw=0$  sphaleron (first found by Dashen et al. 
\cite{Dashen:1974ck} and later generalized by Witten \cite{Witten:1976ck}). However,  the Higgs field is not the same 
as for the sphaleron but rather in the form suggested by  Nambu \cite{Nambu:1977ag}. Within the notation to be used below, here 
is the Higgs field for the sphaleron and for the monopole:
 \be               \label{Higgs0}
 \Phi_{\rm sph}=\phi\, \exp\left(i\,\xi\, \vec{n}\vec{\tau}/2 \right)
 \begin{pmatrix}
0  \\
1
\end{pmatrix},~~~~~~~
\Phi_{\rm mon}=\phi\, \exp\left(i\,\xi/2\right) \begin{pmatrix}
- \sin\frac{\vartheta}{2}\,e^{-i{\varphi}}\ \\
\cos\frac{\vartheta}{2}
\end{pmatrix}.
\ee
Here $\phi,\xi$ depend on the radial coordinate $r$ whereas  $\vec{n}=\vec{x}/r$ is the radial unit vector and $\vec{\tau}$ are the Pauli matrices.
Inserting these expressions  into the field equations and using the same form of  the SU(2)$\times$U(1) gauge field in both cases,
the angular variables separate only for $\thetaw=0$
in the sphaleron case and for any $\thetaw$ in the monopole case. 

It seems that the latter fact has gone more or less unnoticed and it remains largely 
unknown that, apart from the CM monopole, the electroweak theory
admits a whole sector of spherically symmetric solutions, static or time-dependent. Below we shall describe this sector by 
combining 
$\Phi_{\rm mon}$ with the generalized Witten ansatz for the SU(2) gauge field and with the U(1) field. 
We find that this sector contains magnetic monopoles, their dyon generalizations
including the electric field, and also the oscillating solutions. 
Postponing the discussion of the latter to a separate publication,
we shall concentrate below  on magnetic monopoles. 

Before considering  spherically symmetric monopoles, 
one should   say that the Nambu form for the Higgs field in \eqref{Higgs0} was originally proposed 
within a slightly different context:  to describe monopoles connected to a vortex \cite{Nambu:1977ag}.   
Specifically, the Higgs field $\Phi_{\rm mon}$ in \eqref{Higgs0} 
does not have a $\vartheta\to \pi$ limit, and to cure this one can assume that $\phi$ vanishes in this limit, thereby producing  a vortex 
(similar to the Z-string \cite{Vachaspati:1992fi})
that starts on the monopole and extends  along the negative $z$-axis.
If the vortex is (semi)infinite, then the resulting system is called the Nambu monopole and it cannot be static, since the vortex will be pulling the monopole. 
The vortex may  also have a finite length and terminate
some distance away  on an antimonopole, then the resulting  monopole-antimonopole pair will be spinning around the common center of mass 
\cite{Urrestilla:2001dd}. 

A different possibility to interpret  the Nambu form of the Higgs field is to apply the same procedure  as for the Dirac monopole and 
assume that $\Phi_{\rm mon}$ in \eqref{Higgs0} should  be used only in the upper hemisphere where it is regular, while in the lower
hemisphere one uses its gauge-transformed version $\tilde{\Phi}_{\rm mon}=e^{i\varphi}\Phi_{\rm mon}$ 
which is regular for $\vartheta\to \pi$. 
The U(1) gauge transformation $e^{i\varphi}$ relating the two gauges is regular in the equatorial transition region. This 
provides a globally regular description of a static and spherically symmetric monopole, 
and it is this approach that we shall adopt. 

We find in what follows that the electroweak theory admits in the spherically symmetric sector 
 an infinite number of solutions describing Abelian magnetic monopoles with the magnetic charge $n=\pm 1, \pm 2,\pm 3, \ldots $ 
 in unites of ${\bm \hbar} {\bm c}/(2{\bm  e})$. These are the Dirac monopoles embedded into the electroweak theory. 
 In addition, there is 
 one solution describing the non-Abelian monopole  with $n=\pm 2$, 
which   is  the CM monopole. 
These solutions were know previously, 
but this shows that in the spherically symmetric sector there are no other monopoles. 
Finally, we discover spherically symmetric oscillating solutions which have a finite energy and 
can be purely electric or neutral.

Our main interest  in this text is to analyze the stability of the electroweak monopoles.
To the best of our knowledge, this problem has not been addressed before, even  for the Abelian electroweak monopoles, 
although the stability of Dirac monopoles within the pure Yang-Mills theory has been considered  \cite{Yoneya:1977yi,Brandt:1979kk}. 
The mathematical aspects of the CM monopole have been studied by Yang who gave the
existence proof  for this solution \cite{yang2014solitons}, but there are no topological or some other arguments to insure that 
the CM monopole corresponds to an energy  minimum. 
At best, one may argue that the CM monopole minimizes the energy in the spherically symmetric sector, 
but this does not guarantee that it minimizes the energy 
  with respect to arbitrary deformations as well. At the same time, even though there are no topological arguments for its stability, 
the CM monopole may be stable dynamically. 
This issue can be clarified   by applying the perturbation theory, 
as was  the case for the t'Hooft-Polyakov  monopole away from the Bogomol'nyi limit \cite{Baacke:1990at}. 

In what follows we analyze the stability of the electroweak monopoles 
with respect to arbitrary perturbations within the linear perturbation theory. 
We use the method of  complex spacetime tetrad and spin-weighted spherical harmonics 
to separate variables in the perturbation equations, assuming the harmonic time dependence $e^{\pm i\omega t}$ 
for the perturbations.  This yields a system of 20 ordinary differential equations 
for 20 functions of the radial coordinate, which describe perturbations of the SU(2)$\times$U(1) gauge field 
and of the Higgs field. Imposing the background gauge condition reduces the number of equations to 16, which 
further  split into two independent parity groups. In each parity group the equations assume the form 
of a symmetric multi-channel eigenvalue problem to determine  $\omega^2$. If there are bound state solutions 
 with $\omega^2<0$ (negative modes), then the background monopole configuration is unstable. 

By applying  the Jacobi criterion, we check that the spectra of perturbations around the CM monopole do not contain 
negative modes in sectors with angular momentum $j=0,1,2,3$. 
Since an instability in larger $j$ sectors is unlikely due to the high centrifugal barrier, 
this strongly indicates that this solutions is stable. 
Of course, this only concerns stability with respect to {\it small} perturbations. 
We also find that the $n=\pm 1$ Abelian 
monopole is stable, but all Abelian monopole with $|n|>1$ are unstable with respect to perturbations in the sector with $j=|n|/2-1$.
Since the Abelian $|n|=2$ monopole is unstable only in the $j=0$ sector while the  CM monopole is stable and also has $|n|=2$, 
it is conceivable that the CM monopole can be viewed as a stable remnant of the decay of the Abelian monopole. 
One may similarly conjecture that stable remnants exist also for monopoles with $|n|>2$, hence the CM monopole may be just the 
first member of a  sequence of non-Abelian monopole solutions labeled by their magnetic charge $n$. 
 Only the CM monopole is spherically symmetric while the non-Abelian monopoles with $|n|>2$ should not be rotationally invariant. 
 
 The rest of the text is organized as follows. After describing the electroweak theory in Section II, we present the 
 spherically symmetric ansatz in Section III,  describe the spherically symmetric solutions in Section IV, 
 and discuss spherically symmetric  perturbations in Section~V.  
 Generic perturbations and separation of variables in the perturbation equations are described in Section VI, 
 while Section VII presents the gauge fixing procedure, the discussion of the residual gauge freedom, and splitting into 
 two parity groups. The results of the stability analysis are presented in Section VIII and summarized in 
 Section IX. The desingularization procedure for the spherically symmetric ansatz is described in Appendix A, 
 the electroweak equations in the spherically symmetric case are derived in Appendix B, 
 while the perturbation equations after the variable separation and their Shroedinger form are shown in Appendix C and Appendix D.

\section{ELECTROWEAK THEORY}
\setcounter{equation}{0}

The dimensionful action of the bosonic part of the electroweak theory of Weinberg and Salam (WS) can be represented in the form 
\be                                     \label{00}
{\bf S}=\frac{1}{\bm{ c\,g}_0^2} \int {\cal L}_{\rm WS}\,\sqrt{-{\rm g}}\, d^4 x
\ee
with the Lagrangian 
\be                             \label{L}
{\cal L}_{\rm WS}=
-\frac{1}{4g^2}\,\WW^a_{\mu\nu}\WW^{a\mu\nu}
-\frac{1}{4g^{\prime 2}}\,{\F}_{\mu\nu}{\F}^{\mu\nu}
-(D_\mu\Phi)^\dagger D^\mu\Phi
-\frac{\beta}{8}\left(\Phi^\dagger\Phi-1\right)^2,
\ee
where  all fields and couplings  as well as the spacetime coordinates $x^\mu$ and metric ${\rm g}_{\mu\nu}$ 
are rendered dimensionless by rescaling. 
The Abelian U(1) and non-Abelian SU(2) field strengths are 
\begin{align}
{\F}_{\mu\nu}=\partial_\mu{\A}_\nu
-\partial_\nu{\A}_\mu\, ,~~~~~~
\WW^a_{\mu\nu}=\partial_\mu\WW^a_\nu
-\partial_\nu \WW^a_\mu
+\epsilon_{abc}\WW^b_\mu\WW^c_\nu\, ,~~~~~
\end{align}
while  Higgs field $\Phi$
is in the fundamental 
representation of SU(2) with the covariant derivative 
\begin{align}
D_\mu\Phi
&=\left(\partial_\mu-\frac{i}{2}\,{\A}_\mu
-\frac{i}{2}\,\tau^a \WW^a_\mu\right)\Phi\,, \label{unbold} 
\end{align}
where $\tau^a$ are the Pauli matrices.
The two coupling constants are 
$g=\cos\thetaw$ and
$g^\prime=\sin\thetaw$ where the physical value of the Weinberg angle is
$
g^{\prime 2}=\sin^2\thetaw=0.23. 
$

The dimensionful (boldfaced) parameters appearing in the action \eqref{00} are the speed of light ${\bm c}$ and also ${\bm g}_0$ 
related to the electron charge ${\bm e}$, 
\be
\frac{\bm e^2}{4\pi\bm \hbar \bm c}=\frac{\bm \hbar {\bm c}}{4\pi}\,{\left(gg^\prime{\bm g}_0\right)^2}
\approx \frac{1}{137}.
\ee
The dimensionful fields often used in the literature   are 
${\bm \A}_\mu=({\mbox{\boldmath $\Phi$}_0}/g^\prime)\A_\mu$,  
${\bm W}^a_\mu=({\mbox{\boldmath $\Phi$}_0}/g)\WW^a_\mu$, and 
${\mbox{\boldmath $\Phi$}}={\mbox{\boldmath $\Phi$}_0}\Phi$ where 
$\mbox{\boldmath $\Phi$}_0=246$ GeV 
is the Higgs field vacuum expectation value. The dimensionful coordinates are 
${\bm x}^\mu={\bm L}_{\rm WS}\,x^\mu$  with  the electroweak  length scale  ${\bm L}_{\rm WS}=1/({\bm g}_0{\bm \Phi}_0)=1.52\times 10^{-16}$ cm.  

The theory  is invariant under 
SU(2)$\times$U(1) gauge transformations
\be                               \label{gauge}
\Phi\to {\rm U}\Phi,~~~~~~~~
{\cal W}\to {\rm U}{\cal W}{\rm U}^{-1}
+i{\rm U}\partial_\mu {\rm U}^{-1}dx^\mu\,,
\ee
with 
\be                            \label{U}
{\cal W}=
\frac12\, (B_\mu+\tau^a\WW^a_\mu)\, dx^\mu\,,~~~~~~~~~
{\rm U}=\exp\left(\frac{i}{2}\,\Theta+\frac{i}{2}\,\tau^a\theta^a\right),
\ee
where $\Theta$ and $\theta^a$ are functions of $x^\mu$. 
Varying the action 
gives the equations,
\begin{align}
\nabla^\mu {B}_{\mu\nu}&=g^{\prime 2}\,\frac{i}{2}\,
(\Phi^\dagger D_\nu\Phi -(D_\nu\Phi)^\dagger\Phi
)\equiv g^{\prime 2}J_\nu,\nn 
\\
{\cal D}^\mu \WW^a_{\mu\nu}
&=g^{2}\,\frac{i}{2}\,
(
\Phi^\dagger\tau^a D_\nu\Phi
-(D_\nu\Phi)^\dagger\tau^a\Phi
)
\equiv g^{2} J^{a}_{\nu}, \nn 
\\
D_\mu D^\mu\Phi&-\frac{\beta}{4}\,(\Phi^\dagger\Phi-1)\Phi=0,      \label{P2}
\end{align}
with ${\cal D}_\mu\WW^a_{\alpha\beta}=\nabla_\mu \WW^a_{\alpha\beta}
+\epsilon_{abc}\WW^b_\mu\WW^c_{\alpha\beta}$ where $\nabla_\mu$ is the geometrical covariant derivative with respect 
to the spacetime metric. 
Varying the action with respect to the latter determines the energy-momentum tensor 
\be                      \label{TT}
T_{\mu\nu}=
\frac{1}{g^2}\,\WW^a_{~\mu\sigma}\WW^{a~\sigma}_{~\nu}
+\frac{1}{g^{\prime\,2}}B_{\mu\sigma}B_\nu^{~\sigma}
+(D_\mu\Phi)^\dagger D_\nu\Phi
+(D_\nu\Phi)^\dagger D_\mu\Phi+g_{\mu\nu}\mathcal{L}_{\rm WS}\,.
\ee

The vacuum is defined as the configuration with $T_{\mu\nu}=0$. Modulo gauge transformations, it can be chosen as 
\be
\WW^a_\mu=\A_\mu=0,~~~~~~ 
 \Phi=\begin{pmatrix}
0  \\
1
\end{pmatrix}.
\ee 
Allowing for small fluctuations around the vacuum and 
linearising the field equations 
with respect to the fluctuations gives the perturbative mass spectrum 
containing the massless photon 
and the massive Z, W and Higgs bosons with dimensionless masses
\be                                   \label{masses}
\mz=\frac{1}{\sqrt{2}},~~~~~~
\mw=g\mz,~~~~~~
\mh=\sqrt{\beta}\,\mz.
\ee
 Multiplying  these by 
${\bm e}\mbox{\boldmath $\Phi$}_0/(gg^\prime)$
gives the dimensionfull masses, 
for example one has
$
{\bm m}_{\mbox{\tiny Z}}{\bm c}^2=
{\bm e}\mbox{\boldmath $\Phi$}_0/(\sqrt{2}gg^\prime)
\approx 91~ {\rm GeV}. 
$
Using the Higgs  mass 
${\bm m}_{\mbox{\tiny H}}{\bm c}^2\approx 125$ GeV
yields the value $\beta\approx 1.88$. 

Summarizing, the dimensionless parameters in the equations are 
\be
g^{\prime 2}=0.23,~~~~g^2=1-g^{\prime 2},~~~~\beta=1.88. 
\ee
We shall adopt  the definition  of Nambu for the 
electromagnetic and Z fields \cite{Nambu:1977ag}, 
\be                                  \label{Nambu}
F_{\mu\nu}=\frac{g}{g^\prime}\,  
\A_{\mu\nu}+\frac{g^{\prime}}{g}\,n^a\WW^a_{\mu\nu}\,,~~~~~~
{Z}_{\mu\nu}=\A_{\mu\nu}-n^a\WW^a_{\mu\nu}\,,
\ee
where 
$
n^a=\Phi^\dagger\tau^a\Phi/(\Phi^\dagger\Phi). 
$
This definition is not unique \cite{Coleman:1985rnk}
but it gives more satisfactory results \cite{Hindmarsh:1993aw}
than the other known definitions \cite{tHooft:1974kcl}. 
In general, away from the Higgs vacuum, 
the 2-forms \eqref{Nambu} are  not closed and do not admit potentials, 
however, there is no reason 
why the Maxwell equations should hold off the Higgs vacuum.

\section{SPHERICAL SYMMETRY}
\setcounter{equation}{0}
To describe spherically symmetric fields, 
it is convenient to represent the background Minkowski metric as
\be                   \label{metr}
{\rm g}_{\mu\nu}dx^\mu dx^\nu =-dt^2+dr^2+r^2\left(d\vartheta^2+\sin^2\vartheta d\varphi^2\right),
\ee
which is invariant under the action of the SO(3) spatial rotations. 

The spherically symmetric  gauge fields should be  invariant under the combined action of the spatial rotations and 
gauge transformations \cite{Forgacs:1979zs}. 
Let $\T_a=\frac12\, \tau_a$ be the SU(2) gauge group generators such that $[\T_a,\T_b]=i\epsilon_{abc}\T_c$. 
The SO(3)-invariant SU(2) gauge field $W=\T_a W^a_\mu dx^\mu$ can be represented in the form 
that generalizes the well-known ansatz of Witten \cite{Witten:1976ck},
 \be                \label{W2}
W=\T_3\,(a_0\, dt+a_1\, dr)&+&\left(w_2\,T_1+w_1 \T_2\right) d\vartheta \nn \\
&+&\nu\left(w_2\,\T_2-w_1 T_1\right)\sin\vartheta \,d\varphi+\T_3\,\nu \cos\vartheta d\varphi.~~~~~~~
\ee
Here $a_0,a_1,w_1,w_2$ are functions of $t,r$ and $\nu$ is a constant parameter. Written in this gauge, 
the field is singular at the $z$-axis (we call it singular gauge),  but the singularity can be removed 
if $\nu\in \mathbb{Z}$, as explained in Appendix A. In addition, 
if $w_1=w_2=0$ then half-integer values of $\nu$ are allowed as well, 
$2\nu\in \mathbb{Z}$. One could  in principle always work in the regular gauge described by Eq.\eqref{W1} in Appendix A,
but the gauge \eqref{W2} is much easier to use because nothing depends on the azimuthal angle $\varphi$. 
Since the field equations are gauge invariant, it suffices to use \eqref{W2}  and 
show that its singularity 
 can be gauged away, as explained in Appendix A.
 

The spherically symmetric 
 U(1) gauge field is 
\be              \label{B}
B_\mu dx^\mu=b_0\, dt+ b_1\,dr+P\cos\vartheta\, d\varphi\,,
\ee 
where $b_0$ and $b_1$ depend on $t,r$ and $P$ is a constant. 
This field is also singular at the $z$-axis, but its singularity can also be gauged away, as shown 
in Appendix A. 

Finally, the spherically symmetric Higgs field is 
\be               \label{Higgs}
 \Phi=\begin{pmatrix}
0 \\
\phi\, e^{i\xi/2} 
\end{pmatrix}=
\phi\, e^{i\xi/2} 
\begin{pmatrix}
0 \\
1
\end{pmatrix}, 
\ee
where $\phi$ and $\xi$ depend on $t,r$. This is the gauge-transformed version of the monopole field in \eqref{Higgs0}.
Specifically, Eq.\eqref{Higgs0} corresponds to the gauge where the SU(2) field is regular and given by Eq.\eqref{W1}, while 
\eqref{Higgs} shows the same thing when the SU(2) field is written in the singular gauge \eqref{W2} (see Appendix A). 
For comparison, the similarly  gauge-transformed version of the  sphaleron  field  from  \eqref{Higgs0} 
is given by $\Phi_{\rm sph}^{\rm sing}$ in 
\eqref{Higgsa}, in which case the angular dependence separates only for  $\thetaw=0$ when  the U(1) field  is absent.

On the other hand, injecting \eqref{W2}--\eqref{Higgs} 
to the WS equations \eqref{P2}, the angular dependence separates for arbitrary $\thetaw$ yielding 
differential equations for functions $a_0,a_1,w_1,w_2,b_0,b_1,\phi,\xi$
depending on $t,r$. One also obtains  conditions for the parameters $\nu$ and $P$. All these are shown in Appendix~B.  

The equations can be simplified by noting that 
the ansatz  \eqref{W2}--\eqref{Higgs} preserves its form 
under gauge transformations \eqref{gauge} generated by 
\be
{\rm U}=\exp\left( \frac{i}{2} \,\lambda+\frac{i}{2}\,\gamma\,\tau_3     \right)
\ee
where $\lambda,\gamma$ depend on $t,r$. Its 
effect on the 8 field amplitudes is 
\bea               \label{gt}
&&w_1\to w_1\,\cos\gamma-w_2\,\sin\gamma,~~~~w_2\to w_1\,\sin\gamma+w_2\,\cos\gamma, ~~~a_0\to a_0+\dot{\gamma},\\
&& a_1\to a_1+\gamma^\prime,~~~
b_0\to b_0+\dot{\lambda},~~~~b_1\to b_1+\lambda^\prime,~~~~\phi\to\phi,~~~~~{\xi\to \xi+\lambda-\gamma},~~~\phi\to\phi.   \nn
\eea
Setting $w_1=f\,\cos\alpha$ and $w_2=f\,\sin\alpha$ so that $f\to f$, $\alpha\to \alpha+\gamma$, 
the following 4 combinations 
\be           \label{gt1}
\Omega_0=a_0-\dot{\alpha},~~~~\Omega_1=a_1-\alpha^\prime,~~~~\Theta_0=a_0-b_0+\dot{\xi},~~~~\Theta_1=a_1-b_1+\xi^\prime,~~~~~
\ee
and also $f,\phi$ do not change under gauge transformations. 
As shown in Appendix B, the equations can be expressed only in terms of these 6 gauge-invariant amplitudes, while $\alpha,\xi$ 
correspond to pure gauge degrees of freedom and drop out form the equations. 

 One obtains in this way second order differential equations 
for functions depending on $t,r$, but also first order differential constraints and
algebraic constraints (see Appendix B). 
The first algebraic constraints reads 
\be
(P-\nu)\,\phi^2=0, 
\ee
and 
since the Higgs amplitude $\phi$ cannot vanish identically because it should approach the unit value at infinity (Higgs vacuum),
it follows that one should set 
\be               \label{Pnu}
{P=\nu.}
\ee
After this the second algebraic constraint reduces to 
\be     \label{PP}
(\nu^2-1) f=0,
\ee
hence either $\nu^2=1$ or $f=0$. Let us first consider the option in which  the $f$-amplitude can be non-trivial,
\be         \label{PP1}
{P=\nu=\pm 1.}
\ee 
The equations for the $f$ and $\phi$ then read 
 \bea          \label{fp}
f^{\prime\prime}
-\ddot{f}
&=&\left(\frac{f^2-1}{r^2}
+\Omega_1^2-\Omega_0^2
+\frac{g^2}{2}\,\phi^2 \right)f\,,\nn \\
\frac{1}{r^2}\,(r^2\phi^\prime)^\prime
-\ddot{\phi}
&=&\frac{1}{4} \left(\beta(\phi^2-1)+\Theta_1^2-\Theta_0^2+\frac{2}{r^2}\, f^2\right) \phi\,,
\eea
while the equations for the electric amplitudes are 
\bea             \label{el}
\left(r^2(\Omega_0^\prime -\dot{\Omega}_1)\right)^\prime 
&=&
2 f^2 \,\Omega_0+\frac{g^2}{2}\,  r^2\phi^2\Theta_0 \,,\nn \\
r^2\left(\Omega_0^\prime -\dot{\Omega}_1\right)^{\mbox{$\cdot$}}
&=&
2 f^2\Omega_1+\frac{g^2}{2}\,r^2\phi^2 \Theta_1 \,,\nn \\
\left(r^2(\Theta_0^\prime -\dot{\Theta}_1)\right)^\prime 
&=&2 f^2 \Omega_0+
\frac12\, r^2 \phi^2 \Theta_0  \,,\nn \\
r^2\left(\Theta_0^\prime -\dot{\Theta}_1\right)^{\mbox{$\cdot$}}
&=&2 f^2 \Omega_1+
\frac12\, r^2 \phi^2 \Theta_1 \,.
\eea
In addition there are two first order differential conditions 
\bea           \label{con}
\left(f^2 \Omega_1\right)^\prime=\left(f^2\Omega_0\right)^{\mbox{$\cdot$}}\,,~~~~~~~~
\left(r^2 \phi^2 \Theta_1\right)^\prime=r^2\left(\phi^2 \Theta_0\right)^{\mbox{$\cdot$}}\,, 
\eea
but these are not independent and follow form \eqref{el}
since the $t$-derivatives of the first and third equations in \eqref{el} should coincide 
with the $r$-derivatives 
of the second and fourth equations. 

To the best of our knowledge, Eqs.\eqref{fp}--\eqref{con} have never been described in the literature. 
Although $\nu$ does not appear explicitly in these equations, they are valid only for $\nu=\pm 1$ if $f\neq 0$. 
 If $\nu\neq\pm 1$ then one should according to \eqref{PP} set everywhere  $f=0$ 
 (the $\nu=\pm 1$ values  are then also allowed).

We shall also need the expression for the energy, 
\be
E=4\pi \int_0^\infty r^2 T_{00} \,dr=4\pi \int_0^\infty \left({\cal E}_\phi
+\frac{{\cal E}_f }{g^2}\,
+\frac{{\cal E}_0}{g^{\prime 2}}\,+{\cal E}_\nu
\right)\,dr\,,
\ee
where 
\bea        \label{E}
{\cal E}_f&=&U_f+U_\Omega
+\frac{r^2}{2}\left(\Omega_0^\prime-\dot{\Omega}_1\right)^2\,,\nn \\
{\cal E}_\phi&=&r^2(U_\phi+U_\Theta)+\frac{f^2\phi^2}{2}+\frac{\beta\,r^2}{8}(\phi^2-1)^2\,,\nn \\
{\cal E}_0&=&\frac{r^2}{2}\left(\Omega_0^\prime-\dot{\Omega}_1-\Theta_0^\prime+\dot{\Theta}_1\right)^2\,, \nn \\
{\cal E}_\nu&=&\frac{1}{2g^2 r^2}\left(\frac{\nu^2}{g^{\prime 2}}-1+(f^2-1)^2 \right), 
\eea
and also 
\bea             \label{Ua}
U_f&=&f^{\prime 2}+\dot{f}^2\,,~~
~~~~~~~~~~~~~U_\phi=\phi^{\prime 2}+\dot{\phi}^2\,,\nn \\
U_\Omega&=&\left(\Omega_1^2+\Omega_0^2\right)f^2\,,
~~~~~~U_\Theta=\left(\Theta_1^2+\Theta_0^2\right)\frac{\phi^2}{4}.
\eea
We notice that  $\nu$ appears explicitly  
 in the ${\cal E}_\nu$ term in \eqref{E}. 
According to \eqref{PP}, one should set $\nu=\pm 1$ if $f\neq 0$, in which case
\be
{\cal E}_\nu=\frac{(f^2-1)^2}{2 g^2 r^2}+\frac{1}{2g^{\prime 2} r^2}. 
\ee
If $\nu\neq\pm 1$ then $f=0$ and 
\be
{\cal E}_\nu=\frac{\nu^2}{2 g^2 g^{\prime 2} r^2}.
\ee
The ${\cal E}_\nu$-term renders the energy  divergent, unless if $\nu=0$. This divergence is 
due to the presence of a pointlike magnetic charge. 

Let us see what is known about solutions of these equations.

\section{SOLUTIONS}
\setcounter{equation}{0}

Let us assume that nothing depends on time. Then the constraints \eqref{con} imply that
\be
\Omega_1 =\frac{C_1}{f^2},~~~~~~~~~\Theta_1=\frac{C_2}{\phi^2 r^2}\,,
\ee
where $C_1,C_2$ are integration constants. 
Injecting this into the second and forth equations in \eqref{el}, the left hand sides of the equations vanish,
while the right hand sides reduce to 
\be
C_1+\frac{g^2}{4}\, C_2=0,~~~~~~C_1+\frac14\, C_2=0,
\ee
hence $C_1=C_2=0$ and $\Omega_1=\Theta_1=0$. 
 Eqs.\eqref{fp},\eqref{el} reduce to 
\bea           \label{e01}
f^{\prime\prime}&=&\left(\frac{f^2-1}{r^2}-\Omega_0^2+\frac{g^2}{2}\,\phi^2 \right)f\,, \nn \\
\frac{1}{r^2}\,(r^2\phi^\prime)^\prime&=&\frac14 \left(\beta(\phi^2-1)-\Theta_0^2+\frac{2 f^2}{r^2}\right)\phi \,, \nn \\
\frac{1}{r^2}\left( r^2 \Omega_0^\prime\right)^\prime &=&\frac{2f^2}{r^2}\,\Omega_0+\frac{g^2\phi^2}{2}\,\Theta_0\,, \nn \\
\frac{1}{r^2}\left( r^2 \Theta_0^\prime\right)^\prime &=&\frac{2f^2}{r^2}\,\Omega_0+\frac{\phi^2}{2}\,\Theta_0\,.
\eea

We shall be mainly interested in 
the purely magnetic case when the electric field is absent. Setting $\Omega_0=\Theta_0=0$, the equations reduce to 
\bea           \label{e1}
f^{\prime\prime}&=&\left(\frac{f^2-1}{r^2}+\frac{g^2}{2}\,\phi^2 \right)f\,, \nn \\
\frac{1}{r^2}\,(r^2\phi^\prime)^\prime&=&\frac14 \left(\beta(\phi^2-1)+\frac{2 f^2}{r^2}\right)\phi \,
\eea
with the energy 
\be
E=4\pi \int_0^\infty \left(
{\cal B}+\frac12\,{\cal A}
\right) dr
\ee
where 
\be           \label{e2}
{\cal B}=\frac{1}{g^2} f^{\prime 2} +r^2\phi^{\prime 2},~~~~~
{\cal A}=\frac{(f^2-1)^2}{g^2\, r^2}+ f^2\phi^2 +\frac{\beta r^2}{4}\,(\phi^2-1)^2
+\frac{\nu^2-g^{\prime 2}}{g^2 g^{\prime 2} r^2}. 
\ee

\subsection{Abelian electroweak monopoles}
The simplest solution of \eqref{e1} is given by 
\be
f=0,~~~~\phi=1,~~~~~2\nu\in \mathbb{Z}.
\ee
Notice that, since $f=0$, the parameter $\nu$ can assume both integer and half-integer values
(see Appendix A).
The U(1) and SU(2)  gauge fields  are 
\be
B=\nu\cos\vartheta\, d\varphi,~~~~~W=\T_3\, B,
\ee
which define the electromagnetic field \eqref{Nambu} whose potential
(passing for a moment to the dimensionful quantities like ${\bf x^\mu}=x^\mu/({\bm \Phi}_0{\bf g}_0)$) reads 
\be
{\bm A}={\bm A}_\mu d{\bf x^\mu}&=&{\bm \Phi}_0 \left(\frac{g}{g^\prime}\,B_\mu+\frac{g^\prime }{g}\, W^3_\mu\right) 
\frac{dx^\mu}{\bm \Phi_0 {\bm g}_0} \nn \\
&=&\frac{\nu}{gg^\prime{\bm g}_0}\, \cos\vartheta\, d\varphi=\frac{\hbar \bm c}{\bm e}\, \nu \cos\vartheta\,d\varphi\equiv {\bm P}\cos\vartheta\,d\varphi\,.
\ee
This is the potential of the Dirac monopole with the  magnetic charge
\be
{\bm P}=\frac{\hbar \bm c}{\bm e}\,\nu\,.
\ee
Applying the gauge transformations \eqref{toreg}, the potential ${\bm A}$ can be transformed  into two 
locally regular forms ${\bm A}_\pm$. 
Since $\nu$ can be integer or half-integer, 
\be       \label{nu}
n\equiv 2\nu\in \mathbb{Z}, 
\ee
the magnetic charge automatically fulfills 
the Dirac quantization condition 
\be
\frac{\bm{ e P}}{\hbar \bm c}=\frac{n}{2}~~~~\text{with}~~~~n=\mathbb{Z}. 
\ee
Of course, this is not a coincidence but the consequence of \eqref{nu}, which in turn follows from the standard argument 
leading to the charge quantization described  by Eqs.\eqref{W2a}--\eqref{W2c} in Appendix A. 

As a result, we  obtain  the Abelian Dirac  monopoles embedded into the electroweak theory. 
These solutions can be generalized to include an electric field, since Eqs.\eqref{e01} are solved by 
$f=\Theta_0=0$, $\phi=1$, $\Omega_1=Q/r$, hence 
\be
B=\frac{Q}{r}\, dt+\nu\cos\vartheta d\varphi,~~~~~W=\T_3\, B, 
\ee
which corresponds to the dyon with the electric charge $Q$ and magnetic charge $n=2\nu$.

\subsection{The non-Abelian monopole of Cho and Maison}

\begin{figure}
    \centering

    \includegraphics[scale=0.75]{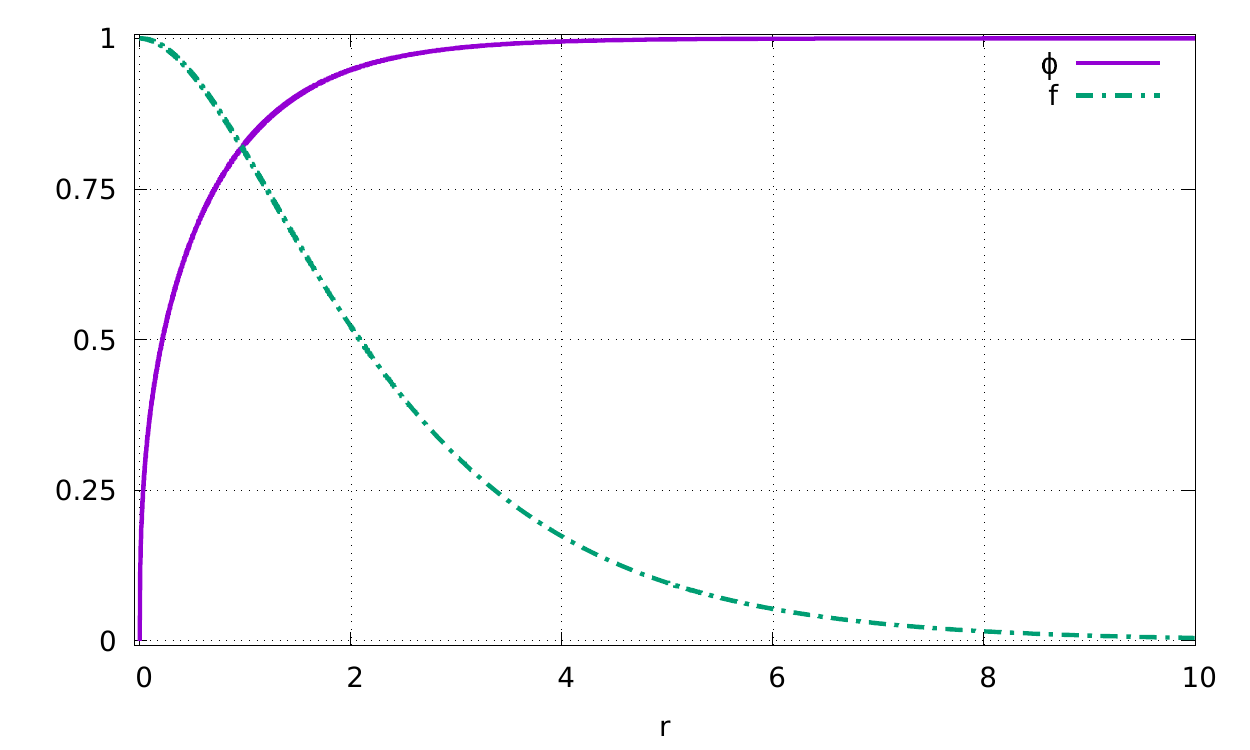}

    \caption{Profiles of $f(r)$ and $\phi(r)$ for the CM monopole.}
    \label{cho}
\end{figure}

A less trivial  situation  arises for $\nu=\pm 1$, hence for the magnetic charge $n=\pm 2$. 
Eqs.\eqref{e1} then admit a smooth solution for which the amplitudes $f,\phi$ interpolate between 
the following asymptotic values:
\be
f=1+{\cal O}(r^2),~~~~~ \phi={\cal O}(r^\delta)~~~~~~\text{for}~~~~~r\to 0
\ee
with $\delta=(\sqrt{3}-1)/2$ and 
\be
f={\cal O}\left(e^{-\mw r} \right),~~~~~ \phi=1+{\cal O}\left(e^{-\mh r} \right)~~~~~~\text{for}~~~~~r\to \infty.
\ee
This solution can be obtained numerically, see Fig.\ref{cho}, and its existence was proven in \cite{yang2014solitons}. 
At infinity the fields approach those for the magnetic monopole with $n=\pm 2$, while at the 
origin the non-Abelian fields are regular and their contribution to the energy is finite.
However, the U(1) contribution to the energy is infinite. 
The energy can be represented as $E=E_0+E_1$ where $E_1$ is finite but $E_0$ diverges, 
\be               \label{Energy}
    E_0&=&\frac{2\pi}{g'^2}\int_0^\infty{\frac{dr}{r^2}},\nn \\
    E_1&=&4\pi\int_0^\infty dr\left(\frac{1}{g^2}\left(\frac{(f^2-1)^2}{2r^2}+f'^2\right)+(r\phi')^2+\frac{r^2\beta}{8}(\phi^2-1)^2+\frac{1}{2}(f\phi)^2\right). 
\ee
This solution can be generalized  by adding the electric amplitudes $\Omega_0$ 
and $\Theta_0$ and solving Eqs.\eqref{e01}.

\subsection{Oscillating  solutions}
The only possibility to obtain  finite energy solutions is to choose $\nu=0$, in which case the singular at the origin 
term ${\cal E}_\nu$ in \eqref{E} vanishes. One should then set $f=0$ and Eqs.\eqref{fp},\eqref{el}
reduce to 
 \bea          \label{fp1}
\frac{1}{r^2}\,(r^2\phi^\prime)^\prime
-\ddot{\phi}
&=&\frac{1}{4} \left(\beta(\phi^2-1)+\Theta_1^2-\Theta_0^2\right) \phi\,,\nn \\
\frac{1}{r^2}\left(r^2(\Omega_0^\prime -\dot{\Omega}_1)\right)^\prime 
&=&
\frac{g^2}{2}\,  \phi^2\Theta_0 \,,~~~~~
\left(\Omega_0^\prime -\dot{\Omega}_1\right)^{\mbox{$\cdot$}}
=
\frac{g^2}{2}\,\phi^2 \Theta_1 \,,\nn \\
\frac{1}{r^2}\left(r^2(\Theta_0^\prime -\dot{\Theta}_1)\right)^\prime 
&=&
\frac12\, \phi^2 \Theta_0  \,,~~~~~~
\left(\Theta_0^\prime -\dot{\Theta}_1\right)^{\mbox{$\cdot$}}
=
\frac12\, \phi^2 \Theta_1 \,.
\eea
One can show that non-trivial static solutions of these equations have infinite energy.
However, there are  time-dependent solutions with a finite energy. 
Solutions of this type  
were previously studied assuming the sphaleron anstaz \eqref{Higgs0} for the Higgs field, 
in which case the fields are not spherically symmetric, unless for $\thetaw=0$ \cite{Ratra:1987dp,Farhi:2005rz,Graham:2006vy,Graham:2007ds}. 
On the other hand, Eqs.\eqref{fp1} describe spherically symmetric systems for any $\thetaw$.  

Let us set the  gauge fields to zero, $\Omega_0=\Omega_1=\Theta_0=\Theta_1=0$. 
Rescaling the spacetime coordinates via   $t\to 2/\sqrt{\beta}\, t$ and $r\to 2/\sqrt{\beta}\, r$, the remaining 
Higgs equation reduces to 
\be
\frac{1}{r^2}\,(r^2\phi^\prime)^\prime
-\ddot{\phi}=(\phi^2-1)\phi. 
\ee
This equation has been extensively studied, and it is known that it describes {\it oscillons} --  oscillating 
quaziperiodic configurations with  a finite energy \cite{Copeland:1995fq,Honda:2001xg,Fodor:2006zs}. 
Oscillons are  well-localized in space during a certain period of time but finally they 
decay into a pure radiation. However, their  lifetime, that is the period when they remain localized, can be very large, depending on the initial values. 
In principle, their lifetime can be comparable with the age of the universe. 
We therefore discover that the electroweak theory admits spherically symmetric oscillons, which, to the best of our knowledge, 
has never been observed before. We shall present separately their detailed  analysis 
as well as more general solutions 
of \eqref{fp1} \cite{GGV}.

\section{SPERICALLY SYMMETRIC PERTURBATIONS}
Consider small perturbations around a static and purely magnetic background 
\bea              \label{p1}
f(r)\to {f}(r)+\delta f(t,r),~~~\phi(r)\to {\phi}(r)+\delta \phi(t,r),~
\eea
and also 
\bea              \label{p2}
\Omega_0(t,r)&=&\delta \Omega_0(t,r),~~~~
\Omega_1(t,r)=\delta \Omega_1(t,r),~~\nn \\
\Theta_0(t,r)&=&\delta \Theta_0(t,r),~~~~
\Theta_1(t,r)=\delta \Theta_1(t,r). 
\eea
Injecting this to Eqs.\eqref{fp}--\eqref{el} and linearizing with respect to the perturbations,
the linearized equations split into two independent groups containing, respectively, the perturbation 
described by \eqref{p1}
and those  in \eqref{p2}.

\subsection{Perturbations  of the CM monopole} 

Setting
\be
\delta f(t,r) = e^{i\omega t}\, \psi_1(r),~~~~~~\delta \phi(t,r)=\frac{1}{g r}\, e^{i\omega t}\, \psi_2(r),
\ee
injecting to \eqref{fp} and linearizing yields the two-channel eigenvalue problem 
\bea                \label{eig}
\left(-\frac{d^2}{dr^2}+\hat{V}\right)\Psi=\omega^2 \Psi~~~~~\text{with}~~~~\Psi=\begin{pmatrix}
\psi_1  \\
\psi_2
\end{pmatrix},
\eea
where the potential $\hat{V}$ is a symmetric $2\times 2$ matrix with components 
\be                 \label{eig1}
V_{11}&=&\frac{g^2}{2}\,\phi^2+\frac{3f^2-1}{r^2},~~~~~
V_{22}=
\frac{\beta}{4}\, (3\phi^2-1)+\frac{f^2}{2r^2}, ~~~~~
V_{12}=V_{21}=g \frac{f\phi}{r}\,. 
\ee
Linearizing similarly Eqs.\eqref{el},\eqref{con} with respect to perturbations and 
setting 
\be           \label{XXXr}
\delta\Omega_1(t,r)&=&\,\frac{g\, e^{i\omega t}}{f}\,\psi_1(r),~~~~~~~
~~~~~~\delta\Theta_1(t,r)=\frac{2\, e^{i\omega t}}{r \phi}\,\psi_2(r), \nn \\
\delta\Omega_0(t,r)&=&\frac{g\, e^{i\omega t}}{i\omega}\,\frac{1}{f^2}\left( f\psi_1\right)^\prime,~~~~~~
\delta\Theta_0(t,r)=\frac{2\, e^{i\omega t}}{i\omega}\,\frac{1}{r^2\phi^2}\left( r\phi\, \psi_2\right)^\prime
\ee
yields again the eigenvalue problem of the form \eqref{eig} with the following components of the potential matrix
\bea                \label{eig2}
V_{11}&=&\frac{f^2+1}{r^2}+2\,\frac{f^{\prime 2}}{f^2}-\frac{g^2\phi^2}{2}\,, ~~~~~
V_{12}=V_{21}=g\frac{ f\phi}{r}\,,\nn \\
V_{22}&=&2\left(\frac{(r\phi)^\prime}{r\phi}\right)^2+\frac{\phi^2}{2}+\frac{\beta}{4}\,(1-\phi)^2-\frac{f^2}{2r^2}\,.
\eea
Let us assume $f,\phi$ to be those for the CM monopole.   One should study 
the 2-channel eigenvalue problems \eqref{eig} with the potential given by either \eqref{eig1} or by \eqref{eig2}
to see if there is a negative part of spectra of $\omega^2$. As will be shown in Section \ref{stab} below, 
the spectra of $\omega^2$ are positive in both cases. 

As a result, the CM monopole is stable with respect to fluctuations within the spherically 
symmetric sector. However, it may be unstable with respect to more general 
perturbations. For example, since it has the magnetic charge 
$n=\pm 2$, it could  in principle split  into two monopoles with $n=\pm 1$ via developing an instability 
in the  axially-symmetric sector. 
Therefore, one should study the more general  perturbations, but this
requires a much more 
involved  analysis to be described in the following Sections.

\subsection{Perturbations  of the Abelian monopole}

For the Abelian monopole  with $\nu=\pm 1$ and $f=0$, $\phi=1$ the above perturbation equations 
do not apply because \eqref{XXXr} contains $f$ in the denominator. In addition, using 
 the function $f=\sqrt{w_1^2+w_2^2}$ would make no sense within the 
linear perturbation theory for small $w_1$ and $w_2$. We shall describe in Section \ref{stab} below how to 
handle the problem in a  gauge-invariant way, but for the time being let us return 
to the original amplitudes
$w_1,w_2,a_0,a_1,b_0,b_1,\xi$.  They all vanish  (modulo gauge transformations) for the monopole background,
hence they are small for small fluctuations, 
while  the Higgs field is $\phi=1+\delta \phi$. 

Linearizing Eqs.\eqref{app3}--\eqref{app5},  one obtains three decoupled from each other equations 
\be           \label{BOT0}
w_1^{\prime\prime}-\ddot{w}_1=\left(\frac{g^2}{2}-\frac{1}{r^2} \right)w_1\,,~~~
w_2^{\prime\prime}-\ddot{w}_2=\left(\frac{g^2}{2}-\frac{1}{r^2} \right)w_2\,,~~~
\frac{1}{r^2}\left(r^2 \delta\phi^\prime\right)^\prime-\delta\ddot{\phi}=\frac{\beta}{2}\,\delta\phi,~~~~~
\ee
and five coupled equations 
\be                 \label{BOT00}
\left( r^2(a_0^\prime-\dot{a}_1)\right)^\prime-
\frac{g^2 r^2}{2}\left(a_0-b_0+\dot{\xi} \right)=0\,,&&~~~~
\left( r^2(a_0^\prime-\dot{a}_1)\right)^{\mbox{$\cdot$}}-
\frac{g^2 r^2 }{2}\left(a_1-b_1+\xi^\prime \right)=0\,,\nn \\
 \left( r^2(b_0^\prime-\dot{b}_1)\right)^\prime+\frac{g^{\prime 2} r^2}{2}\left(a_0-b_0+\dot{\xi} \right)=0\,,&&~~~
\left( r^2(b_0^\prime-\dot{b}_1)\right)^{\mbox{$\cdot$}}+\frac{g^{\prime 2} r^2}{2}\left(a_1-b_1+\xi^\prime \right) =0\,,\nn \\
  \left( r^2(a_1-b_1+\xi^\prime) \right)^\prime&=&\left( r^2(a_0-b_0+\dot{\xi}) \right)^{\mbox{$\cdot$}}\,.~~~~~~~~
 \ee
The last equation here can be fulfilled by setting 
\be        \label{BOT}
r^2(a_1-b_1+\xi^\prime) =\dot{\Psi},~~~~~~r^2(a_0-b_0+\dot{\xi})=\Psi^\prime\,,
\ee
which implies that the other four equations in \eqref{BOT00} can  be integrated yielding 
\be
r^2(a_0^\prime-\dot{a}_1)=\frac{g^2}{2}\,\Psi,~~~~~~~r^2(b_0^\prime-\dot{b}_1)=-\frac{g^{\prime 2}}{2}\,\Psi\,,
\ee
hence 
\be       \label{BOT1}
(a_0-b_0)^\prime-(a_1-b_1)^{\mbox{$\cdot$}}=\frac{\Psi}{2\,r^2}\,.
\ee
On the other hand, one obtains from \eqref{BOT} 
\be
a_0-b_0=\frac{\Psi^\prime}{r^2}-\dot{\xi},~~~~~~~a_1-b_1=\frac{\dot{\Psi}}{r^2}-\xi^\prime,
\ee
injecting which to \eqref{BOT1} gives
\be
\left(\frac{\Psi^\prime}{r^2} \right)^\prime-\left(\frac{\dot{\Psi}}{r^2} \right)^{\mbox{$\cdot$}}=\frac{\Psi}{2\,r^2}\,,
\ee
which reduces upon setting $\Psi=e^{i\omega t}\,r\,\psi(r)$ to 
\be           \label{B1}
\left(-\frac{d^2}{dr^2}+\frac{2}{r^2}+\frac12\right)\psi=\omega^2 \psi.
\ee
Setting 
$w_a=e^{i\omega t}\psi_a(r)$ with  $a=1,2$ and  $\delta\phi=e^{i\omega t}\psi_3(r)/r$ 
the three decoupled equations in \eqref{BOT0} reduce to 
\be          \label{B2}
\left(-\frac{d}{dr^2}-\frac{1}{r^2}+\frac{g^2}{2}\right)\psi_a=\omega^2 \psi_a,~~~~a=1,2,
\ee
and to 
\be          \label{B3}
\left(-\frac{d^2}{dr^2}+\frac{\beta}{2}     \right)\psi_3=\omega^2 \psi_3\, .
\ee
As a result, spherically symmetric perturbations of the Abelian monopole split into four independent channels
containing the $Z$-boson described by \eqref{B1},  the complex-valued $W$-boson described by two real equations \eqref{B2},
and the Higgs boson described by \eqref{B3}. 

One might think that only one of the two equations \eqref{B2} describing $w_1$ and $w_2$ should be counted, since 
the local gauge symmetry 
$w_1\to \tilde{w}_1=w_1\,\cos\gamma-w_2\,\sin\gamma$ and 
$
w_2\to \tilde{w}_2= w_1\,\sin\gamma+w_2\,\cos\gamma
$
contained in  \eqref{gt}
 can be used to set $\tilde{w}_2=0$. However, this would require choosing the gauge parameter $\gamma$ such that 
 $\tan\gamma=-w_2/w_1$. Since  $w_1$ and $w_2$ oscillate   around zero, the derivatives of $\gamma$ would then 
 be unbounded. 
 The gauge transformations of the other amplitudes in \eqref{gt}, as for example 
$a_0\to a_0+\dot{\gamma}$ and 
$a_1\to a_1+\gamma^\prime$, would be unbounded too.
 However, within the linear perturbation theory all amplitudes should be small, hence such gauge transformations are 
 not allowed. Therefore, one cannot gauge away one of the two amplitudes $w_1$ and $w_2$ and each of them should be counted. 
 This  will be independently confirmed  by the analysis in Section \ref{stab}.

Notice finally that the perturbation potentials in the $Z$ and Higgs channells are positive, hence $\omega^2>0$. However, 
in the W-channel the potential admits infinitely many bound states with negative 
and arbitrarily large $\omega^2$. 
This issue will be discussed in more detail in Section \ref{stab}. 
As a result,  the 
Abelian  monopole with $n=2\nu= \pm 2$ is unstable with respect to fluctuations 
in the spherically symmetric sector. This  conclusion does not apply to monopoles 
with $\nu\neq\pm 1$ since in that case one has $w_1=w_2=0$ and 
the spherically symmetric $W$-perturbation channel is closed.

\section{GENERIC PERTURBATIONS}

Let us consider small fluctuations around a background configuration $(W^a_\mu,B_\mu,\Phi)$,

\begin{equation}
    W^a_\mu\rightarrow W^a_\mu+\delta W^a_\mu,\quad\quad B_\mu\rightarrow B_\mu+\delta B_\mu,\quad\quad\Phi\rightarrow\Phi+\delta\Phi.
\end{equation}

Inserting to \eqref{P2} and linearizing with respect to $\delta W^a_\mu$, $\delta B_\mu$, $\delta\Phi$ gives the perturbation equations
\begin{align*}
    \nabla_\mu\nabla^\mu\delta B_\nu&-\frac{g'^2}{2}\left(\Phi^\dagger\left(\delta B_\nu+\tau_a \delta W^a_\nu\right)\Phi+2i\left(\delta\Phi^\dagger D_\nu\Phi-(D_\nu\Phi)^\dagger\delta\Phi\right)\right)\\
    &=\nabla_\nu\left(\nabla_\mu\delta B^\mu-\frac{ig'^2}{2}\left(\delta\Phi^\dagger\Phi-\Phi^\dagger\delta\Phi\right)\right),\\
    \mathcal{D}_\mu\mathcal{D}^\mu\delta W^a_\nu&+2\epsilon_{abc}W^b_{\nu\mu}\delta W^{c\mu}-\frac{g^2}{2}\left(\Phi^\dagger\left(\tau_a\delta B_\nu+\delta W^a_\nu\right)\Phi+2i\left(\delta\Phi^\dagger\tau_a D_\nu\Phi-(D_\nu\Phi)^\dagger\tau_a\delta\Phi\right)\right)\\
    &=\mathcal{D}_\nu\left(\mathcal{D}_\mu\delta W^{a\mu}-\frac{ig^2}{2}\left(\delta\Phi^\dagger\tau_a\Phi-\Phi^\dagger\tau_a\delta\Phi\right)\right),\\
    D_\mu D^\mu\delta\Phi&-i\left(\delta B_\mu+\tau_a\delta W^a_\mu\right)D^\mu\Phi-\frac{\beta}{4}\left(\delta\Phi^\dagger\Phi+\Phi^\dagger\delta\Phi\right)-\frac{\beta}{4}\left(\Phi^\dagger\Phi-1\right)\delta\Phi\\
    \label{perteq}
    &=\frac{i}{2}\left(\nabla_\mu\delta B^\mu+\tau_a\mathcal{D}_\mu\delta W^a_\mu\right)\Phi.\numberthis{}
\end{align*}

The gauge symmetry is now expressed  by the linearized version of transformations \eqref{gauge}, that is,  Eqs. \eqref{perteq} are invariant under the replacement
\begin{equation}
\label{lingaugetrans}
    \delta\Phi\rightarrow\delta\Phi+\frac{i}{2}\left(\delta\Theta+\tau_a \delta\theta^a\right)\Phi,\quad\delta B_\mu\rightarrow\delta B_\mu+\partial_\mu\delta\Theta,\quad\delta W^a_\mu\rightarrow\delta W^a_\mu+\mathcal{D}_\mu\delta\theta^a,
\end{equation}
where $\delta\Theta$, $\delta\theta^a$ are functions of $t$, $r$, $\vartheta$, $\varphi$.

\subsection{Separation of variables} 

Let us assume the background fields  to be static, spherically symmetric and purely magnetic.  It is
described by Eqs.\eqref{W2}--\eqref{Higgs} where one sets $\Omega_0=\Omega_1=\Theta_0=\Theta_1=0$
to eliminate the electric fields and for simplicity sets to zero also the pure gauge functions $\alpha,\xi$, which yields 
\be                \label{W222}
W=f(r)\left[ \T_2\, d\vartheta
-\nu\, T_1\,\sin\vartheta \,d\varphi\right]+\T_3\,\nu \cos\vartheta d\varphi,~~~~
B=\nu\cos\vartheta\, d\varphi\,,~~~~
 \Phi=\begin{pmatrix}
0 \\
\phi(r)
\end{pmatrix}.~~~~~~
\ee
The parameter  $\nu$ can be integer or half-integer and fulfills $(\nu^2-1)f=0$. 
Injecting this to the  perturbation equations \eqref{perteq} and assuming the perturbations to depend on $t,r,\vartheta,\varphi$, 
the first step is to separate the variables. This  is a non-trivial task because 
different component of perturbations have different 
spin and different isospin, and since spin, isospin and the orbital angular momentum are all coupled, this leads to a rather complex 
angular dependence. 

Fortunately, there is a powerful method to treat similar  problems, originally 
proposed within the context of the Newman-Penrose approach in the General Relativity \cite{Newman:1961qr}.
The method is based on introducing 
 the complex-valued spacetime tetrad $\theta^\alpha=\theta^\alpha_{~\mu}\, dx^\mu$ consisting of 1-forms 
\begin{equation}
    \theta^0=dt,\quad\quad\theta^1=dr,\quad\quad\theta^2=\frac{r}{\sqrt{2}}\,(d\vartheta-i\sin\vartheta d\varphi),\quad\quad\theta^3=\left(\theta^2\right)^\ast,
\end{equation}
whose scalar products determine the tetrad metric $\eta^{\alpha\beta}=(\theta^\alpha,\theta^\beta)$ with the 
only non-vanishing elements $\eta_{00}=-1$, $\eta_{11}=\eta_{23}=\eta_{32}=1$. 
It is worth noting that $\theta^2$ and $\theta^3$ are null, $(\theta^2,\theta^2)=(\theta^3,\theta^3)=0$. 
In addition, instead of the Lie-algebra basis ${\rm T}_a$ used in  \eqref{W2} one uses new generators 
${\rm L}_1={\rm T}_1-i{\rm T}_2$, ${\rm L}_2={\rm T}_1+i{\rm T}_2$, ${\rm L}_3={\rm T}_3$. 
The perturbations are decomposed as 
\be
\delta B_\mu=b_\alpha \,\theta_{~\mu}^\alpha ,~~~~~~~\delta W^a_\mu {\rm T}_a=w^b_\alpha \,\theta^\alpha_{~\mu} {\rm L}_b\,,
\ee
and it turns out that the angular dependence of the tetrad projections $b_\alpha$ and $w^b_\alpha$ as well as that of the two  Higgs
components $\delta\Phi^A$ ($A=1,2$) is given in terms of the spin-weighted spherical harmonics $_s Y_{jm}(\vartheta,\varphi)$
\cite{Goldberg:1966uu}.
Here the indices $j,m$ are the usual orbital and azimuthal quantum numbers, while the spin weight $s$ can be integer or half-integer 
and  encodes information 
about the spin and isospin of perturbations.  One has $j=|s|,|s|+1,\ldots$ and $m=-j,-j+1,\ldots ,j$. 
The functions $_0 Y_{jm}(\vartheta,\varphi)$
coincide with the usual spherical harmonics $Y_{jm}(\vartheta,\varphi)$, while for non-zero and integer values of $s$ they 
can be generated by the lowering ${\cal L}^{-}$ and raising ${\cal L}^{+}$ operators defined by 
\be
{{\cal L}}^{\mp}{_sY}_{jm}\equiv  \left(\frac{\partial}{\partial\vartheta}\mp \frac{i}{\sin\vartheta}\,\frac{\partial}{\partial\varphi}\pm s\,\cot\vartheta\right) {_sY}_{jm}=
\pm\sqrt{(j\pm s)(j\mp s+1)}\, {_{s\mp 1}}Y_{jm}.
\ee
Acting on $_s Y_{jm}$ first with ${\cal L}^{-}$ and then with ${\cal L}^{+}$ yields the differential equation 
\be                \label{spin}
\left(
\frac{1}{\sin\vartheta}\frac{\partial}{\partial\vartheta}\sin\vartheta\frac{\partial}{\partial \vartheta}+\frac{1}{\sin^2\vartheta}\frac{\partial^2}{\partial\varphi^2}
+2is\,\frac{\cos\vartheta}{\sin^2\vartheta}\,\frac{\partial}{\partial\varphi}-\frac{s^2}{\sin^2\vartheta}+j(j+1)
\right) {_{s}}Y_{jm}=0. 
\ee
The  $\varphi$-dependence of the harmonics is  $ {_{s}}Y_{jm}\sim e^{im\varphi}$ hence 
\be              
 {_{s}}Y_{j,m}(\vartheta,\varphi) ~~~\text{and}~~~~\left({_{-s}}Y_{j,m}(\vartheta,\varphi)\right)^\ast
\ee
fulfill the same differential equation. From the practical viewpoint, the analysis can be simplified 
by noting that the perturbations should not depend
on the azimuthal quantum number $m$, while for $m=j$ one has
\be                \label{Y1}
{_{s}}Y(\vartheta,\varphi)_{j,j}=const.\times  (\sin\vartheta)^j \left(\tan\frac{\vartheta}{2}\right)^s e^{i\,j\varphi}. 
\ee
As a result, to determine the angular dependence of all amplitudes, it suffices to consider the following two solutions of \eqref{spin}, 
$ _s{\Sigma}_j(\vartheta)\, e^{+ij\varphi}$  and $_{-s}{\Sigma}_j(\vartheta)\, e^{-ij\varphi}$, both 
with the same spin weight $s$, where
\be  \label{Y2}
_s{\Sigma}_j(\vartheta)=\left(\sin\vartheta\right)^j\left(\tan\frac{\vartheta}{2}\right)^{s}.
\ee
To separate the variables, we set 
\be                      \label{bbb}
    b_\alpha=b^{+}_{\alpha}(r)\,{_s}{\Sigma}_j(\vartheta)\e^{+i(\omega t+j\varphi)}
    +b^{-}_{\alpha}(r)\,{_{-s}}{\Sigma}_j(\vartheta)\e^{-i(\omega t+j\varphi)}  
    \ee
 where $b^{+}_{\alpha}(r)$ and $b^{-}_{\alpha}(r)$ are complex-valued radial functions.
 Both terms on the right here have the same spin weight $s$ whose value depends 
  on value of the index 
 $\alpha$ and on the winding number $\nu$. Similar expressions are used for 
$w^b_\alpha$ and for $\delta\Phi,\delta\Phi^\dagger$. The perturbations contain the  $``+"$ part proportional to 
$\e^{+i(\omega t+j\varphi)}$ and the  $``-"$  part proportional to 
$\e^{-i(\omega t+j\varphi)}$. One cannot retain only one of these parts because they are intermixed in the equations
which 
contain both $\delta\Phi$ and $\delta \Phi^\dagger$.  This doubles the 
number of the radial amplitudes: for example 4 components of $b_\alpha$  in \eqref{bbb} contain 8
radial functions
$ b^{+}_{\alpha}(r)$ and $ b^{-}_{\alpha}(r)$. A similar doubling occurs for radial functions in $w^b_\alpha$ and $\delta\Phi,\delta\Phi^\dagger$. 
As a result, injecting everything to the equations and setting separately to zero the parts of the equations proportional to 
$\e^{+i(\omega t+j\varphi)}$ and those proportional to $\e^{-i(\omega t+j\varphi)}$ yields 40 differential equations.
These equations contain 40 
complex functions of $r$ contained in 20 components of $b_\alpha,w^b_\alpha,\delta\Phi,\delta\Phi^\dagger$
and also functions $_{\pm s}\Sigma_j(\vartheta)$ with the component-dependent value of $s$. 

A direct inspection of the equations then reveals that one can adjust the values of the spin weight $s$ for all components such that 
the $\vartheta$-dependence in all 40 equations factorizes. This yields 40 ordinary differential equations for 40 complex functions of $r$. 
A further inspection shows that, imposing simple linear relations between the functions of $r$ contained in the ``$+$" part  and those 
contained in the ``$-$" part reduces twice the number of independent equations. This yields   20 
equations for 20 complex functions of $r$. At the end of the day, one imposes the reality conditions: the spacetime components 
$\delta B_\mu$ and $\delta W^a_\mu$ should be real, while components of $\delta \Phi$ and $\delta \Phi^\dagger$ should be 
mutually complex conjugated. This renders all 20 functions of $r$ real-valued. 

Skipping the  details, here is the resulting ansatz that provides  
the complete separation of the angular and temporal variables. Defining $\Om=\omega t+j\varphi$, 
the 4 tetrad components $b_\alpha$ and their spin weights $s$ are given by  
\be
b_0&=&+2\cos(\Om)\, {_0}\Sigma_{j}(\vartheta)\,\frac{S_0(r)}{r}\,, \hspace{6 cm} (s=0)\nn \\
b_1&=&-2\sin(\Om)\, {_0}\Sigma_{j}(\vartheta)\,\frac{S_1(r)}{r}\,,\hspace{6 cm} (s=0)\nn \\
b_2&=&i\, e^{+i\Om} \, {_{+1}}{\Sigma}_{j}(\vartheta)\,\frac{S_2(r)}{r}-i \,e^{-i\Om} \,{_{-1}} {\Sigma}_{j}(\vartheta)\,\frac{S_3(r)}{r},\hspace{2 cm} (s=+1)\nn  \\
b_3&=&i\, e^{+i\Om} \, {_{-1}}{\Sigma}_{j}(\vartheta)\,\frac{S_3(r)}{r}
-i \,e^{-i\Om} \,  {_{+1}}{ \Sigma}_{j}(\vartheta)\,\frac{S_2(r)}{r},   \hspace{2 cm} (s=-1)       \label{anz1}
\ee
the tetrad components $w^1_\alpha$ are 
\be
w^1_0&=&e^{i\Om}\, {_{-\nu}}{\Sigma}_{j}(\vartheta)\,\frac{X_{10}(r)}{r}+e^{-i\Om}\, {_{+\nu}}{\Sigma}_{j}(\vartheta)\,\frac{X_{20}(r)}{r}\, ,\hspace{2 cm} (s=-\nu)\nn \\
w^1_1&=&i\, e^{i\Om}\, {_{-\nu}}{\Sigma}_{j}(\vartheta)\,\frac{X_{11}(r)}{r}-i\,e^{-i\Om}\, {_{+\nu}}{\Sigma}_{j}(\vartheta)\,\frac{X_{21}(r)}{r}\,, \hspace{1.8 cm} (s=-\nu),\nn \\
w^1_2&=&i\, e^{i\Om}\, {_{+1-\nu}}{\Sigma}_{j}(\vartheta)\,\frac{X_{12}(r)}{r}-i\,e^{-i\Om}\, {_{+\nu-1}}{\Sigma}_{j}(\vartheta)\,\frac{X_{23}(r)}{r}\, ,\hspace{0.8 cm} (s=1-\nu)\nn \\
w^1_3&=&i\, e^{i\Om}\, {_{-1-\nu}}{\Sigma}_{j}(\vartheta)\,\frac{X_{13}(r)}{r}-i\,e^{-i\Om}\, {_{+\nu+1}}{\Sigma}_{j}(\vartheta)\,\frac{X_{22}(r)}{r}\, ,\hspace{0.7 cm} (s=-1-\nu)      \label{anz2}
\ee
the tetrad components $w^2_\alpha$ are 
\be
w^2_0&=&e^{i\Om}\, {_{+\nu}}{\Sigma}_{j}(\vartheta)\,\frac{X_{20}(r)}{r}+e^{-i\Om}\, {_{-\nu}}{\Sigma}_{j}(\vartheta)\,\frac{X_{10}(r)}{r}\, ,\hspace{2.4 cm} (s=\nu)\nn \\
w^2_1&=&i\, e^{i\Om}\, {_{+\nu}}{\Sigma}_{j}(\vartheta)\,\frac{X_{21}(r)}{r}-i\,e^{-i\Om}\, {_{-\nu}}{\Sigma}_{j}(\vartheta)\,\frac{X_{11}(r)}{r}\, ,\hspace{2 cm} (s=\nu)\nn \\
w^2_2&=&i\, e^{i\Om}\, {_{+1+\nu}}{\Sigma}_{j}(\vartheta)\,\frac{X_{22}(r)}{r}-i\,e^{-i\Om}\, {_{-\nu-1}}{\Sigma}_{j}(\vartheta)\,\frac{X_{13}(r)}{r}\, ,\hspace{0.7 cm} (s=1+\nu)\nn \\
w^2_3&=&i\, e^{i\Om}\, {_{-1+\nu}}{\Sigma}_{j}(\vartheta)\,\frac{X_{23}(r)}{r}-i\,e^{-i\Om}\, {_{-\nu+1}}{\Sigma}_{j}(\vartheta)\,\frac{X_{12}(r)}{r}\, ,\hspace{0.7 cm} (s=-1+\nu)      \label{anz3}
\ee
the tetrad components $w^3_\alpha$ read 
\be
w^3_0&=&2\cos(\Om)\, {_{0}}{\Sigma}_{j}(\vartheta)\,\frac{X_{30}(r)}{r}\, ,\hspace{6.2 cm} (s=0)\nn \\
w^3_1&=&-2\sin(\Om)\, {_{0}}{\Sigma}_{j}(\vartheta)\,\frac{X_{31}(r)}{r}\, ,\hspace{6 cm} (s=0)\nn \\
w^3_2&=&i\, e^{i\Om}\, {_{+1}}{\Sigma}_{j}(\vartheta)\,\frac{X_{32}(r)}{r}-i\,e^{-i\Om}\, {_{-1}}{\Sigma}_{j}(\vartheta)\,\frac{X_{33}(r)}{r}\, ,\hspace{2 cm} (s=1)\nn \\
w^3_3&=&i\, e^{i\Om}\, {_{-1}}{\Sigma}_{j}(\vartheta)\,\frac{X_{33}(r)}{r}-i\,e^{-i\Om}\, {_{+1}}{\Sigma}_{j}(\vartheta)\,\frac{X_{32}(r)}{r}\, ,\hspace{2 cm} (s=-1)\       \label{anz4}
\ee
while the perturbations of the Higgs field are described by 
\be
\delta\Phi_1&=&e^{i\Om}\, {_{-\nu}}{\Sigma}_{j}(\vartheta)\,\frac{H_1(r)}{r}+e^{-i\Om}\, {_{+\nu}}{\Sigma}_{j}(\vartheta)\,\frac{H_2(r)}{r}\,, \hspace{1.5 cm} (s=-\nu)\nn \\
\delta\Phi_2&=&e^{i\Om}\, {_{0}}{\Sigma}_{j}(\vartheta)\,\frac{H_3(r)}{r}+e^{-i\Om}\, {_{0}}{\Sigma}_{j}(\vartheta)\,\frac{H_4(r)}{r}\,. \hspace{2 cm} (s=0)      \label{anz5}
\ee
Here the 20 radial functions $S_0(r),S_1(r),\ldots , H_4(r)$ are all real-valued. 
Inserting \eqref{anz1}--\eqref{anz5} into the perturbation equations \eqref{perteq}, the variables separate yielding 20 ordinary differential equations for the 20 radial  functions. 
These are Eqs.\eqref{10a}--\eqref{hd}  shown in Appendix C. The variables separate if only the background condition
$(\nu^2-1) f=0$  is fulfilled. Therefore Eqs.\eqref{10a}--\eqref{hd} make sense if only either $\nu=\pm 1$ when  $f(r)$ can be arbitrary 
or if $f(r)=0$ when $\nu$ can be 
arbitrary. 

Notice finally that the above separation ansatz can be extended to arbitrary values of the azimuthal quantum number $m$ 
via replacing everywhere in \eqref{anz1}--\eqref{anz5}
\be
{_{s}}\Sigma_j(\vartheta) e^{ij\varphi}\to {_{s}}Y_{jm}(\vartheta,\varphi)\,,~~~~~~~~~
{_{-s}}\Sigma_j(\vartheta) e^{-ij\varphi}\to \left({_{-s}}Y_{jm}(\vartheta,\varphi)\right)^\ast\,.
\ee

\section{GAUGE FIXING}
The radial equations \eqref{10a}--\eqref{hd} 
admit the gauge symmetry generated by the gauge transformations \eqref{lingaugetrans} with the gauge parameters $\delta\Theta$ and 
$\delta \theta^a$ chosen as 
\be           \label{residual}
\delta\Theta&=&\sin(\Omega)\,{_{0}}{\Sigma}_{j}(\vartheta)\,\frac{u(r)}{r}\,, ~~~~~\delta\theta^3=\sin(\Omega)\,{_{0}}{\Sigma}_{j}(\vartheta)\,\frac{u_3(r)}{r}\,, \nn \\
\delta\theta_1&=&\sin(\Omega)\,\left(  {_{\nu}}{\Sigma}_{j}(\vartheta)\,\frac{u_1(r)}{r} + {_{-\nu}}{\Sigma}_{j}(\vartheta)\,\frac{u_2(r)}{r}\right)\,, \nn \\
\delta\theta_2&=&-\cos(\Omega)\,\left(  {_{\nu}}{\Sigma}_{j}(\vartheta)\,\frac{u_1(r)}{r} - {_{-\nu}}{\Sigma}_{j}(\vartheta)\,\frac{u_2(r)}{r}\right)\, .
\ee
This choice of the gauge parameters is compatible with the considered above separation of variable and 
generates the following transformations of the radial functions:
\be
&&S_0\to S_0+\frac{\omega}{2}\, u,~~~~~
S_1\to S_1-\frac{r}{2}\left(\frac{u}{r}\right)^\prime,~~~~\nn \\
&&S_2\to S_2+\frac{j\,u}{2\sqrt{2}\, r},~~~~~
S_3\to S_3-\frac{j\,u}{2\sqrt{2}\, r},       \label{gauge1}
\ee
\be
&&X_{10}\to X_{10}+\omega\, u_2,~~~~~~~~~
X_{11}\to X_{11}-r\left(\frac{u_2}{r}\right)^\prime,~~\nn \\
&&X_{12}\to X_{12}+\frac{1}{\sqrt{2}\, r}\left((\nu+j)\, u_2-f\,\delta_{0,1-\nu}\, u_3\right),~~\nn \\
&&X_{13}\to X_{13}+\frac{1}{\sqrt{2}\, r}\left((\nu-j)\, u_2-f\,\delta_{0,1+\nu}\, u_3\right),~     \label{gauge2}
\ee
\be
&&X_{20}\to X_{20}+\omega\, u_1,~~~~~~
X_{21}\to X_{21}-r\left(\frac{u_1}{r}\right)^\prime,~~\nn \\
&&X_{22}\to X_{22}+\frac{1}{\sqrt{2}\, r}\left((j-\nu)\, u_1-f\,\delta_{0,1+\nu}\, u_3\right),~~~\nn \\
&&X_{23}\to X_{23}-\frac{1}{\sqrt{2}\, r}\left((j+\nu)\, u_1+f\,\delta_{0,1-\nu}\, u_3\right),~~~     \label{gauge3}
\ee
\be
&&X_{30}\to X_{30}+\frac{\omega}{2}\, u_3,~~~~~~~~
X_{31}\to X_{31}-\frac{r}{2}\left(\frac{u_3}{r}\right)^\prime,~\nn \\
&&X_{32}\to X_{32}+\frac{1}{\sqrt{2}\, r}\left( \frac{j}{2}\, u_3+f\,\delta_{0,1-\nu}\, u_1 +f\,\delta_{0,1+\nu}\, u_2\right),~\nn\\
&&X_{33}\to X_{33}+\frac{1}{\sqrt{2}\, r}\left( -\frac{j}{2}\, u_3+f\,\delta_{0,1+\nu}\, u_1 +f\,\delta_{0,1-\nu}\, u_2\right),~     \label{gauge4}
\ee
\be
&&H_1\to H_1+\phi\,\frac{u_2}{2}\,,~~~~~~~~~H_2\to H_2-\phi\,\frac{u_1}{2},~~\nn \\
&& H_3\to H_3+\phi\,\frac{u-u_3}{4}, ~~~~~~H_4\to H_4+\phi\,\frac{u_3-u}{4} .      \label{gauge5}
\ee
A straightforward verification shows that these transformations leave the radial equations \eqref{10a}--\eqref{hd} invariant,
provided that the background equations for $f(r)$ and $\phi(r)$ are fulfilled. 
The four arbitrary functions $u(r), u_1(r), u_2(r), u_3(r)$ generating these transformations can be adjusted to impose 
four gauge conditions. A convenient gauge choice is 
\be
{\cal C}_1=0,~~~~{\cal C}_2=0,~~~~{\cal C}_3=0,~~~~{\cal C}_4=0,~~~~
\ee
where ${\cal C}_A$ $(A=1,2,3,4)$ are defined by Eqs.\eqref{Ca}--\eqref{Cd} in Appendix C.
This gauge choice implies that 
  \be
  \nabla_\mu\delta B^\mu-\frac{ig'^2}{2}\left(\delta\Phi^\dagger\Phi-\Phi^\dagger\delta\Phi\right)=0,~~~~~~
   \mathcal{D}_\mu\delta W^{a\mu}-\frac{ig^2}{2}\left(\delta\Phi^\dagger\tau_a\Phi-\Phi^\dagger\tau_a\delta\Phi\right)=0,
   \ee
and hence the right hand sides of perturbation equations for $\delta B_\mu$ and $ \delta W^a_\mu$ in \eqref{perteq} 
and of the radial equations 
\eqref{10a}--\eqref{hd} vanish. The radial equations then split into two independent groups because the 4 
equations \eqref{10a},\eqref{20a},\eqref{30a},\eqref{Sa} will contain only the temporal components of the fields
and reduce  to a coupled system, 
\be               \label{temp}
&&\hat{{\cal D}}X_{10}+\frac{2\nu\, j\, f}{r^2}\, X_{30}=0\,,~~~~~~
\hat{{\cal D}}X_{20}-\frac{2\nu\, j\, f}{r^2}\, X_{30}=0\,,~~~~~~\hat{{\cal D}}_s\, S_0+\frac{g^{\prime 2}\phi^2}{2}\, X_{30}=0\,,
\nn \\
&&\left(\hat{{\cal D}} -\frac{f^2+\nu^2}{r^2}\right) X_{30}
+\frac{\nu\,(j+1)\, f}{r^2}\,(X_{10}-X_{20})+\frac{g^2\phi^2}{2}\, S_0=0\,, 
\ee
with $\hat{{\cal D}}$ defined in \eqref{dif}. 
The remaining 16 equations in \eqref{10a}--\eqref{hd}  will contain only the remaining 16 field components. 
The 4 constraints ${\cal C}_A=0$  can be  resolved with respect to the 4 temporal components, 
\be
X_{10}&&=\frac{1}{\omega}\left(\frac{\left( r X_{11}\right)^\prime}{r} 
+\frac{j-\nu_-}{\sqrt{2}\, r}\, X_{12}
-\frac{j+\nu_+}{\sqrt{2}\, r}\, X_{13}
-\frac{\nu_-\, f}{\sqrt{2}\, r}\,  X_{32}
+\frac{\nu_+\, f}{\sqrt{2}\, r}\, X_{33}
+g^2\phi\, H_1\right) \,,\nn \\
X_{20}&&=\frac{1}{\omega}\left(\frac{\left( r X_{21}\right)^\prime}{r}
+\frac{j+\nu_+}{\sqrt{2}\, r}\, X_{22}
-\frac{j-\nu_-}{\sqrt{2}\, r}\, X_{23}
+\frac{\nu_+\, f}{\sqrt{2}\, r}\,  X_{32}
-\frac{\nu_-\, f}{\sqrt{2}\, r}\, X_{33}
-g^2\phi\, H_2\right) \,,\nn \\
X_{30}&&=\frac{1}{\omega}\left(\frac{\left( r X_{31}\right)^\prime}{r}  
+\frac{\nu_-\, f}{2\sqrt{2}\, r}\,  (X_{13}+X_{22})
-\frac{\nu_+\, f}{2\sqrt{2}\, r}\,  (X_{12}+X_{23})\right.\nn \\
&&\hspace{7 cm} \left.+\frac{j+1}{\sqrt{2}\, r}\, (X_{32}-X_{33}) 
+\frac{g^2}{2}\,\phi\, (H_4-H_3)\right) \,, \nn \\
S_{0}&&=\frac{1}{\omega}\left(\frac{\left( r S_{1}\right)^\prime}{r}     \label{solve}
+\frac{j+1}{\sqrt{2}\, r}\, (S_{2}-S_3)
+\frac{g^{\prime 2}}{2}\,\phi\, (H_3-H_4)\right) \,.
\ee
Injecting this  to the 4 equations \eqref{temp} and using the background equations yields identities by virtue of the 16 equations
for the 16 functions that appear on the right in \eqref{solve}. 
Therefore, the number of independent radial functions reduces to $16=20-4$. This corresponds to the 12 
physical degrees of freedom of the fields plus additional 4 gauge degrees of freedom. The additional gauge degrees  appear because 
the gauge conditions ${\cal C}_A=0$ do not fix the gauge completely. 
Specifically, the gauge conditions ${\cal C}_A=0$  will be invariant under gauge transformations \eqref{gauge1}--\eqref{gauge5} 
if the parameters $u(r),u_1(r),u_2(r),u_3(r)$ of the latter fulfill equations that have exactly the same structure as Eqs.\eqref{temp},
up to the replacement 
\be
X_{10}\to u_2,~~~~X_{20}\to u_1,~~~~X_{30}\to \frac{u_3}{2},~~~S_0\to \frac{u}{2}. 
\ee
Defining $u_\pm=u_1\pm u_2$, these equations can be represented as one independent equation
\be                 \label{temps1}
\hat{{\cal D}}\,u_{+}=0,
\ee
and three coupled equations 
\be               \label{temp2}
&&\hat{{\cal D}}\, u_{-}-\frac{2\nu\, j\, f}{r^2}\, u_3=0,~~~~~~
\hat{{\cal D}}_s\, u+\frac{g^{\prime 2}\phi^2}{2}\, u_3=0,
\nn \\
&&\left(\hat{{\cal D}} -\frac{f^2+\nu^2}{r^2}\right) u_3
-\frac{2\nu\,(j+1)\, f}{r^2}\,u_{-}+\frac{g^2\phi^2}{2}\, u=0,
\ee
with $\hat{{\cal D}}$ and $\hat{{\cal D}}_s$ defined in \eqref{dif},\eqref{dif1}. 
Solutions of these equations generate gauge transformations which preserve the ${\cal C}_A=0$ conditions. 
One possibility to use this residual symmetry is to impose extra gauge conditions and pass to the temporal gauge by setting 
$X_{10}=X_{20}=X_{30}=S_0=0$. This is possible because the temporal field components in \eqref{temp} and the 
gauge parameters $u,u_1,u_2,u_3$  satisfy the same equations. In this case the relations \eqref{solve} would become 
constraints that would further reduce to 12 the number of independent functions. However, explicitly resolving these constraints leads 
to very complicated expressions. Therefore, instead of using the temporal gauge condition we prefer a different method. 

We notice  that defining 
\be
\kappa=\sqrt{\frac{j}{j+1}},~~~~~~~~~\kappa_\pm=\sqrt{\frac{j\pm \nu}{j+1\mp\nu}},
\ee
one can express the 16 functions in terms of 7 amplitudes $Y_1(r),\ldots , Y_7(r)$ and 9 amplitudes $Z_1(r),\ldots , Z_9(r)$ as 
\be
X_{11}&=&\frac{\kappa}{2}\left[\, Y_2(r)+Z_2(r)\,\right]\,,~~~~~~~~~X_{12}=\frac{\kappa\,\kappa_+}{2}\left[\, Y_1(r)+Z_1(r)\,\right]\,,\nn \\
X_{13}&=&\frac{\kappa\,\kappa_-}{2}\left[\, Y_4(r)+Z_4(r)\,\right]\,,~~~~~~~X_{21}=\frac{\kappa}{2}\left[\, Y_2(r)-Z_2(r)\,\right]\,, \nn \\
X_{22}&=&\frac{\kappa\,\kappa_-}{2}\left[\, -Y_4(r)+Z_4(r)\,\right]\,,~~~~~X_{23}=\frac{\kappa\,\kappa_+}{2}\left[\, -Y_1(r)+Z_1(r)\,\right]\,,\nn \\
X_{31}&=&\frac{1}{2}\,\left[\, -g\, Z_6(r)+g^\prime Z_8(r)\,\right]\,, ~~~~~S_1=\frac{g^{\prime}}{2g}\,\left[\, g^{\prime}\,Z_6(r)+g\, Z_8(r)\,\right]\,,\nn \\
X_{32}&=&\left.\left.\frac{\kappa}{2\sqrt{2}}\right(-g\,\left[\,Y_5(r)+Z_7(r)\,\right]+g^\prime\,\left[\,Y_6(r)+Z_9(r)\,\right]\right)\,,\nn \\
X_{33}&=&\left.\left.\frac{\kappa}{2\sqrt{2}}\right(g\,\left[\,-Y_5(r)+Z_7(r)\,\right]+g^\prime\,\left[\,Y_6(r)-Z_9(r)\,\right]\right)\,,\nn \\
S_{2}&=&\left.\left.\frac{\kappa\,g^\prime}{2\sqrt{2}\, g}\right(g^\prime \,\left[\,Y_5(r)+Z_7(r)\,\right]+g\,\left[\,Y_6(r)+Z_9(r)\,\right]\right)\,,\nn \\
S_{3}&=&\left.\left.\frac{\kappa\,g^\prime}{2\sqrt{2}\, g}\right(g^\prime \,\left[\,Y_5(r)-Z_7(r)\,\right]+g\,\left[\,Y_6(r)-Z_9(r)\,\right]\right)\,,\nn \\
H_1&=&\frac{\kappa}{2\sqrt{2}\,g}\,\left[\,Y_3(r)-Z_3(r)\,\right]\,,~~~~~~~H_2=-\frac{\kappa}{2\sqrt{2}\,g}\,\left[\,Y_3(r)+Z_3(r)\,\right]\,,~\nn \\
H_3&=&\frac{1}{2\sqrt{2}\,g}\,\left[\,Y_7(r)-Z_5(r)\,\right]\,,~~~~~~~H_4=\frac{1}{2\sqrt{2}\,g}\,\left[\,Y_7(r)+Z_5(r)\,\right]\,. \label{YZ}
\ee
Injecting this to the radial equations \eqref{10a}--\eqref{hd}, where one should set ${\cal C}_A=0$  and omit the 4 equations for the temporal 
components, the 7 equations for $Y_1(r),\ldots , Y_7(r)$  decouple from the 9 equations for $Z_1(r),\ldots , Z_9(r)$. The reason for this is the 
difference in their  behaviour under  the parity reflection. 
To see this, one should restore the full dependence on the spacetime coordinates. For example, the perturbations of the U(1) field are 
\be
\delta B_\mu dx^\mu =b_\alpha\, \theta^\alpha_{~\mu}dx^\mu=b_0\, dt+ b_1\, dr
+ \frac{r\, b_2}{\sqrt{2}}\,(d\vartheta-i\sin\vartheta\ d\varphi)
+ \frac{r\, b_3}{\sqrt{2}}\,(d\vartheta+i\sin\vartheta\ d\varphi),~~~~~~
\ee
where $b_\alpha$ are determined by \eqref{anz1} in terms of $t,r,\vartheta,\varphi$ with  $S_\alpha(r)$ 
expressed in terms of  $Y_A(r)$ and $Z_B(r)$ by \eqref{YZ} (notice that $b_0,b_1\in \mathbb{R}$ while $b_3=(b_2)^\ast$). 
It turns out that the angle-dependent coefficients in front of the 
$Y_A(r)$ amplitudes and those in front of the $Z_B(r)$ amplitudes behave differently under the parity reflection
\be
\vartheta\to \pi-\vartheta,~~~~~~\varphi\to \pi+\varphi.
\ee
If the  coefficients in front of the 
$Y$-amplitudes stay invariant under parity, then those  in front of the 
$Z$-amplitudes change sign, or the other way round, depending on values 
of $j,\nu$. This explains the separation of the $Y$-equations from the $Z$-equations. 

Introducing the 7-component vector $\Psi_Y=[Y_1(r),\ldots , Y_7(r)]^{\rm tr}$ and the 
9-component vector $\Psi_Z=[Z_1(r),\ldots , Z_9(r)]^{\rm tr}$, 
the 16 perturbation equations assume the form of  two independent Schroedinger-type systems 
\be               \label{Schrod}
\left(-\frac{d^2}{dr^2}+\hat{U}\right)\Psi_Y=\omega^2 \Psi_Y\,,~~~~~
\left(-\frac{d^2}{dr^2}+\hat{V}\right)\Psi_Z=\omega^2 \Psi_Z\,,~~~~~
\ee
where $\hat{U}$ and $\hat{V}$ are symmetric matrices whose components are shown in Appendix D. 
In the $r\to\infty$ limit, where $\phi\to 1$ and $f\to 0$, these potentials reduce to 
\be
\hat{U}&=&{\rm diag}\left(\mw^2,\mw^2,\mw^2,\mw^2,\mz^2,0,\mh^2\right),\nn \\
\hat{V}&=&{\rm diag}\left(\mw^2,\mw^2,\mw^2,\mw^2,\mz^2,\mz^2,\mz^2,0,0\right),
\ee
with the field masses given by \eqref{masses}. One can see that the $Y$-sector
contains four polarization states of the W-boson, one of which is a gauge mode, a $Z$-boson polarization,
a photon polarization, and the Higgs mode. The $Y$-sector
contains four polarization states of the W-boson one of which is a gauge mode, three $Z$-boson polarizations two of which are gauge modes, 
and two photon polarizations one of which is a gauge mode. 

The gauge modes can be excluded, which can be nicely illustrated 
for the $Y$-amplitudes. These amplitudes are gauge-dependent, but their transformations  
under the residual gauge symmetry depend only on the $u_{+}$ gauge 
amplitude subject to \eqref{temps1}, 
\be                  \label{residualY}
Y_1&\to& Y_1+\frac{(j+\nu)\,u_{+}}{\sqrt{2}\,\kappa\kappa_{+}\, r}\,,~~~~~
Y_2\to Y_2 -\frac{r}{\kappa}\left(\frac{u_{+}}{r}\right)^\prime\,, ~~~~~
Y_3\to Y_3+\frac{g\,\phi\, u_{+}}{\sqrt{2}\,\kappa}\,,~~~~ \\
Y_4&\to& Y_4 -\frac{(j-\nu)\, u_{+}}{\sqrt{2}\,\kappa\kappa_{-}\, r}\,,~~~~~~
Y_5\to Y_5-\frac{g\, f}{\kappa\, r}\, u_{+}\,,~~~~~~
Y_6\to Y_6+\frac{g^\prime \, f}{\kappa\, r}\, u_{+}\,,~~~~
Y_7\to Y_7. \nn
\ee
This implies that the following 6 ``calligraphic" combinations are gauge-invariant, 
\be
{\cal Y}_1&=&Y_1+\sqrt{\frac{(j+\nu)(j+1-\nu)}{(j-\nu)(j+1+\nu)}}\, Y_4\,,~~~~~~
{\cal Y}_2=Y_2-\frac{\sqrt{2}\, r\, Y_4^\prime }{ \sqrt{(j-\nu)(j+1+\nu)}}\, ,~~~~~\nn \\
{\cal Y}_3&=&Y_3+\frac{g\, r\,\phi\, Y_4}{ \sqrt{(j-\nu)(j+1+\nu)}}\,,~~~~~~~~~
{\cal Y}_5=Y_5-\frac{\sqrt{2}\, g\,f\, Y_4}{\sqrt{(j-\nu)(j+1+\nu)}}\, \,,~~~\nn \\
{\cal Y}_6&=&Y_6+\frac{\sqrt{2}\, g^\prime \,f\, Y_4}{\sqrt{(j-\nu)(j+1+\nu)}}\, \,,~~~~
{\cal Y}_4=Y_7\,. 
\ee
Injecting this to the equations \eqref{Schrod} 
reveals that these 6  gauge invariant amplitudes fulfill a closed system of 6 equations.
The remaining gauge-dependent amplitude $Y_4$ filfills (for $\nu=1$) the inhomogeneous equation
\be               
\hat{{\cal D}}\, (r\,Y_4)=\left.\left.\frac{\sqrt{2(j-1)(j+2)}}{r}\, \right(f\left(\,g\,{\cal Y}_5 -g^\prime\,{\cal Y}_6\,\right) -{\cal Y}_2\right).
\ee
If there are two solutions of this equation which differ from each other by a gauge transformation, $Y_4$ and $\tilde{Y}_4$, then their difference fulfills
the homogeneous equation 
\be
\hat{{\cal D}}\, (r\,(Y_4-\tilde{Y}_4))=0, 
\ee
which is compatible with the fact that $r( Y_4-\tilde{Y}_4)\propto  u_{+}$ where $\hat{{\cal D}}\,u_{+}=0$. 

In the $Z$-sector the residual gauge symmetry is generated by three functions $u_{-},u,u_3$ subject to \eqref{temp2}. 
Defining $v_{\pm}=u\pm u_3$ one has 
\be            \label{residualZ}
Z_1&\to& Z_1 + \frac{f\,\delta_{0,1-\nu}\,(v_{-}-v_{+})-(j+\nu)\, u_{-}}{\sqrt{2}\,\kappa\kappa_{+}\,r}\,,~~~~~
Z_2\to Z_2+\frac{r}{\kappa}\left(\frac{u_{-}}{r} \right)^\prime\,, ~~~~
Z_{3}\to Z_{3}+\frac{g\,\phi\, u_{-}}{\sqrt{2}\,\kappa}\,,~~~\nn \\
Z_4&\to& Z_4 + \frac{f\,\delta_{0,1+\nu}\,(v_{-}-v_{+})+(j-\nu)\, u_{-}}{\sqrt{2}\,\kappa\kappa_{-}\,r}\,,~~~
Z_5\to Z_5-\frac{g\,\phi\, v_{-}}{\sqrt{2}}\,,~~~~~~
Z_6\to Z_6-g\,r\left(\frac{v_{-}}{r} \right)^\prime\,, ~~~\nn \\
Z_7&\to&  Z_7 + g\,\frac{f\,(\delta_{0,1+\nu}-\delta_{0,1-\nu})\,u_{-}+j\,v_{-}}{\kappa\,r}\,,~~~~
Z_8\to Z_8-\frac{g^2-g^{\prime 2}}{2 g^\prime }\, r \left(\frac{v_{-}}{r} \right)^\prime
-\frac{r}{2 g^\prime}\left(\frac{v_{+}}{r} \right)^\prime\,, \nn \\
Z_9&\to& Z_9-\frac{g^\prime\, f}{\kappa\, r}\,(\delta_{0,1+\nu}-\delta_{0,1-\nu})u_{-}
+\frac{j}{2g^\prime\kappa}\,((g^2-g^{\prime 2})\,v_{-}+v_{+}).
\ee
Using these expressions one can build 6 gauge invariant amplitudes. 
It turns out, however, that instead of expressing everything in terms of gauge invariant variables, 
it is sometimes  easier to work with equations \eqref{Schrod} containing gauge degrees of freedom 

\section{STABILITY ANALYSIS \label{stab}} 

We are now ready to carry out the stability analysis of the monopole solutions. 
For this one should study the Schroedinger problems \eqref{Schrod} to see if they admit 
bound state solutions with $\omega^2<0$. Such negative modes are most likely to appear in 
sectors with lower angular momentum $j$, therefore one should analyze first of all 
the sectors with $j=0,1$. 

In order to consider these sectors one makes use of the following important fact:
the spin-weighted harmonics ${_s}Y_{jm}(\vartheta,\varphi)$ separating the angular variables vanish if 
$j<|s|$. Physically this corresponds to the 
fact that modes with angular momentum less than spin are non-dynamical. 
It can also be seen that the angular functions in \eqref{Y1},\eqref{Y2} are defined only for $j\geq |s|$ 
since for $j<|s|$ they would be unbounded. 

Therefore, for low values of $j$ all field amplitudes whose spin weight $s$ is too large should be set to zero. 
Remarkably, this eliminates precisely those amplitudes whose coefficients in the equations 
become undefined for small $j$ because they 
contain in the denominators $\kappa=\sqrt{j/(j+1)}$ and $\kappa_\pm=\sqrt{(j\pm \nu)/(j+1\mp\nu)}$. 
Let us see how this works.

\subsection{$j=0$ sector}
The spin weights $s$ of all field amplitudes are shown in \eqref{anz1}-\eqref{anz5}. 
If $j=0$ then only amplitudes with $s=0$ are allowed. For example, one can see in Eq.\eqref{anz1}  that $b_0$ and $b_1$ 
are allowed, hence the radial amplitudes $S_0(r)$ and $S_1(r)$ should be kept, but $b_2$ and $b_3$ should vanish because
${_{\pm 1}}\Sigma_0=0$ hence one should set $S_2=S_3=0$. The other non-vanishing amplitudes are $X_{30}, X_{31}, H_3, H_4$.
In addition, one takes into account amplitudes $X_{12}$ and $X_{23}$ if $\nu=1$ since they have spin weights  $s=\pm(1-\nu)=0$,  
whereas for $\nu=-1$ 
one should keep $X_{22},X_{13}$. 
Using the definitions of the $Y$ and $Z$ amplitudes in \eqref{YZ}, we see that the non-vanishing amplitudes are 
$Y_7,Z_5,Z_6,Z_8$ and also $Y_1,Z_1$ if $\nu=1$ (or $Y_4,Z_4$ if $\nu=-1$). All other amplitudes in Eqs.\eqref{Schrod} 
should be set to zero. 

\subsubsection{$Y$-sector}

The equations \eqref{Schrod} in the  $Y$-sector then reduce for  $\nu=1$ to
\be                    \label{magn}
\left(-\frac{d^2}{dr^2}-\omega^2+\frac{3f^2-1}{r^2}+\frac{g^2\phi^2}{2}\right)Y_1=\frac{gf\,\phi}{r}\, Y_7\,,\nn \\
\left(-\frac{d^2}{dr^2}-\omega^2+\frac{f^2}{2r^2}+\frac{\beta}{4}\,(3\phi^2-1)\right)Y_7=\frac{gf\,\phi}{r}\, Y_1\,.
\ee
Same equations up to $Y_1\to Y_4$ are obtained for $\nu=-1$. 
This agrees with the equations previously obtained in \eqref{eig},\eqref{eig1} 
(replacing $Y_1\to \psi_1$ and $Y_7\to-\psi_2$). These equations are gauge-invariant since the residual 
gauge transformations \eqref{residualY} in the $Y$-sector are generated by $u_{+}=u_1+u_2$, however, as seen in 
\eqref{residual}, the $u_1$ and $u_2$ gauge amplitudes have spin weights $s=\pm 1$ and hence vanish for $j=0$. 
Therefore, there is no residual gauge freedom in this case. 
For $\nu\neq \pm 1$ 
one sets  $f=0$, $\phi=1$ and  $Y_1=0$ (also $Y_4=0$) and obtains a single equation 
\be          \label{magn1}
\left(-\frac{d^2}{dr^2}+\frac{\beta}{2}\,\right)Y_7=\omega^2\, Y_7\,.
\ee
Eqs.\eqref{magn},\eqref{magn1} should be studied to see if there are bounded solutions with $\omega^2<0$ which would 
correspond to unstable modes. With the latter equation the situation is simple because  it describes a free Higgs mode 
with $\omega^2\geq \beta/2=\mh^2$. Therefore, all Abelian monopoles with $\nu\neq \pm 1$  are stable in this sector. 

Consider now Eqs.\eqref{magn} for $\nu=\pm 1$, first in the case when the background is Abelian, $f=0$, $\phi=1$. 
Then the two equations in \eqref{magn} decouple from each other. 
The second one reduces  to \eqref{magn1} and  is precisely the same as Eq.\eqref{B3} derived above. 
The first equation in \eqref{magn} becomes  the same as one of the two equations in \eqref{B2}, 
\be                \label{Ybound}
\left(-\frac{d^2}{dr^2}-\frac{1}{r^2}+\frac{g^2}{2}\right)Y_1=\omega^2 Y_1\,.
\ee
This provides  a  classical example of an equation admitting infinitely many bound states. Indeed, 
for $\omega=0$ the solution in the $r\ll 1$ region where one can neglect the $g^2/2$ term reads 
\be
Y_1={\cal A}\,\sqrt{r}\,\cos\left(\frac{\sqrt{3}}{2}\,\ln\frac{r}{r_0}\right),
\ee
where ${\cal A},r_0$ are integration constants. 
This solution  oscillates infinitely many times as $r\to 0$.  
According to the well-known Jacobi criterion \cite{gelfand2000calculus}, 
since the zero-energy solution oscillates, 
there are infinitely many bound states with $\omega^2<0$. As a result, 
the $\nu=1$ Abelian monopole in unstable in this sector and the  same is true for $\nu=-1$  
when one obtains the same equation up to $Y_1\to Y_4$. 

Summarizing, we have recovered in  \eqref{magn1}, \eqref{Ybound}
two of the four equations \eqref{B1}--\eqref{B3} previously obtained in the Abelian case. 

\begin{figure}
    \centering

    \includegraphics[scale=0.85]{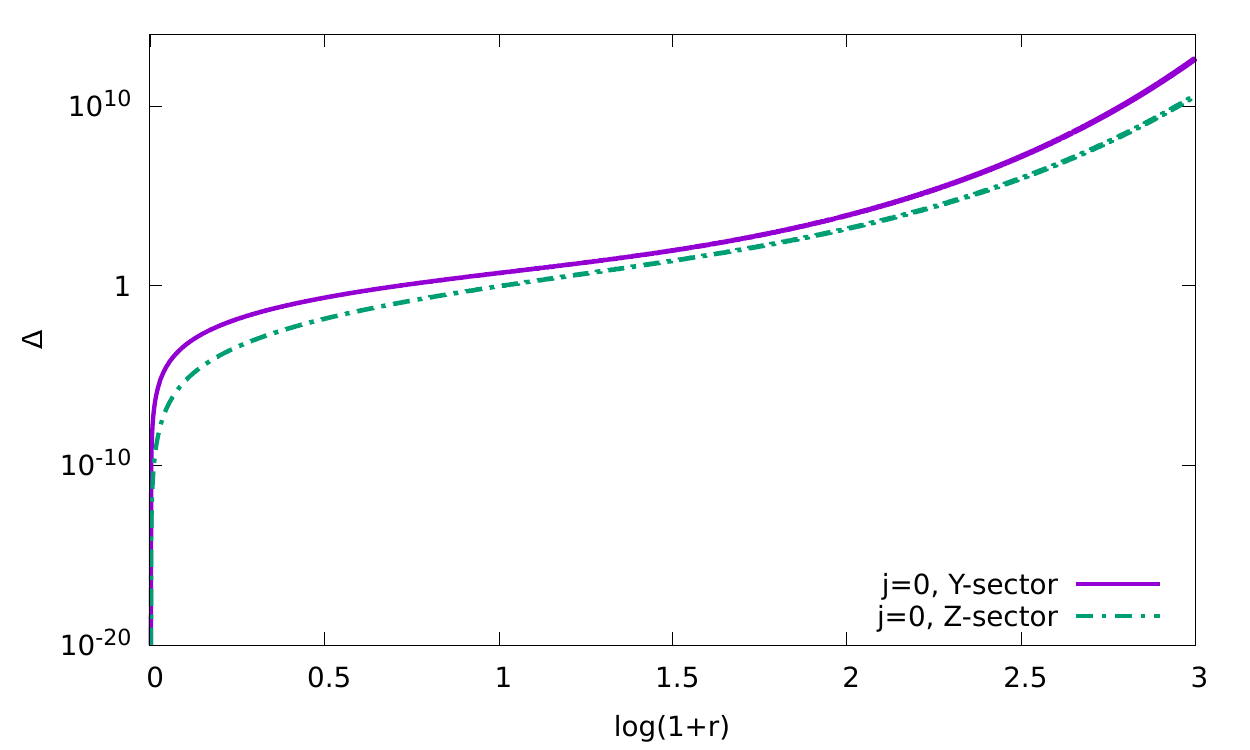}

    \caption{The Jacobi determinant for the CM monopole perturbations in the $Y,Z$  sectors with $j=0$.}
    \label{Jc1}
\end{figure}

Let us now return Eqs.\eqref{magn}  and consider  the non-Abelian CM background 
with $f\neq 0$, $\phi\neq 1$. In this case one can apply the generalization of the Jacobi criterion  for multi-channel 
systems   \cite{gelfand2000calculus}. 
Setting again $\omega=0$, one can check that in the  $r\to 0$ limit, where $f\to 1$ and $\phi\to 0$, 
the regular at the origin solutions of  Eqs.\eqref{magn}   comprise a two-parameter set
labeled by two integration constants $C_1$ and $C_2$,
\be
Y_1=C_1\times (r^2+\ldots),~~~~~Y_7=C_2\times \left( r^\frac{1+\sqrt{3}}{2}+\ldots\right),
\ee
where the dots denote subleading terms. This determines two independent solutions of the equations: $\left(Y_1^{(1)},Y_7^{(1)}\right)$
obtained for  $(C_1,C_2)=(1,0)$,   and $\left(Y_1^{(2)},Y_7^{(2)}\right)$ 
obtained for $(C_1,C_2)=(0,1)$.   Integrating the equations numerically then allows one to compute the 
Jacobi determinant 
\be             \label{Jac1}
\Delta(r)=
\left|
\begin{matrix}
Y_1^{(1)}(r)  & ~Y_7^{(1)}(r)  \\
Y_1^{(2)}(r)  & ~Y_7^{(2)}(r) 
\end{matrix}
\right|,
 \ee
 and if $\Delta(r)$ crosses zero for $r>0$, then there are bound states with $\omega^2<0$. 
 However, as seen in Fig.\ref{Jc1}, 
 $\Delta(r)$  never vanishes, hence the CM monopole is stable in this sector. 
 
\subsubsection{$Z$-sector}

Setting again $\nu=1$, the $Z$-equations \eqref{Schrod} reduce for $j=0$ to a 4-channel system for the non-vanishing
amplitudes $Z_1,Z_5,Z_6,Z_8$. This time there is a non-trivial residual gauge symmetry \eqref{residualZ}
generated by the  gauge parameters $v_{\pm}=u\pm u_3$ which have spin weignt $s=0$ and do not vanish for $j=0$. 
Using the fact that $\kappa\kappa_{+}=1$ for $\nu=1$, the gauge transformations \eqref{residualZ} reduce to 
\be                 \label{gauZ}
Z_1&\to& Z_1+\frac{f}{\sqrt{2} r}\,(v_{-}-v_{+})\,,~~~~~~Z_5\to Z_5-\frac{g\phi}{\sqrt{2}}\, v_{-}\,, \nn \\
Z_6&\to& Z_6-g\,r\left(\frac{v_{-}}{r}\right)^\prime\,,~~~~
Z_8\to Z_8 +\frac{r}{2g^\prime}\left(\frac{v_{-}-v_{+}}{r}\right)^\prime-\frac{g^2}{g^\prime}\, r\left(\frac{v_{-}}{r}\right)^\prime\,,
\ee
and this implies that the following two ``calligraphic" combinations
\be
{\cal Z}_1=Z_8-\frac{g}{g^\prime}\, Z_6-\frac{r}{\sqrt{2}\,g^\prime}\left(\frac{Z_1}{f}\right)^\prime\,,~~~~~~~~
{\cal Z}_2=Z_6-\sqrt{2}\,r\,\left(\frac{Z_5}{r\phi} \right)^\prime \,,~~~~
\ee
are gauge invariant. Using Eqs.\eqref{Schrod} one finds  that equations for ${\cal Z}_1$ and ${\cal Z}_2$
decouple from the rest. Setting
\be
{\cal Z}_1=\frac{r\,\psi_1}{2g^\prime f},~~~~~~~
{\cal Z}_2=-\frac{\psi_2}{\phi},~~~~~
\ee
the two decoupled gauge-invariant equations read 
\be                   \label{el0}
\left(-\frac{d^2}{dr^2}-\omega^2+\frac{f^2+1}{r^2}+\frac{2 f^{\prime 2}}{f^2}-\frac{g^2\phi^2}{2}\right)\psi_1+\frac{g\phi\,f}{r}\,\psi_2&=&0,\nn \\
\left(-\frac{d^2}{dr^2}-\omega^2
+
2\left(\frac{(r\phi)^\prime}{r\phi}\right)^2+\frac{\phi^2}{2}+\frac{\beta}{4}\,(1-\phi)^2-\frac{f^2}{2r^2}
\right)\psi_2+\frac{g\phi\,f}{r}\,\psi_1&=&0. 
\ee
This agrees with Eqs.\eqref{eig2} derived above. 
Assuming $f,\phi$ to correspond to the CM background, we 
integrate these equations to obtain  the determinant $\Delta(r)$ constructed from 
two linearly independent and regular at the origin solutions similar to  \eqref{Jac1}. 
As seen in Fig.\ref{Jc1}, the determinant is positive, 
 hence the CM 
monopole is stable in this perturbation sector as well. 

The Abelian case where $f=0$ and $\phi=1$ should be considered separately. The gauge transformations \eqref{gauZ} imply that the 
gauge-invariant amplitudes in this case are 
\be              \label{Z1}
{\cal Z}_1=Z_1,~~~~~~~~~{\cal Z}_2=Z_6-\sqrt{2}\,r\,\left(\frac{Z_5}{r} \right)^\prime \,,~~~~
\ee
and these fulfill equations 
\be            \label{Z2}
\left(-\frac{d^2}{dr^2}-\frac{1}{r^2}+\frac{g^2}{2}\right){\cal Z}_1=\omega^2 {\cal Z}_1\,,~~~~~
\left(-\frac{d^2}{dr^2}+\frac{2}{r^2}+\frac{1}{2}\right){\cal Z}_2=\omega^2 {\cal Z}_2\,.~~~~~
\ee
These are precisely the second of the $W$-equations \eqref{B2} and the $Z$-equation \eqref{B1}. 
Together with \eqref{magn1},\eqref{Ybound},  this reproduces 
all four equations \eqref{B1}--\eqref{B3} previously obtained in the Abelian case. 
The spectrum in the ${\cal Z}_2$-channel is positive 
but the ${\cal Z}_1$-equation admits infinitely many bound states. 

The above discussion corresponds to $\nu=1$ but for $\nu=-1$ one obtains the same equations, up to the replacement $Z_1\to Z_4$. 
Let us now consider Abelian backgrounds with $f=0$, $\phi=1$ and for $\nu\neq \pm 1$. In this case one should set $Z_1=0$ since 
its spin weight $s=\nu-1\neq 0$, hence  there remains only the ${\cal Z}_2$  amplitude defined in \eqref{Z1}.
It fulfills the same equation as in \eqref{Z2} so that there are no negative modes in this case. 

Summarizing, the $\nu=\pm 1$ Abelian monopoles show infinitely many instabilities in the $j=0$ sector while 
the non-Abelian CM monopole and all Abelian monopoles 
with $\nu\neq \pm 1$ are stable in this sector.

\begin{center}
\begin{table}            \label{Tab1}
 \begin{tabular}{|c | m{3em} | m{3em} | m{3em} | m{5em}  | m{3em} | m{3em} | m{3em} | m{3em} | m{3em} |}
\hline
$k$ &$~~~1$  &  $~~~2$  & $~~~3$   &  $~~~~~4$  & $~~~5$  & $~~~6$    & $~~~7$    & $~~~8$   &  $~~~9$ \\
\hline
~~$s$ for $Y_k:~$ & $~1-\nu$ & $~-\nu$ &  $~-\nu$ &  $~\pm (1+\nu)$ &  $~~~1$ & $~~~1$ & $~~~0$ &  & \\
\hline
~~$s$ for $Z_k:~$ & $~1-\nu$ & $~-\nu$ &  $~-\nu$ &  $~\pm (1+\nu)$ &  $~~~0$ & $~~~0$ & $~~\pm 1$ & $~~~0$ & $~~\pm 1$\\
\hline
\end{tabular}
\caption{Values of the spin weight $s$ for $Y_k$  and $Z_k$.  }
\end{table}
\end{center}

\subsection{$j=1$ sector}
Table I shows spin weights $s$ for amplitudes $Y_k$ ($k=1,\ldots, 7$) and $Z_k$ ($k=1,\ldots, 9$). 
In some cases $s$ is defined only up to sign, 
as for example $Y_4=\kappa\kappa_{-}(X_{13}-X_{22})$ where  $X_{13}$ has spin wight $-1-\nu$ and 
$X_{22}$ has spin wight $1+\nu$. However, only the absolute value $|s|$ is important in what follows. 

Let us set $\nu=1$. Then for $Y_4$ and $Z_4$ one has $|s|=2$ hence these amplitudes should be set to zero if $j=1$. 
The perturbation equations \eqref{Schrod} then become 
\be             \label{eee}
-Y_k^{\prime\prime}+\sum_{m=1}^7 U_{km}Y_m=\omega^2 Y_k\,,~~~~~~
-Z_k^{\prime\prime}+\sum_{m=1}^9 V_{km}Z_m=\omega^2 Z_k\,,~~~~~~~k,m\neq 4\,,
\ee
where $U_{km}$ and $V_{km}$ are defined by the expressions shown in
Appendix D and restricted  to  $\nu=j=1$. There are 6 and 8 equations, respectively,  in the $Y$ and $Z$ sectors.
These equations admit residual gauge symmetry expressed by \eqref{residualY} and \eqref{residualZ}. 
One can pass to gauge-invariant amplitudes similarly to what was done above, and then both the $Y$ and $Z$ sectors 
will contain only 5 equations for 5 gauge-invariant amplitudes. 
However, to check the stability of the CM background is actually easier by using  Eqs.\eqref{eee}. 
These equations describe  both physical modes and gauge modes and if they do not show instabilities then 
the physical modes are all stable.

\begin{figure}
    \centering

    \includegraphics[scale=0.85]{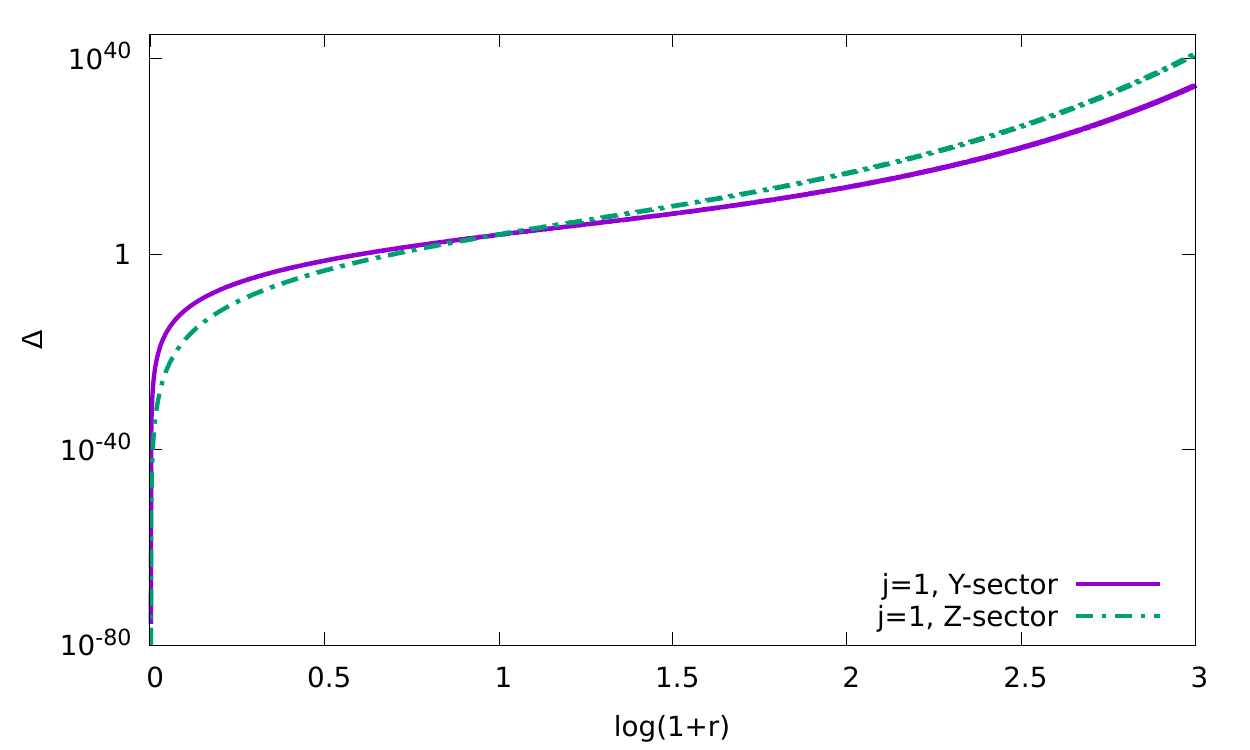}

    \caption{The Jacobi determinant for the CM monopole perturbations in the $Y$,$Z$  sectors with $j=1$.}
    \label{Fig2a}
\end{figure}

We  apply the Jacobi criterion to \eqref{eee}. Let us illustrate the procedure within the $Y$-channel 
containing 6 equations. To apply the criterion, we set $\omega=0$ and look for 6 independent solutions $Y_k^{(a)}$
($a=1,\ldots 6$) that are regular at the origin. The equations imply that near $r=0$ all solutions should have the structure 
$
Y_k(r)=c_k\, r^\alpha (1+\ldots)
$
where the 6 constants $c_k$ should fulfill a system of linear homogeneous algebraic equations.  
Its determinant vanishes for 12 values of $\alpha$ among which 6 are strictly positive. For example, 
if $g^2=3/4$, which is close to $\cos^2\thetaw=0.77$, then one finds 
$\alpha=1,2,3,3,(1+\sqrt{3})/2,(1+\sqrt{15})/2$. This gives rise to 6 regular at the origin solutions, and extending them
numerically allows one to compute the Jacobi determinant 
\be
\Delta(r)=\left| Y_k^{(a)}(r)\right|~~~~~~\text{where}~~~a=1,\ldots 6,~~k=1,2,3,5,6,7.
\ee
A similar procedure is used in the $Z$-sector. As seen in Fig.\ref{Fig2a}, the Jacobi determinant is positive both in the $Y$ and $Z$ sectors,
hence the CM monopole is stable with respect to $j=1$ perturbations. This is an important conclusion since, contrary to what one 
may expect, this $n=2$ monopole is stable under dipole deformations and does not split into two $n=1$ monopoles. 

There remains to study the stability of the Abelian monopoles with $f=0$, $\phi=1$. For $\nu=1$ the equations 
are provided by \eqref{eee} and it turns out that this Abelian monopole is also stable under
dipole deformations, although it is unstable in the $j=0$ sector. For $\nu>2$ one should return to the generic equations \eqref{Schrod}. 
For $\nu=2$, as seen in Table I, the amplitudes $Y_2,Y_3,Y_4,Z_2,Z_3,Z_4$ should be set to zero if $j=1$ since their spin weight 
is too large. All equations then decouple from each other and show no instabilities,  apart from the $Y_1$ and $Z_1$ 
channels where one finds the equation with infinitely many bound states, 
\be                \label{Ybound1}
\left(-\frac{d^2}{dr^2}-\frac{2}{r^2}+\frac{g^2}{2}\right)Y_1=\omega^2 Y_1\,,
\ee
and exactly the same equation for $Z_1$. 
Therefore, the $\nu=2$ monopole is unstable. The same conclusion applies for $\nu=-2$ (one sets $Y_1=Y_2=Y_3=0$, $Z_1=Z_2=Z_3=0$ 
in this case). 
No instabilities 
in the $j=1$ sector are found for $|\nu|>2$. 

Summarizing, the non-Abelian CM monopole  and all Abelian monopoles with an integer $\nu$ are stable with respect to dipole
deformations, apart from the $\nu=\pm 2$ Abelian monopole which shows infinitely many negative modes in the $j=1$ perturbation sector. 

\subsection{Higher values of $j$}

\begin{figure}
    \centering

    \includegraphics[scale=0.85]{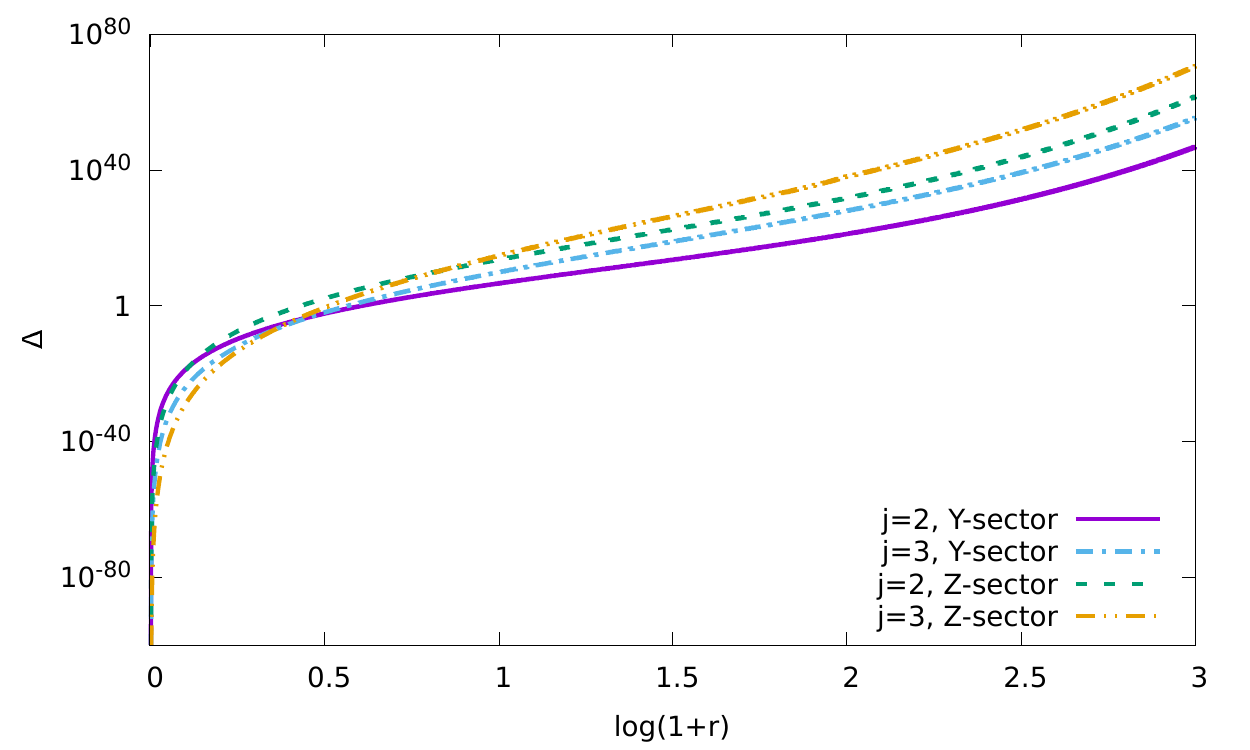}

    \caption{The Jacobi determinant for the CM monopole perturbations in the $Y$,$Z$  sectors with $j=2,3$.}
    \label{Fig22}
\end{figure}

If $j\geq 2$ and $\nu=1$ then, as seen in Table I, all amplitudes $Y_k$, $Z_k$ have spin weights $|s|\leq j$ and hence none of them vanish,
so that one should consider the general equations \eqref{Schrod}. 
Appliyng the Jacobi criterion, we find 
that they do not admit negative modes for $j=2,3$ (see Fig.\ref{Fig22}). 
One could repeat the procedure also for higher values of $j$, 
but since the centrifugal barrier grows with growing $j$, the existence of negative modes becomes  implausible. 
Therefore, since the  CM monopole has no negative modes for $j=0,1,2,3$, this strongly indicates  that it is stable with respect to all 
perturbations. 

For the Abelian monopoles the conclusion is different. We already know that the $\nu=1$ monopole is unstable in the $j=0$ sector 
whereas  the $\nu=2$ monopole is unstable in the $j=1$ sector. It turns out that
for any given  $|\nu|>2$ the corresponding monopole is  
unstable in the sector with 
$j=|\nu-1|$ and is stable with respect to all other perturbations. The instability is due to the particular 
term in the potentials $\hat{U}$ and $\hat{V}$ (see Appendix D),
\be
\frac{j(j+1)-\nu^2}{r^2}\,.
\ee
The magnetic charge makes the negative contribution to this term, which  is the well-known effect for  monopoles \cite{Coleman:1982cx}.
If this negative  contribution overcomes the positive centrifugal term, then the potential becomes attractive, which  produces the instability. 
For most channels this does not happen since $j$ should be larger than the spin weight determined by $\nu$. However,
if $|\nu|>1$ and $j=|\nu|-1$ then 
\be
\frac{j(j+1)-\nu^2}{r^2}=-\frac{|\nu|}{r^2}\,.
\ee
If $|\nu|>1$ and $j=|\nu|-1$ then $j<|\nu|$ and hence, 
as seen in Table I, 
one should set  $Y_2=Y_3=Z_2=Z_3=0$ and either $Y_4=Z_4=0$ if $\nu>0$ or 
$Y_1=Z_1=0$ if $\nu<0$. 
For example, if $\nu<0$ then the spin weight of $Y_1, Z_1$ is $s=1-\nu<0$ hence $|s|=|\nu|-1=j$ so that these amplitudes do not vanish,
whereas for $Y_2,Y_3,Z_2,Z_3$ one has $|s|=|\nu|>j$  and these amplitudes should be set to zero. 
As a result, the $Y_k$ and $Z_k$ equations with $k=1,2,3,4$ reduce to 
\be                \label{YYbound}
\left(-\frac{d^2}{dr^2}-\frac{|\nu| }{r^2}+\frac{g^2}{2}\right){\cal Y}=\omega^2 {\cal Y}\,,
\ee
where ${\cal Y}=Y_1$ or ${\cal Y}=Z_1$ if $\nu>0$, and ${\cal Y}=Y_4$ or ${\cal Y}=Z_4$ if $\nu<0$. 
This equation admits infinitely many bound states with $\omega^2<0$. 
The remaining $Y_k$  equations with $k=5,6,7$ and the $Z_k$ equations with $k=5,6,7,8,9$ decouple and show no instability. 
The conclusion is that the  Abelian monopole with $n=2\nu$ and  $|\nu|\geq 1$
has infinitely many instabilities  in the perturbation channel with $j=|\nu|-1$ but is stable with respect to all other perturbations. 

Up to now we have assumed $\nu$ to be integer, but  for the Abelian monopoles $\nu$ can assume also half-integer vales. 
The spin weights $s$ of $Y_k$ and $Z_k$ amplitudes with $k=1,2,3,4$ 
are then also half-integer and therefore their  total angular momentum  $j=|s|, |s|+1,\ldots$ is half-integer as well.
It is well-known that $j$ can assume half-integer values in the presence of monopoles \cite{Brandt:1979kk}. 
The azimuthal quantum number $m=-j,-j+1,\ldots ,j$ is also half-integer, but 
after passing to the regular gauge with the  transformation \eqref{toreg}, the perturbations become proportional to 
$e^{i(m\pm \nu)\varphi}$ hence  they are single-valued functions of $\varphi$. The above analysis then directly applies
with the same conclusion: for $j=|\nu|-1$ there are unstable modes both in the $Y$ and $Z$ sectors which fulfill 
Eq.\eqref{YYbound}.
Therefore, monopoles with a half-integer $\nu=3/2,5/2,\ldots$ are unstable as well. However, this  conclusion does not apply to the 
fundamental monopole with $\nu=\pm 1/2$ because $j=|\nu|-1$ should be non-negative. Therefore, the decay channel 
is closed and  the fundamental monopole is stable.

\section{SUMMARY OF RESULTS AND CONCLUDING REMARKS} 

To summarize our results, we have studied spherically symmetric sector  of the 
electroweak theory. This sector admits 
Abelian magnetic monopoles with magnetic charges $n=\pm 1, \pm 2,\pm 3, \ldots $ and  one solution describing 
the non-Abelian monopole of Cho-Maison with $n=\pm 2$. All these monopoles have infinite energy 
due to the Colombian behaviour of the U(1) magnetic 
field at the origin. These solutions were known before but 
 a systematic analysis of 
the electroweak spherically symmetric sector  has been lacking so far.  In addition, 
we discover spherically symmetric  electroweak oscillons which were not known before, but this issue will 
be discussed  separately   \cite{GGV}.

Our principal  result is the analysis of the perturbative stability of the electroweak monopoles. To the best of our knowledge, 
this has not been done before. Our  main thrust  was on studying stability  of the non-Abelian CM monopole,
and we have found that this solution is stable. The fundamental Abelian monopole with $n=\pm 1$ is also stable, but all other 
 Abelian monopoles are unstable. The Abelian monopole with the magnetic charge $|n|>1$ 
decays within the perturbative channel with angular momentum $j=|n/2|-1$. It is interesting that these properties 
of electroweak Abelian monopoles  are very similar to those for the Abelian monopoles embedded into the pure Yang-Mills theory \cite{Yoneya:1977yi,Brandt:1979kk}.

One may wonder  what the unstable monopoles decay into when perturbed.  
The simplest would be to think that  the Abelian monopole with $|n|>1$ 
decays into $|n|$ fundamental monopoles, which mutually repel  and fly away from each other. However, this hypothesis encounters problems 
already in the simplest case of $n=\pm 2$. Indeed, for the $|n|=2$ monopole to split in two, it has to be unstable with respect to 
dipole perturbations  within the $j=1$ sector. However, it is unstable only within the $j=0$ sector and 
only its spherically symmetric perturbations grow in time. 

Since all unstable modes of the $|n|=2$ Abelian monopole are spherically symmetric, 
a part of the growing perturbation will form a spherical wave and escape to infinity. 
The monopole will be loosing energy and will evolve 
 toward a less energetic  spherically symmetric  state containing  a non-Abelian condensate in its center. 
 It should finally approach the stable CM configuration, which also has $|n|=2$. Even though the CM monopole has infinite energy,
 it is less energetic than the Abelian monopole  since in both cases the energy is $E=E_0+E_1$ where $E_0$ diverges 
 but $E_1$ is finite for the CM monopole and diverges for the Abelian monopole (see Eq.\eqref{Energy}). 
 Therefore, the CM monopole can be viewed as a stable remnant of the decay of the  $|n|=2$  Abelian monopole. 
 
\begin{figure}
    \centering

   \includegraphics[scale=0.4]{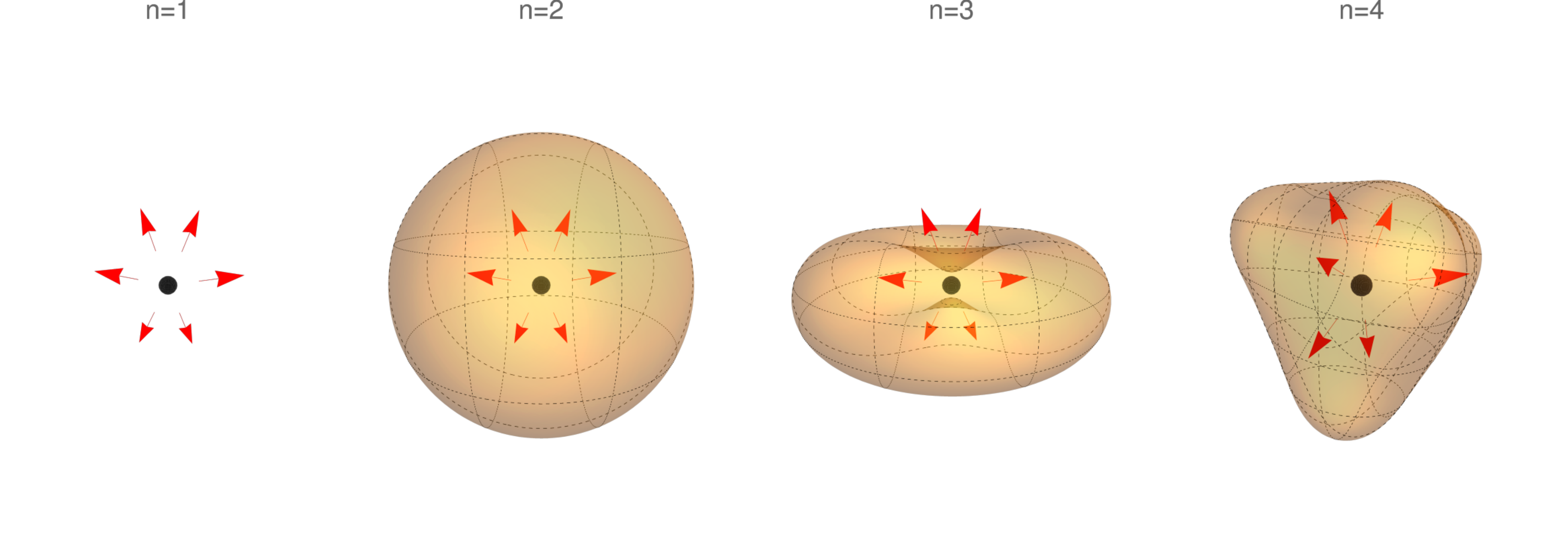}

    \caption{A schematic sketch of electroweak monopoles. The $n=1$ monopole is Abelian while the others have a non-Abelian core
    containing the pointlike singularity of the U(1) field in the center. 
    Monopole solutions with $|n|=3,4,\ldots $ (not yet constructed)  are not  spherically symmetric. }
    \label{Fig33}
\end{figure}

 One may conjecture that for the Abelian monopoles with $|n|>2$ the situation is similar 
 and every momopole decays by emitting radiation and condensing to an equilibrium 
 non-Abelian  state. However, the monopole with $|n|>2$ decays within the sector with $j=|n/2|-1>0$, hence the final 
 state of its evolution will be non-spherically symmetric. This suggests that the CM monopole may be just the first member of a
 sequence of more general and hitherto unknown non-Abelian solutions labeled by their magnetic charge $n$. 
 Only the CM monopole is spherically symmetric while non-Abelian  monopoles with $|n|>2$ should not be rotationally invariant.
 An attempt to show this  is schematically  presented in  Fig.\ref{Fig33}.

 Of course, this conjecture should be verified by explicitly constructing non-spherically symmetric solutions with $|n|>2$
 (an  axially symmetric solution  with $n=1$ was reported in  \cite{Teh:2014xva},
 but so far this has not been confirmed independently).
 At the same time, the conjecture  is corroborated by the evidence obtained via taking 
 gravity into account.
 Unless for quantum renormalization effects (which are unlikely to apply if $|n|\gg 1$), 
 the energy of all flat space monopoles is infinite. Gravity provides a natural cutoff by creating an event horizon, such that all Abelian 
 monopoles are described by the black hole geometry of Reissner-Nordstrom. It turns out that large black holes are  dynamically stable,
 but instabilities appear for smaller black holes.  Varying the horizon size, one can detect the moment
 when the instability just starts to appear and when it is not yet a negative mode but a zero mode \cite{GVin}. 
 Such zero modes provide 
 the perturbative approximation for the new non-Abelian solution branches bifurcating with the Abelian branch.  Starting 
 at the bifurcation point and iteratively decreasing the event horizon size, one can construct magnetically charged  hairy black holes 
 with non-Abelian fields in the exterior region.  If one wishes, one can finally send
 the Newton constant to zero to obtain non-Abelian  solutions in flat space. 
 
 A similar observation of the stability change for magnetically charged black holes has been made 
 within the context of a theory which 
 is similar although not exactly identical to the electroweak theory \cite{Ridgway:1994sm,Ridgway:1995ke,Ridgway:1995ac}. 
 It was similarly suggested that the phenomenon could be used to 
 construct new non-Abelian solutions without spherical symmetry. 
 However, such a construction has never been carried out  at the fully non-perturbative level. 
 
 The viewpoint that magnetic monopoles in the electroweak theory should be generically non-spherically symmetric was recently 
 advocated by Maldacena \cite{Maldacena:2020skw}, who argued that the spherically symmetric 
 magnetic field is homogeneous on the sphere and hence should be unstable with respect to segmentation  into vortices
 due to the Meissner effect. 
 One may finally note that t'Hooft-Polyakov monopoles for $|n|>2$  are also not spherically symmetric 
 \cite{Forgacs:1980ym}.

 \acknowledgements
 We thank Eugen Radu and  Yasha Shnir for discussions. 

\section*{APENDIX A: SPHERICALLY SYMMETRIC ELECTROWEAK FIELDS  \label{AppA}}
\renewcommand{\theequation}{A.\arabic{equation}}
\setcounter{equation}{0}

In this Appendix we discuss equivalent forms of the spherically symmetric fields expressed by 
Eqs.\eqref{W2},\eqref{B},\eqref{Higgs} in the main text. Those expressions are convenient for calculations, 
but one should have in mind that both the SU(2) field \eqref{W2} and U(1) field \eqref{B} are 
singular at the symmetry-axis where $\vartheta=0,\pi$
because  their $\varphi$-components do not vanish there. 
It is important to check that the singularity can be removed by passing to a regular gauge, local or global, 
since otherwise Eqs.\eqref{W2},\eqref{B},\eqref{Higgs} would make no sense. It turns out that the singularity can indeed be 
removed, but only for special values of the parameter $\nu$ in \eqref{W2}. 

The singularity in the SU(2) field \eqref{W2} can be removed by the gauge transformation 
\eqref{gauge}
generated by \cite{Arafune:1974uy}
 \be                   \label{Ureg} 
 {\rm U}=
 \exp\left(- i\nu\varphi \T_3\right)
 \exp\left(- i\vartheta \T_2\right)
 = 
\begin{pmatrix}
\cos\frac{\vartheta}{2}\,e^{-i{\nu \varphi/2}}  & ~-\sin\frac{\vartheta}{2}\,e^{-i{\nu \varphi/2}}  \\
\sin\frac{\vartheta}{2}\,e^{i{\nu \varphi/2}}  ~& ~\cos\frac{\vartheta}{2}\, e^{i{\nu \varphi/2}} 
\end{pmatrix}.
 \ee
This transformation is singular itself since ${\rm U}$ has no limit for $\vartheta\to 0,\pi$ but 
the transformed SU(2) field becomes  regular in the new gauge, 
 \be                \label{W1}
W=(a_0\, dt+a_1\, dr)\T_r+\left(w_2\,T_\vartheta-(1-w_1) \T_\varphi\right) d\vartheta
+\nu\left(w_2\,\T_\varphi+(1-w_1) T_\vartheta\right)\sin\vartheta \,d\varphi.~~~~~~~
\ee
Here the angle-dependent generators 
\be           \label{TTa}
\T_r =n^a \T_a,~~~~~~\T_\vartheta=\partial_\vartheta \T_r,
~~~~~~\T_\varphi=\frac{1}{\nu\sin\vartheta}\,\partial_\varphi \T_r
\ee
are expressed in terms of the unit vector 
\be            \label{nn}
n^a=\left[\,\sin\vartheta\cos(\nu\varphi),\sin\vartheta\sin(\nu\varphi),\cos\vartheta \,\right],
\ee
and it is clear that the parameter $\nu$ should be integer, since otherwise $n^a$ is not single-valued. 
For $\nu=0$ generators \eqref{TTa} are not defined, but in this case  one can use the original form  \eqref{W2}
which  becomes regular. 

For $\nu=1$, using the Cartesian coordinates $x^a=r\,n^a$, the field \eqref{W1} can be represented in the form 
originally used by Witten   \cite{Witten:1976ck}, 
\be                  \label{W}
W=\T_a\, n^a\, a_0 dt+
\T_a\left(a_1\,n^a n_i  +\frac{w_2}{r}\,\delta^a_i +\frac{1-w_1}{r}\,\epsilon_{a ij }\,n^j \right) dx^i\,.
\ee
The gauge transformation \eqref{Ureg} does not affect the U(1) field \eqref{B}  while the Higgs 
 field \eqref{Higgs} changes, so that  in the new gauge 
\be              \label{B22}
B=B_\mu dx^\mu=b_0\, dt+ b_1\,dr+P\cos\vartheta\, d\varphi,~~~~~
\Phi=\phi\, e^{i\xi/2} \begin{pmatrix}
 -\sin\frac{\vartheta}{2}\,e^{-i{\nu \varphi/2}}  \\
\cos\frac{\vartheta}{2}\,e^{i{\nu \varphi/2}} 
\end{pmatrix}. 
\ee 
The $B$-field is singular at $\vartheta=0,\pi$ while 
the Higgs field has no definite value at the negative $z$-axis, for $\vartheta=\pi$, unless if $\nu=0$. However, 
performing an extra U(1) gauge transformation generated by 
\be
{\rm U}=\exp\left(-\frac{i\nu\varphi}{2}\right),
\ee
the SU(2) field \eqref{W1} does not change  while  the U(1) field and the Higgs field become, 
assuming that $P=\nu$ according to \eqref{Pnu}, 
\be              \label{B22a}
B_{+}=b_0\, dt+ b_1\,dr+\nu\, (\cos\vartheta-1)\, d\varphi,~~~~~
\Phi_{+}=\phi\, e^{i\xi/2} \begin{pmatrix}
 -\sin\frac{\vartheta}{2}\,e^{-i{\nu \varphi}}  \\
\cos\frac{\vartheta}{2}\,
\end{pmatrix}. 
\ee 
Both the B-field and Higgs field are now regular for $\vartheta=0$, but there is still the Dirac string singularity of the B-field at 
the negative $z$-axis, where $\vartheta=\pi$, whereas the Higgs field still has no limit there. 
Therefore, this gauge can be used only in the upper part of the sphere, for $\vartheta\in[0,\pi-\epsilon)$. 
However, 
after a further U(1) gauge transformation generated by 
\be            \label{trans}
{\rm U}=\exp\left(+i\nu\varphi\right)
\ee
one obtains 
\be              \label{B22b}
B_{-}=b_0\, dt+ b_1\,dr+\nu\,(\cos\vartheta+ 1)\, d\varphi,~~~~~
\Phi_{-}=\phi\, e^{i\xi/2} \begin{pmatrix}
 -\sin\frac{\vartheta}{2}\ \\
\cos\frac{\vartheta}{2}\,e^{+i{\nu \varphi}} 
\end{pmatrix},
\ee 
such that the singularity is now moved to the positive $z$-axis, $\vartheta=0$, while for $\vartheta=\pi$
everything is regular.  This gauge provides the regular description 
in the lower  part of the sphere  for 
$\vartheta\in(\epsilon,\pi]$. 
Therefore,  the $B$ and $\Phi$ fields will be completely regular if 
one uses two  local gauges: the gauge $B_{+},\Phi_{+}$ given by \eqref{B22a} in the upper part of the sphere and 
the gauge $B_{-},\Phi_{-}$ given by 
\eqref{B22b} in the lower part of the sphere. The transition from one local gauge to the other 
is performed in the equatorial region where $\epsilon<\vartheta<\pi-\epsilon$
and provided by the function \eqref{trans} which is 
regular and single-valued in this region.

Summarizing, the regular form of the fields is given by \eqref{W1} and by two local gauges \eqref{B22a} or \eqref{B22b} used,
respectively,  for $\vartheta\in [0,\pi-\epsilon)$ and for $\vartheta\in (\epsilon,\pi]$.

Returning back to Eqs.\eqref{W2},\eqref{B},\eqref{Higgs}, let us finally discuss the case where $w_1=w_2=0$.
Setting $P=\nu$, one has 
\be                \label{W2a}
W=\T_3\,(a_0\, dt+a_1\, dr)+\T_3\,\nu \cos\vartheta d\varphi,~~~~~~~
B=b_0\, dt+ b_1\,dr+\nu\cos\vartheta\, d\varphi\,. 
\ee 
Both SU(2) and U(1) fields are singular at the symmetry axis, but after the gauge transformation 
generated by 
\be               \label{toreg}
{\rm U}_\pm=\exp\left(\mp\frac{i\nu\varphi}{2}(1+\tau_3)\right)
\ee
they become
\be                \label{W2b}
W_\pm&=&\T_3\,(a_0\, dt+a_1\, dr)+\T_3\,\nu (\cos\vartheta\mp 1) d\varphi,~~~~~~~\nn \\
B_\pm&=&b_0\, dt+ b_1\,dr+\nu(\cos\vartheta\mp 1)\, d\varphi\,, 
\ee 
while the Higgs field \eqref{Higgs} does not change (because it has only the lower component). 
The fields $W_{+}, B_{+}$ are regular at $\vartheta=0$ and can be used in the northern hemisphere while 
$W_{-}, B_{-}$ are regular at $\vartheta=\pi$ and can be used in the southern hemisphere. Using these two local gauges 
provides a completely regular description. The transition from   $W_{+}, B_{+}$ to $W_{-}, B_{-}$ is provided by the
gauge transformation in the equatorial region with 
\be             \label{W2c}
{\rm U}=\exp\left(i\nu\,\varphi(1+\tau_3)\right)= 
\begin{pmatrix}
\exp\left( 2i\,\nu\varphi\right)  & 0  \\
0 ~& ~1
\end{pmatrix},
\ee
which is single-valued if $\nu$ is integer or {\it  half-integer}. The latter conclusion is very important, since the value $\nu=1/2$ is 
needed to describe the magnetic monopole with the lowest charge, which arises precisely when $w_1= w_2=0$. 

The upshot of our discussion is that the fields  \eqref{W2},\eqref{B},\eqref{Higgs} can be made regular by gauge 
transformations if $\nu\in \mathbb{Z}$ or when $2\nu\in \mathbb{Z}$ if  $w_1=w_2=0$. 

Before finishing this section, let us compare with 
the spherically symmetric sphaleron field.  
Depending on weather the SU(2) gauge field is chosen in the regular gauge \eqref{W1}  
or in the singular gauge  \eqref{W2}, the sphaleron Higgs field is given, respectively, by 
 \be               \label{Higgsa}
 \Phi_{\rm sph}^{\rm reg}=\phi\, \exp\left(i\xi\, \T_r \right)
 \begin{pmatrix}
0  \\
1
\end{pmatrix}, ~~~~~~
 \Phi_{\rm sph}^{\rm sing}=\phi\, \exp\left(i\xi\, \T_3 \right)
 \begin{pmatrix}
-\sin\frac{\theta}{2}\, e^{-\nu\phi/2}  \\
\cos\frac{\theta}{2}\, e^{\nu\phi/2} 
\end{pmatrix}.
\ee
This is compatible with the spherical symmetry if only the Weinberg angle vanishes and the U(1) field decouples. 
The difference with the monopole field can also be   seen in the matrix element 
$\langle \Phi_{\rm sph}^{\rm reg}|\tau^a|\Phi_{\rm sph}^{\rm reg}\rangle=-\phi^2\, \delta^a_3$ because for the monopole field 
$\Phi$  given in the regular gauge by  \eqref{B22a} or \eqref{B22b}
one has $\langle \Phi|\tau^a|\Phi\rangle=-\phi^2\, n^a$ where $n^a$ is the unit vector defined in \eqref{nn}.

\section*{APENDIX B:  ELECTROWEAK EQUATIONS IN THE SPHERICALLY SYMMETRIC  SECTOR  \label{AppB}}
\renewcommand{\theequation}{B.\arabic{equation}}
\setcounter{equation}{0}

Injecting the ansatz  \eqref{metr}--\eqref{Higgs}
to the WS equations \eqref{P2}, the angular dependence separates yielding  second order differential equations 
for functions depending on $t,r$, but also first order differential and algebraic 
constraints. The first algebraic constraint   reads
$
(P-\nu)\,\phi^2=0
$
and since $\phi\neq 0$,  one should set 
$P=\nu$. 
Taking this into account, the other algebraic constraints read
\be            \label{app0}
(\nu^2-1)w_a=0,~~~~~~~
(\nu^2-1)(w_1^2+w_2^2-1)\,w_a=0,~~~~a=1,2\,,
\ee
while the first order differential constraints are 
\be          \label{app2}
(\nu^2-1)(\dot{w}_1+a_0\,w_2)=0\,,~~~~~(\nu^2-1)(w^\prime_1+a_1\,w_2)=0\,,\nn \\
(\nu^2-1)(\dot{w}_2-a_0\,w_1)=0\,,~~~~~(\nu^2-1)(w^\prime_2-a_1\,w_1)=0\,. 
\ee
The second order equations come from the SU(2) sector, 
 \be         \label{app3}
 w_1^{\prime\prime}-\ddot{w}_1&=&\left(\frac{(w_1^2+w_2^1-1)}{r^2}+\frac{g^2\phi^2}{2}+a_1^2-a_0^2 \right) w_1
 +2\,(a_0\,\dot{w}_2-a_1 w_2^\prime)+(\dot{a}_0-a_1^\prime)\, w_2\, ,\nn \\
 w_2^{\prime\prime}-\ddot{w}_2&=&\left(\frac{(w_1^2+w_2^1-1)}{r^2}+\frac{g^2\phi^2}{2}+a_1^2-a_0^2 \right) w_2
 -2\,(a_0\,\dot{w}_1-a_1 w_1^\prime)-(\dot{a}_0-a_1^\prime)\, w_1\, ,\nn \\
\left( r^2(a_0^\prime-\dot{a}_1)\right)^\prime&=&
\frac{g^2 r^2 \phi^2}{2}\left(a_0-b_0+\dot{\xi} \right)
+2\left[(w_1^2+w_2^2)\,a_0
+\dot{w}_1\, w_2-\dot{w}_2\, w_1\right]\,,\nn \\
\left( r^2(a_0^\prime-\dot{a}_1)\right)^{\mbox{$\cdot$}}&=&
\frac{g^2 r^2 \phi^2}{2}\left(a_1-b_1+\xi^\prime \right)
+2\left[(w_1^2+w_2^2)\,a_1
+w^\prime_1\, w_2-w^\prime_2\, w_1\right]\,,
 \ee
 from the U(1) sector, 
 \be           \label{app4}
 \left( r^2(b_0^\prime-\dot{b}_1)\right)^\prime+\frac{g^{\prime 2} r^2 \phi^2}{2}\left(a_0-b_0+\dot{\xi} \right)&=&0\,,\nn \\
\left( r^2(b_0^\prime-\dot{b}_1)\right)^{\mbox{$\cdot$}}+\frac{g^{\prime 2} r^2 \phi^2}{2}\left(a_1-b_1+\xi^\prime \right) &=&0\,,
 \ee
and from the Higgs sector, 
 \be       \label{app5}
 &&\frac{1}{r^2}\left(r^2 \phi^\prime\right)^\prime-\ddot{\phi}=\left(\beta\,(\phi^2-1)
 +\left(a_1-b_1+\xi^\prime \right)^2 -
 \left(a_0-b_0+\dot{\xi} \right)^2+\frac{2}{r^2}\,(w_1^2+w_2^2)\right)\frac{\phi}{4}\,,\nn\\
 &&\left( r^2\phi^2 (a_1-b_1+\xi^\prime) \right)^\prime=\left( r^2\phi^2 (a_0-b_0+\dot{\xi}) \right)^{\mbox{$\cdot$}}\,.~~~~~~~~
 \ee
 These equations can be simplified by setting 
\bea                \label{fields-appa}
w_1&=&f\,\cos\alpha,~~~w_2=f\,\sin\alpha,~~~~
a_0=\Omega_0+\dot{\alpha},~~~a_1=\Omega_1+\alpha^\prime,~~~\nn \\
b_0&=&\dot{\xi}+a_0-\Theta_0,~~~~~
b_1={\xi}^\prime +a_1
-\Theta_1\,,
\eea
where $f,\alpha,\Omega_0,\Omega_1,\Theta_0,\Theta_1$ are functions of $t,r$. The amplitudes $\alpha,\xi$ correspond to pure gauge 
degrees of freedom and 
drop out from the equations, which leads to Eqs.\eqref{fp}--\eqref{con} in the main text. 
It is not difficult to see that the algebraic and first order differential constraints \eqref{app0},\eqref{app2} 
leave only two options: either $\nu^2=1$ or $w_1=w_2=0$, 
which can be expressed by the condition
$(\nu^2-1)\, f=0$ shown in Eq.\eqref{PP} in the main text.

\section*{APENDIX C:  PERTURBATION EQUATIONS AFTER VARIABLE SEPARATION  \label{AppC} }
\renewcommand{\theequation}{C.\arabic{equation}}
\setcounter{equation}{0}

In this Appendix we show the radial equations obtained by 
inserting the ansatz \eqref{anz1}--\eqref{anz5} into the perturbation equations \eqref{perteq}
and separating the angular and temporal variables. 
There are altogether 20 coupled ordinary differential equations for the 20 radial amplitudes $S_0(r),
\ldots,$ $X_{10}(r),\ldots,$ $H_1(r),\ldots, H_4(r)$. 
Defining the quantities 
\begin{subequations}
\begin{align}
{\cal C}_1&=\frac{\left( r X_{11}\right)^\prime}{r}-\omega\, X_{10} \label{Ca}
+\frac{j-\nu_-}{\sqrt{2}\, r}\, X_{12}
-\frac{j+\nu_+}{\sqrt{2}\, r}\, X_{13}
-\frac{\nu_-\, f}{\sqrt{2}\, r}\,  X_{32}
+\frac{\nu_+\, f}{\sqrt{2}\, r}\, X_{33}
+g^2\phi\, H_1 \,,\\
{\cal C}_2&=\frac{\left( r X_{21}\right)^\prime}{r}-\omega\, X_{20}  \label{Cb}
+\frac{j+\nu_+}{\sqrt{2}\, r}\, X_{22}
-\frac{j-\nu_-}{\sqrt{2}\, r}\, X_{23}
+\frac{\nu_+\, f}{\sqrt{2}\, r}\,  X_{32}
-\frac{\nu_-\, f}{\sqrt{2}\, r}\, X_{33}
-g^2\phi\, H_2 \,, \\
{\cal C}_3&=\frac{\left( r X_{31}\right)^\prime}{r}-\omega\, X_{30}  \label{Cc}
+\frac{\nu_-\, f}{2\sqrt{2}\, r}\,  (X_{13}+X_{22})
-\frac{\nu_+\, f}{2\sqrt{2}\, r}\,  (X_{12}+X_{23})\nn \\
&\hspace{7 cm} +\frac{j+1}{\sqrt{2}\, r}\, (X_{32}-X_{33}) 
+\frac{g^2}{2}\,\phi\, (H_4-H_3) \,, \\
{\cal C}_4&=\frac{\left( r S_{1}\right)^\prime}{r}-\omega\, S_{0}      \label{Cd}
+\frac{j+1}{\sqrt{2}\, r}\, (S_{2}-S_3)
+\frac{g^{\prime 2}}{2}\,\phi\, (H_3-H_4) \,,
\end{align}
\end{subequations}
where $\nu_\pm\equiv \nu\pm 1$, the 4  equations for the SU(2) amplitudes $X_{1\alpha}(r)$ can be represented in the form 
\begin{subequations}
\begin{align}
&~~~\hat{{\cal D}}X_{10}  +\frac{2\nu\, j\, f}{r^2}\,X_{30}=-\omega\,{\cal C}_1\,, \label{10a} \\
&\left(\hat{{\cal D}}-\frac{2}{r^2}\right)X_{11}  
+\frac{2\nu\, j\, f}{r^2}\,X_{31}
-\frac{\sqrt{2}\,(j-\nu_-)}{r^2}\, X_{12} 
+\frac{\sqrt{2}\,(j+\nu_+)}{r^2}\, X_{13}\nn \\
&~~~+\left(\frac{\sqrt{2}\, f}{r}\right)^\prime \left(\nu_+ X_{33}-\nu_-X_{32} \right)
+2g^2 \phi^\prime\, H_1=r\left(\frac{{\cal C}_1}{r}\right)^\prime \,,  \label{11a} \\
&\left(\hat{{\cal D}}-\frac{2\nu f^2}{r^2}\right)X_{12}  
-\frac{\sqrt{2}\,(j+\nu)}{r^2}\,X_{11} 
-\left(\frac{\sqrt{2}\,\nu_+\, f}{r}\right)^\prime X_{31}
 +\frac{2(\nu j+1)\,f}{r^2}\,  X_{32}\nn \\
&~~~+\frac{\nu_+\, g^2 f\,\phi}{\sqrt{2}\, r}\,  H_4=
\frac{\nu_+\, f\, {\cal C}_3 - (j+\nu)\, {\cal C}_1    }{\sqrt{2}\, r}\,,  \label{12a}  \\
&\left(\hat{{\cal D}} +\frac{2\nu f^2}{r^2}\right)X_{13}  
+\frac{ \sqrt{2}\,(j-\nu)}{r^2}\,X_{11} 
+\left(\frac{\sqrt{2}\,\nu_-\,f}{r}\right)^\prime X_{31}
+\frac{2(\nu j-1)\,f}{r^2}\,  X_{33}\nn \\
&~~~ -\frac{\nu_-\, g^2 f\,\phi}{\sqrt{2}\, r}\,  H_4=
\frac{(j-\nu)\, {\cal C}_1   - \nu_-\, f\, {\cal C}_3   }{\sqrt{2}\, r}\,, \label{13a} 
\end{align}
\end{subequations}
with the differential operator
\be                                \label{dif}
\hat{{\cal D}}\equiv \frac{d^2}{dr^2}+\omega^2-\frac{j(j+1)}{r^2}
-\frac{g^2\phi^2}{2}\,+\frac{\nu^2-f^2}{r^2}. 
\ee
The 4  equations for $X_{2\alpha}(r)$ read 
\begin{subequations}
\begin{align}
&~~~~~\hat{{\cal D}}X_{20}  -\frac{2\nu\, j\, f}{r^2}\,X_{30}=-\omega\,{\cal C}_2\,, \label{20a} \\
&\left(\hat{{\cal D}}-\frac{2}{r^2}\right)X_{21}  
-\frac{2\nu\, j\, f}{r^2}\,X_{31}
-\frac{\sqrt{2}\,(j+\nu_+)}{r^2}\, X_{22} 
+\frac{\sqrt{2}\,(j-\nu_-)}{r^2}\, X_{23}\nn \\
&~~~+\left(\frac{\sqrt{2}\, f}{r}\right)^\prime \left(\nu_+ X_{32}-\nu_-X_{33} \right)
-2g^2 \phi^\prime\, H_2=r\left(\frac{{\cal C}_2}{r}\right)^\prime \,, \label{20b} \\\
&\left(\hat{{\cal D}}+\frac{2\nu f^2}{r^2}\right)X_{22}  
-\frac{\sqrt{2}\,(j-\nu)}{r^2}\,X_{21} 
+\left(\frac{\sqrt{2}\,\nu_-\, f}{r}\right)^\prime X_{31}
 -\frac{2(\nu j-1)\,f}{r^2}\,  X_{32}\nn \\
&~~~+\frac{\nu_-\, g^2 f\,\phi}{\sqrt{2}\, r}\,  H_3=
-\frac{(j-\nu)\, {\cal C}_2 + \nu_-\, f\, {\cal C}_3}{\sqrt{2}\, r}\,, \label{20c} \ \\
&\left(\hat{{\cal D}} -\frac{2\nu f^2}{r^2}\right)X_{23}  
+\frac{ \sqrt{2}\,(j+\nu)}{r^2}\,X_{21} 
-\left(\frac{\sqrt{2}\,\nu_+\,f}{r}\right)^\prime X_{31}
-\frac{2(\nu j+1)\,f}{r^2}\,  X_{33}\nn \\
&~~~ -\frac{\nu_+\, g^2 f\,\phi}{\sqrt{2}\, r}\,  H_3=
\frac{(j+\nu)\, {\cal C}_2 + \nu_+\, f\, {\cal C}_3}{\sqrt{2}\, r}\,, \label{20d} 
\end{align}
\end{subequations}
the 4  equations for $X_{3\alpha}(r)$ are 
\begin{subequations}
\begin{align}
&\left(\hat{{\cal D}} -\frac{f^2+\nu^2}{r^2}\right) X_{30}
+\frac{\nu\,(j+1)\, f}{r^2}\,(X_{10}-X_{20})+\frac{g^2\phi^2}{2}\, S_0=-\omega\,{\cal C}_3\,, \label{30a} \\
&\left(\hat{{\cal D}} -\frac{f^2+\nu^2+2}{r^2}\right) X_{31}
+\frac{\nu\,(j+1)\, f}{r^2}\,(X_{11}-X_{21})
-\left(\frac{\nu_+\,f}{\sqrt{2}\,r} \right)^\prime (X_{12}+X_{23})\, \label{30b}  \\
&~+\left(\frac{\nu_-\,f}{\sqrt{2}\,r} \right)^\prime (X_{13}+X_{22})
+\frac{\sqrt{2}\,(j+1)}{r^2}\,(X_{33}-X_{32})
+\frac{g^2\phi^2 }{2}\,S_1
+g^2\,\phi^\prime\, (H_4-H_3)=r\left(\frac{{\cal C}_3}{r}\right)^\prime\,,\nn  \\
&\left(\hat{{\cal D}} -\frac{f^2+\nu^2}{r^2}\right) X_{32}
+\left(\frac{f}{\sqrt{2}\,r} \right)^\prime \left(\nu_+\,X_{21}-\nu_-\,X_{11}\right)
+\frac{[(j+1)\nu-1]\,f}{r^2}\,X_{12}
-\frac{[(j+1)\nu+1]\, f}{r^2}\,X_{22}\, \nn \\
&~~~ 
-\frac{\sqrt{2}\, j}{r^2}\, X_{31}
+\frac{g^2\phi^2}{2}\, S_2
+\frac{g^2\, f\,\phi}{2\sqrt{2}\, r}\,(\nu_-\,H_1+\nu_+\,H_2) 
=
\frac{f\left(\nu_-\,{\cal C}_1-\nu_+\,{\cal C}_2\right)}{2\sqrt{2}\,r}\,-\frac{j\,{\cal C}_3}{\sqrt{2}\,r}\,,\label{30c} \\
&\left(\hat{{\cal D}} -\frac{f^2+\nu^2}{r^2}\right) X_{33}
+\left(\frac{f}{\sqrt{2}\,r} \right)^\prime \left(\nu_+\,X_{11}-\nu_-\,X_{21}\right)
+\frac{(\nu_+\, j+1)\,f}{r^2}\,X_{13}
-\frac{(\nu_+\, j-1)\, f}{r^2}\,X_{23}\, \nn \\
&~~~ 
+\frac{\sqrt{2}\, j}{r^2}\, X_{31}
+\frac{g^2\phi^2}{2}\, S_3
-\frac{g^2\, f\,\phi}{2\sqrt{2}\, r}\,(\nu_+\,H_1+\nu_-\,H_2) 
=
\frac{f\left(\nu_-\,{\cal C}_2-\nu_+\,{\cal C}_1\right)}{2\sqrt{2}\,r}\,+\frac{j\,{\cal C}_3}{\sqrt{2}\,r}\,,\label{30d}
\end{align}
\end{subequations}
the equations for the U(1) amplitudes are 
\begin{subequations}
\begin{align}
&\hat{{\cal D}}_s\, S_0+\frac{g^{\prime 2}\phi^2}{2}\, X_{30}
=-\omega\,{\cal C}_4\,, \label{Sa} \\
&\left(\hat{{\cal D}}_s-\frac{2}{r^2}\right)\, S_1
+\frac{g^{\prime 2}\phi^2}{2}\, X_{31}
+\frac{\sqrt{2}\,(j+1)}{r^2}\,(S_3-S_2) 
+g^{\prime 2}\phi^\prime\,(H_3-H_4)
=r\left(\frac{{\cal C}_4}{r}\right)^\prime  \,,  \label{Sb}  \\
&\hat{{\cal D}}_s\, S_2
+\frac{g^{\prime 2}\phi^2}{2}\, X_{32}
-\frac{\sqrt{2}\, j}{r^2}\, S_1
+\frac{g^{\prime 2}f\,\phi}{2\sqrt{2}\,r}\,(\nu_- H_1+\nu_+ H_2)=-\frac{j\, {\cal C}_4}{\sqrt{2}\, r}\,,  \label{Sc}  \\
&\hat{{\cal D}}_s\, S_3
+\frac{g^{\prime 2}\phi^2}{2}\, X_{33}
+\frac{\sqrt{2}\, j}{r^2}\, S_1
-\frac{g^{\prime 2}f\,\phi}{2\sqrt{2}\,r}\,(\nu_+ H_1+\nu_- H_2)=\frac{j\, {\cal C}_4}{\sqrt{2}\, r}\,,  \label{Sd} 
\end{align}
\end{subequations}
and the equations for the Higgs amplitudes are 
\begin{subequations}
\begin{align}
&\left(\hat{{\cal D}}+\frac{f^2}{2r^2}+\frac{\beta}{4}(1-\phi^2)\right)H_{1}  
-\frac{\nu j f}{r^2} \, H_3 +\phi^\prime X_{11}      \label{ha}  \\
&~~~\hspace{6.5 cm}
+\frac{\nu_-\,f\,\phi }{2\sqrt{2}\,r}\, (S_2+X_{32})
-\frac{\nu_+\,f\,\phi}{2\sqrt{2}\,r}\,  (S_3+X_{33})=-\frac{\phi\,{\cal C}_1}{2}\,, \nn \\
&\left(\hat{{\cal D}}+\frac{f^2}{2r^2}+\frac{\beta}{4}(1-\phi^2)\right)H_{2}  
+\frac{\nu j f}{r^2} \, H_4 
-\phi^\prime X_{21}                   \label{hb}  \\
&~ ~~\hspace{6.5 cm}
+\frac{\nu_+\,f\,\phi }{2\sqrt{2}\,r}\, (S_2+X_{32})
-\frac{\nu_-\,f\,\phi}{2\sqrt{2}\,r}\,  (S_3+X_{33})=\frac{\phi\,{\cal C}_2}{2}\,, \nn \\
&\hat{{\cal D}}_h\, H_3
+\frac{f\,\phi}{2\sqrt{2}\,r}\,(\nu_- X_{22}-\nu_+ X_{23})
-\phi^\prime (X_{31}-S_1)           \label{hc}    \\
&~~~\hspace{6 cm}
  -\frac{\nu\,(j+1)\, f}{r^2}\, H_1+\frac{(1-\beta)\,\phi^2}{4}\, H_4=\frac{\phi}{2}\,({\cal C}_3-{\cal C}_4)\,, \nn \\
&\hat{{\cal D}}_h\, H_4
+\frac{f\,\phi}{2\sqrt{2}\,r}\,(\nu_+ X_{12}-\nu_- X_{13})
+\phi^\prime (X_{31}-S_1)           \label{hd}        \\
&~~~\hspace{6 cm}
+\frac{\nu\,(j+1)\, f}{r^2}\, H_2+\frac{(1-\beta)\,\phi^2}{4}\, H_3=\frac{\phi}{2}\,({\cal C}_4-{\cal C}_3)\,, \nn 
\end{align}
\end{subequations}
with the differential operators 
\be                 \label{dif1}
\hat{{\cal D}}_s&=&\frac{d^2}{dr^2}+\omega^2 -\frac{j(j+1)}{r^2}-\frac{g^{\prime 2}\phi^2}{2},~\nn \\
\hat{{\cal D}}_h&=&\frac{d^2}{dr^2}+\omega^2 -\frac{j(j+1)}{r^2}-\frac{f^2}{2 r^2}
+\frac{\beta}{4}-\frac{1}{4}\,(1+2\beta)\,\phi^2\, .
\ee
These equations are invariant under gauge transformations \eqref{gauge1}--\eqref{gauge5}. 
As discussed in the main text, imposing the harmonic gauge conditions ${\cal C}_1={\cal C}_2={\cal C}_3={\cal C}_4=0$ 
reduces the number of independent equations to 16 which can be 
split into two independent subsystems of 7 and 9 equations. 

\section*{APENDIX D:  PERTURBATION POTENTIALS  \label{AppD}}
\setcounter{equation}{0}
\renewcommand{\theequation}{D.\arabic{equation}}

In this Appendix we show the explicit expression for components of the $7\times 7$ matrix potential $\hat{U}$ and for the 
$9\times 9$ matrix potential $\hat{V}$ in the Schroedinger-type equations \eqref{Schrod}. 
These potentials are symmetric matrices. The components of $\hat{U}$ are 
\be
U_{11}&=&\frac{j(j+1)-\nu^2}{r^2}+\frac{(1+2\nu)\, f^2}{r^2}+\frac{g^2\phi^2}{2}\,,~~
U_{12}=\frac{\sqrt{2}\,(j+\nu)}{\kappa_+\, r^2}\,,~~U_{13}=U_{14}=0\,, \nn \\
U_{15}&=&\frac{g\,\sqrt{2}\,(1+\nu\,j)\,f}{\kappa_+\, r^2}\,, ~~~
U_{16}=-\frac{g^\prime \,\sqrt{2}\,(1+\nu\,j)\,f}{\kappa_+\, r^2}\,, ~~~
U_{17}=-\frac{g\,\nu_+\,f\,\phi}{2\kappa\,\kappa_+\, r}\,,       \nn \\
U_{22}&=&\frac{j(j+1)+2-\nu^2}{r^2}+\frac{f^2}{r^2}+\frac{g^2\phi^2}{2}\,,~~~~~
U_{23}=-g\sqrt{2}\,\phi^\prime\, , ~~~  
U_{24}=-\frac{\sqrt{2}\,\kappa_{-}\,(j+\nu_+)}{r^2}\,,~~~~~ \nn \\
U_{25}&=&2g\, \left(\frac{f}{r}\right)^\prime \,,~~~~~
U_{26}=-2g^\prime \, \left(\frac{f}{r}\right)^\prime \,,~~~~~U_{27}=0\,,\nn 
\ee
\be
U_{33}&=&\frac{j(j+1)-\nu^2}{r^2}+\frac{f^2}{2r^2}+\frac{g^2\phi^2}{2}+\frac{\beta}{4}(\phi^2-1)\,,~~~~U_{34}=0\,,       \nn \\
U_{35}&=&\frac{(g^{\prime 2}-g^2)\,f\,\phi}{\sqrt{2}\,r},~~~
U_{36}=\frac{\sqrt{2}\,gg^\prime \,f\,\phi}{r},~~~~~~
U_{37}=\frac{\nu\,\sqrt{j(j+1)}\, f}{r^2}\,, \nn \\
U_{44}&=&\frac{j(j+1)-\nu^2}{r^2}+\frac{(1-2\nu)\, f^2}{r^2}+\frac{g^2\phi^2}{2}\,,~~~~~\nn \\
U_{45}&=&\frac{\sqrt{2}\,g\,(\nu\,j-1)\,f}{\kappa_{-}\,r^2}\,,~~~
U_{46}=\frac{\sqrt{2}\,g^\prime \,(1-\nu\,j)\,f}{\kappa_{-}\,r^2}\,,~~~
U_{47}=\frac{g\nu_{-}\,f\,\phi}{2\kappa\kappa_{-}\,r}\,,~~~~~~~\nn \\
U_{55}&=&\frac{j(j+1)}{r^2}+\frac{2g^2\,f^2}{r^2}+\frac{\phi^2}{2}\,,~~~
U_{56}=-\frac{2\,gg^\prime\, f^2}{r^2}\,, ~~~~U_{57}=0\,,~~\nn \\
U_{66}&=&\frac{j(j+1)}{r^2}+\frac{2g^{\prime 2}\,f^2}{r^2}\,,~~~~U_{67}=0,~~~~
U_{77}=\frac{j(j+1)}{r^2}+\frac{f^2}{2r^2}+\frac{\beta}{4}\,(3\phi^2-1). 
\ee
The components of $\hat{V}$ read 
\be
V_{11}&=&\frac{j(j+1)-\nu^2}{r^2}+\frac{(1+2\nu)\, f^2}{r^2}+\frac{g^2\phi^2}{2}\,,~~
V_{12}=\frac{\sqrt{2}\,(j+\nu)}{\kappa_+\, r^2}\,,~~V_{13}=V_{14}=0\,, \nn \\
V_{15}&=&-\frac{g\,\nu_+\,f\,\phi}{2\kappa\,\kappa_+\, r}\,, ~~~~
V_{16}=-\frac{\sqrt{2}\, g\, \nu_{+}}{\kappa\kappa_{+}}\, \left(\frac{f}{r}\right)^\prime \,,~~~~~ 
V_{17}=\frac{\sqrt{2}\, g\,(1+\nu j)\, f}{\kappa_{+}\, r^2}     \nn \\
V_{18}&=&\frac{\sqrt{2}\, g^\prime\, \nu_{+}}{\kappa\kappa_{+}}\, \left(\frac{f}{r}\right)^\prime \,,~~~~~ 
V_{19}=-\frac{\sqrt{2}\, g^\prime \,(1+\nu j)\, f}{\kappa_{+}\, r^2} \,, \nn \\
V_{22}&=&\frac{j(j+1)+2-\nu^2}{r^2}+\frac{f^2}{r^2}+\frac{g^2\phi^2}{2}\,,~~~~~
V_{23}=\sqrt{2}\,g\,\phi^\prime\, , ~~~  
V_{24}=-\frac{\sqrt{2}\,\kappa_{-}\,(j+\nu_+)}{r^2}\,,~~~~~ \nn \\
V_{25}&=&0,~~~
V_{26}=\frac{2g\,\kappa\, (j+1)\,\nu f}{r^2},~~~~
V_{27}=-2g \nu\, \left(\frac{f}{r}\right)^\prime \,,~~~~~\nn \\
V_{28}&=&-\frac{2g^\prime\nu\sqrt{j(j+1)}\,f}{r^2}\,,~~~~
V_{29}=2g^\prime \nu\, \left(\frac{f}{r}\right)^\prime , \nn
\ee
\be
V_{33}&=&\frac{j(j+1)-\nu^2}{r^2}+\frac{f^2}{2r^2}+\frac{g^2\phi^2}{2}+\frac{\beta}{4}\,(\phi^2-1)\,,~~~~
V_{34}=0\,, ~~~~
V_{35}=\frac{\nu\,\sqrt{j(j+1)}\, f}{r^2}\,      \nn \\
V_{36}&=&0,~~~
V_{37}=\frac{(g^{\prime 2}-g^2)\,\nu\,f\,\phi}{\sqrt{2}\,r},~~~
V_{38}=0\,,~~~~
V_{39}=\frac{\sqrt{2}\,gg^\prime\nu \,f\,\phi}{r},
 \nn \\
V_{44}&=&\frac{j(j+1)-\nu^2}{r^2}+\frac{(1-2\nu)\, f^2}{r^2}+\frac{g^2\phi^2}{2}\,,~~~~~
V_{45}=\frac{g\nu_{-}f\,\phi}{2\kappa\kappa_{-}\, r}\,,~~~~~
V_{46}=\frac{\sqrt{2}\, g\, \nu_{-}}{\kappa\kappa_{-}}\, \left(\frac{f}{r}\right)^\prime \,,~\nn \\
V_{47}&=&-\frac{\sqrt{2}\,g\,(\nu\,j-1)\,f}{\kappa_{-}\,r^2}\,,~~~
V_{48}=-\frac{\sqrt{2}\, g^\prime\, \nu_{-}}{\kappa\kappa_{-}}\, \left(\frac{f}{r}\right)^\prime \,,~~~~
V_{49}=\frac{\sqrt{2}\,g^\prime\,(\nu\,j-1)\,f}{\kappa_{-}\,r^2}\,,~
~~~~~~\nn 
\ee
\be
V_{55}&=&\frac{j(j+1)}{r^2}+\frac{f^2}{2r^2}+\frac{\phi^2}{2}+\frac{\beta}{4}\,(\phi^2-1),~~~V_{56}=\sqrt{2}\,\phi^\prime,~~~
V_{57}=V_{58}=V_{59}=0\,,\nn \\
V_{66}&=&\frac{j(j+1)+2}{r^2}+\frac{2g^2\,f^2}{r^2}+\frac{\phi^2}{2}\,,~~~V_{67}=\frac{2\sqrt{j(j+1)}}{r^2}\,,~~
V_{68}=-\frac{2\,gg^\prime\, f^2}{r^2}\,, ~~~~V_{69}=0\,,~~\nn \\
V_{77}&=&\frac{j(j+1)}{r^2}+\frac{2g^{2}\,f^2}{r^2}+\frac{\phi^2}{2}\,,~~~~V_{78}=0,~~~~V_{79}=-\frac{2gg^\prime\, f^2}{r^2}\,,   \nn \\
V_{88}&=&\frac{j(j+1)+2}{r^2}+\frac{2g^{\prime 2}\,f^2}{r^2}\,,~~~~V_{89}=\frac{2\sqrt{j(j+1)}}{r^2},~~~~~
V_{99}=\frac{j(j+1)}{r^2}+\frac{2 g^{\prime 2}f^2}{r^2}. 
\ee


\begin{thebibliography}{10}

\bibitem{Dirac:1931kp}
P.~A.~M. Dirac, {\it {Quantised singularities in the electromagnetic field,}},
  {\sl Proc. Roy. Soc. Lond. A} {\bf 133} (1931), no.~821 60--72,
  [\href{http://dx.doi.org/10.1098/rspa.1931.0130}{{\sf
  doi:10.1098/rspa.1931.0130}}].

\bibitem{Wu:1976ge}
T.~T. Wu and C.~N. Yang, {\it {Dirac monopole without strings: monopole
  harmonics}},  {\sl Nucl. Phys. B} {\bf 107} (1976) 365,
  [\href{http://dx.doi.org/10.1016/0550-3213(76)90143-7}{{\sf
  doi:10.1016/0550-3213(76)90143-7}}].

\bibitem{Wu}
T.~T. Wu and C.~N. Yang, {\it {Some solutions of the classical isotopic gauge
  field equations}},  in {\em {Properties of Matter Under Unusual Conditions}}
  ({H. Mark, S. Fernbach}, ed.), pp.~344--354.
\newblock {Wiley-Interscience}, 1969.

\bibitem{tHooft:1974kcl}
G.~'t~Hooft, {\it {Magnetic monopoles in unified gauge theories}},  {\sl Nucl.
  Phys. B} {\bf 79} (1974) 276--284,
  [\href{http://dx.doi.org/10.1016/0550-3213(74)90486-6}{{\sf
  doi:10.1016/0550-3213(74)90486-6}}].

\bibitem{Polyakov:1974ek}
A.~M. Polyakov, {\it {Particle spectrum in quantum field theory}},  {\sl JETP
  Lett.} {\bf 20} (1974) 194--195.

\bibitem{Goddard:1977da}
P.~Goddard and D.~I. Olive, {\it {New developments in the theory of magnetic
  monopoles}},  {\sl Rept. Prog. Phys.} {\bf 41} (1978) 1357,
  [\href{http://dx.doi.org/10.1088/0034-4885/41/9/001}{{\sf
  doi:10.1088/0034-4885/41/9/001}}].

\bibitem{Coleman:1982cx}
S.~R. Coleman, {\it {The magnetic monopole fifty years later}},  in {\em {Les
  Houches Summer School of Theoretical Physics: Laser-Plasma Interactions}},
  pp.~461--552, 6, 1982.

\bibitem{Konishi:2007dn}
K.~Konishi, {\it {The magnetic monopoles seventy-five years later}},  {\sl
  Lect. Notes Phys.} {\bf 737} (2008) 471--521,
  [\href{http://arxiv.org/abs/hep-th/0702102}{{\sf arXiv:hep-th/0702102}}].

\bibitem{Manton:2004tk}
N.~S. Manton and P.~Sutcliffe, {\em {Topological solitons}}.
\newblock Cambridge Monographs on Mathematical Physics. Cambridge University
  Press, 2004.

\bibitem{Shnir:2005vvi}
Y.~M. Shnir, {\em {Magnetic Monopoles}}.
\newblock Text and Monographs in Physics. Springer, Berlin/Heidelberg, 2005.

\bibitem{Chamseddine:1997nm}
A.~H. Chamseddine and M.~S. Volkov, {\it {NonAbelian BPS monopoles in N=4
  gauged supergravity}},  {\sl Phys. Rev. Lett.} {\bf 79} (1997) 3343--3346,
  [\href{http://arxiv.org/abs/hep-th/9707176}{{\sf arXiv:hep-th/9707176}}],
  [\href{http://dx.doi.org/10.1103/PhysRevLett.79.3343}{{\sf
  doi:10.1103/PhysRevLett.79.3343}}].

\bibitem{Forgacs:2003yh}
P.~Forgacs and M.~S. Volkov, {\it {Resonant excitations of the 't
  Hooft-Polyakov monopole}},  {\sl Phys. Rev. Lett.} {\bf 92} (2004) 151802,
  [\href{http://arxiv.org/abs/hep-th/0311062}{{\sf arXiv:hep-th/0311062}}],
  [\href{http://dx.doi.org/10.1103/PhysRevLett.92.151802}{{\sf
  doi:10.1103/PhysRevLett.92.151802}}].

\bibitem{Rajantie:2016paj}
A.~Rajantie, {\it {The search for magnetic monopoles}},  {\sl Phys. Today} {\bf
  69} (2016), no.~10 40--46, [\href{http://dx.doi.org/10.1063/PT.3.3328}{{\sf
  doi:10.1063/PT.3.3328}}].

\bibitem{Mitsou:2019mrs}
V.~A. Mitsou, {\it {Searches for magnetic monopoles: a review}},  {\sl MDPI
  Proc.} {\bf 13} (2019), no.~1 10,
  [\href{http://dx.doi.org/10.3390/proceedings2019013010}{{\sf
  doi:10.3390/proceedings2019013010}}].

\bibitem{Klinkhamer:1984di}
F.~R. Klinkhamer and N.~S. Manton, {\it {A saddle point solution in the
  Weinberg-Salam theory}},  {\sl Phys. Rev. D} {\bf 30} (1984) 2212,
  [\href{http://dx.doi.org/10.1103/PhysRevD.30.2212}{{\sf
  doi:10.1103/PhysRevD.30.2212}}].

\bibitem{Kleihaus:1991ks}
B.~Kleihaus, J.~Kunz, and Y.~Brihaye, {\it {The electroweak sphaleron at
  physical mixing angle}},  {\sl Phys. Lett. B} {\bf 273} (1991) 100--104,
  [\href{http://dx.doi.org/10.1016/0370-2693(91)90560-D}{{\sf
  doi:10.1016/0370-2693(91)90560-D}}].

\bibitem{Dashen:1974ck}
R.~F. Dashen, B.~Hasslacher, and A.~Neveu, {\it {Nonperturbative methods and
  extended hadron models in field theory. 3. Four-dimensional nonabelian
  models}},  {\sl Phys. Rev. D} {\bf 10} (1974) 4138,
  [\href{http://dx.doi.org/10.1103/PhysRevD.10.4138}{{\sf
  doi:10.1103/PhysRevD.10.4138}}].

\bibitem{Yaffe:1989ms}
L.~G. Yaffe, {\it {Static solutions of SU(2) Higgs theory}},  {\sl Phys. Rev.
  D} {\bf 40} (1989) 3463,
  [\href{http://dx.doi.org/10.1103/PhysRevD.40.3463}{{\sf
  doi:10.1103/PhysRevD.40.3463}}].

\bibitem{Ratra:1987dp}
B.~Ratra and L.~G. Yaffe, {\it {Spherically symmetric classical solutions in
  SU(2) gauge theory with a Higgs field}},  {\sl Phys. Lett. B} {\bf 205}
  (1988) 57--61, [\href{http://dx.doi.org/10.1016/0370-2693(88)90398-X}{{\sf
  doi:10.1016/0370-2693(88)90398-X}}].

\bibitem{Farhi:2005rz}
E.~Farhi, N.~Graham, V.~Khemani, R.~Markov, and R.~Rosales, {\it {An oscillon
  in the SU(2) gauged Higgs model}},  {\sl Phys. Rev. D} {\bf 72} (2005)
  101701, [\href{http://arxiv.org/abs/hep-th/0505273}{{\sf
  arXiv:hep-th/0505273}}],
  [\href{http://dx.doi.org/10.1103/PhysRevD.72.101701}{{\sf
  doi:10.1103/PhysRevD.72.101701}}].

\bibitem{Graham:2006vy}
N.~Graham, {\it {An electroweak oscillon}},  {\sl Phys. Rev. Lett.} {\bf 98}
  (2007) 101801, [\href{http://arxiv.org/abs/hep-th/0610267}{{\sf
  arXiv:hep-th/0610267}}],
  [\href{http://dx.doi.org/10.1103/PhysRevLett.98.101801}{{\sf
  doi:10.1103/PhysRevLett.98.101801}}]. [Erratum: Phys.Rev.Lett. 98, 189904
  (2007)].

\bibitem{Graham:2007ds}
N.~Graham, {\it {Numerical simulation of an electroweak oscillon}},  {\sl Phys.
  Rev. D} {\bf 76} (2007) 085017, [\href{http://arxiv.org/abs/0706.4125}{{\sf
  arXiv:0706.4125}}], [\href{http://dx.doi.org/10.1103/PhysRevD.76.085017}{{\sf
  doi:10.1103/PhysRevD.76.085017}}].

\bibitem{Cho:1996qd}
Y.~M. Cho and D.~Maison, {\it {Monopoles in Weinberg-Salam model}},  {\sl Phys.
  Lett. B} {\bf 391} (1997) 360--365,
  [\href{http://arxiv.org/abs/hep-th/9601028}{{\sf arXiv:hep-th/9601028}}],
  [\href{http://dx.doi.org/10.1016/S0370-2693(96)01492-X}{{\sf
  doi:10.1016/S0370-2693(96)01492-X}}].

\bibitem{Cho:2013vba}
Y.~M. Cho, K.~Kim, and J.~H. Yoon, {\it {Finite energy electroweak dyon}},
  {\sl Eur. Phys. J. C} {\bf 75} (2015), no.~2 67,
  [\href{http://arxiv.org/abs/1305.1699}{{\sf arXiv:1305.1699}}],
  [\href{http://dx.doi.org/10.1140/epjc/s10052-015-3290-3}{{\sf
  doi:10.1140/epjc/s10052-015-3290-3}}].

\bibitem{Pak:2013jaa}
D.~G. Pak, P.~M. Zhang, and L.~P. Zou, {\it {On finite energy monopole
  solutions in Weinberg\textendash{}Salam model}},  {\sl Int. J. Mod. Phys. A}
  {\bf 30} (2015), no.~27 1550164, [\href{http://arxiv.org/abs/1311.7567}{{\sf
  arXiv:1311.7567}}], [\href{http://dx.doi.org/10.1142/S0217751X1550164X}{{\sf
  doi:10.1142/S0217751X1550164X}}].

\bibitem{Blaschke:2017pym}
F.~Blaschke and P.~Bene\v{s}, {\it {BPS Cho\textendash{}Maison monopole}},
  {\sl PTEP} {\bf 2018} (2018), no.~7 073B03,
  [\href{http://arxiv.org/abs/1711.04842}{{\sf arXiv:1711.04842}}],
  [\href{http://dx.doi.org/10.1093/ptep/pty071}{{\sf
  doi:10.1093/ptep/pty071}}].

\bibitem{Ellis:2020bpy}
J.~Ellis, P.~Q. Hung, and N.~E. Mavromatos, {\it {An electroweak monopole,
  Dirac quantization and the weak mixing angle}},  {\sl Nucl. Phys. B} {\bf
  969} (2021) 115468, [\href{http://arxiv.org/abs/2008.00464}{{\sf
  arXiv:2008.00464}}],
  [\href{http://dx.doi.org/10.1016/j.nuclphysb.2021.115468}{{\sf
  doi:10.1016/j.nuclphysb.2021.115468}}].

\bibitem{Hung:2020vuo}
P.~Q. Hung, {\it {Topologically stable, finite-energy electroweak-scale
  monopoles}},  {\sl Nucl. Phys. B} {\bf 962} (2021) 115278,
  [\href{http://arxiv.org/abs/2003.02794}{{\sf arXiv:2003.02794}}],
  [\href{http://dx.doi.org/10.1016/j.nuclphysb.2020.115278}{{\sf
  doi:10.1016/j.nuclphysb.2020.115278}}].

\bibitem{Bai:2020ezy}
Y.~Bai and M.~Korwar, {\it {Hairy magnetic and dyonic black holes in the
  Standard Model}},  {\sl JHEP} {\bf 04} (2021) 119,
  [\href{http://arxiv.org/abs/2012.15430}{{\sf arXiv:2012.15430}}],
  [\href{http://dx.doi.org/10.1007/JHEP04(2021)119}{{\sf
  doi:10.1007/JHEP04(2021)119}}].

\bibitem{Witten:1976ck}
E.~Witten, {\it {Some exact multi - instanton solutions of classical Yang-Mills
  theory}},  {\sl Phys. Rev. Lett.} {\bf 38} (1977) 121--124,
  [\href{http://dx.doi.org/10.1103/PhysRevLett.38.121}{{\sf
  doi:10.1103/PhysRevLett.38.121}}].

\bibitem{Nambu:1977ag}
Y.~Nambu, {\it {String-like configurations in the Weinberg-Salam theory}},
  {\sl Nucl. Phys. B} {\bf 130} (1977) 505,
  [\href{http://dx.doi.org/10.1016/0550-3213(77)90252-8}{{\sf
  doi:10.1016/0550-3213(77)90252-8}}].

\bibitem{Vachaspati:1992fi}
T.~Vachaspati, {\it {Vortex solutions in the Weinberg-Salam model}},  {\sl
  Phys. Rev. Lett.} {\bf 68} (1992) 1977--1980,
  [\href{http://dx.doi.org/10.1103/PhysRevLett.68.1977}{{\sf
  doi:10.1103/PhysRevLett.68.1977}}]. [Erratum: Phys.Rev.Lett. 69, 216 (1992)].

\bibitem{Urrestilla:2001dd}
J.~Urrestilla, A.~Achucarro, J.~Borrill, and A.~R. Liddle, {\it {The evolution
  and persistence of dumbbells in electroweak theory}},  {\sl JHEP} {\bf 08}
  (2002) 033, [\href{http://arxiv.org/abs/hep-ph/0106282}{{\sf
  arXiv:hep-ph/0106282}}],
  [\href{http://dx.doi.org/10.1088/1126-6708/2002/08/033}{{\sf
  doi:10.1088/1126-6708/2002/08/033}}].

\bibitem{Yoneya:1977yi}
T.~Yoneya, {\it {Stability and instability of the Wu-Yang solution of
  Yang-Mills field equation}},  {\sl Phys. Rev. D} {\bf 16} (1977) 2567,
  [\href{http://dx.doi.org/10.1103/PhysRevD.16.2567}{{\sf
  doi:10.1103/PhysRevD.16.2567}}].

\bibitem{Brandt:1979kk}
R.~A. Brandt and F.~Neri, {\it {Stability analysis for singular nonabelian
  magnetic monopoles}},  {\sl Nucl. Phys. B} {\bf 161} (1979) 253--282,
  [\href{http://dx.doi.org/10.1016/0550-3213(79)90211-6}{{\sf
  doi:10.1016/0550-3213(79)90211-6}}].

\bibitem{yang2014solitons}
Y.~Yang, {\em Solitons in Field Theory and Nonlinear Analysis}.
\newblock Springer, 2014.

\bibitem{Baacke:1990at}
J.~Baacke, {\it {Fluctuations and stability of the t'Hooft-Polyakov monopole}},
   {\sl Z. Phys. C} {\bf 53} (1992) 399--401,
  [\href{http://dx.doi.org/10.1007/BF01625898}{{\sf doi:10.1007/BF01625898}}].

\bibitem{Coleman:1985rnk}
S.~Coleman, {\em {Aspects of Symmetry}: {Selected Erice Lectures}}.
\newblock Cambridge University Press, Cambridge, U.K., 1985.

\bibitem{Hindmarsh:1993aw}
M.~Hindmarsh and M.~James, {\it {The origin of the sphaleron dipole moment}},
  {\sl Phys. Rev. D} {\bf 49} (1994) 6109--6114,
  [\href{http://arxiv.org/abs/hep-ph/9307205}{{\sf arXiv:hep-ph/9307205}}],
  [\href{http://dx.doi.org/10.1103/PhysRevD.49.6109}{{\sf
  doi:10.1103/PhysRevD.49.6109}}].

\bibitem{Forgacs:1979zs}
P.~Forgacs and N.~S. Manton, {\it {Space-time symmetries in gauge theories}},
  {\sl Commun. Math. Phys.} {\bf 72} (1980) 15,
  [\href{http://dx.doi.org/10.1007/BF01200108}{{\sf doi:10.1007/BF01200108}}].

\bibitem{Copeland:1995fq}
E.~J. Copeland, M.~Gleiser, and H.~R. Muller, {\it {Oscillons: resonant
  configurations during bubble collapse}},  {\sl Phys. Rev. D} {\bf 52} (1995)
  1920--1933, [\href{http://arxiv.org/abs/hep-ph/9503217}{{\sf
  arXiv:hep-ph/9503217}}],
  [\href{http://dx.doi.org/10.1103/PhysRevD.52.1920}{{\sf
  doi:10.1103/PhysRevD.52.1920}}].

\bibitem{Honda:2001xg}
E.~P. Honda and M.~W. Choptuik, {\it {Fine structure of oscillons in the
  spherically symmetric phi**4 Klein-Gordon model}},  {\sl Phys. Rev. D} {\bf
  65} (2002) 084037, [\href{http://arxiv.org/abs/hep-ph/0110065}{{\sf
  arXiv:hep-ph/0110065}}],
  [\href{http://dx.doi.org/10.1103/PhysRevD.65.084037}{{\sf
  doi:10.1103/PhysRevD.65.084037}}].

\bibitem{Fodor:2006zs}
G.~Fodor, P.~Forgacs, P.~Grandclement, and I.~Racz, {\it {Oscillons and
  quasi-breathers in the phi**4 Klein-Gordon model}},  {\sl Phys. Rev. D} {\bf
  74} (2006) 124003, [\href{http://arxiv.org/abs/hep-th/0609023}{{\sf
  arXiv:hep-th/0609023}}],
  [\href{http://dx.doi.org/10.1103/PhysRevD.74.124003}{{\sf
  doi:10.1103/PhysRevD.74.124003}}].

\bibitem{GGV}
J.~Garaud, R.~Gervalle, and M.~S. Volkov, {\it in preparation}.

\bibitem{Newman:1961qr}
E.~Newman and R.~Penrose, {\it {An approach to gravitational radiation by a
  method of spin coefficients}},  {\sl J. Math. Phys.} {\bf 3} (1962) 566--578,
  [\href{http://dx.doi.org/10.1063/1.1724257}{{\sf doi:10.1063/1.1724257}}].

\bibitem{Goldberg:1966uu}
J.~N. Goldberg, A.~J. MacFarlane, E.~T. Newman, F.~Rohrlich, and E.~C.~G.
  Sudarshan, {\it {Spin s spherical harmonics and edth}},  {\sl J. Math. Phys.}
  {\bf 8} (1967) 2155, [\href{http://dx.doi.org/10.1063/1.1705135}{{\sf
  doi:10.1063/1.1705135}}].

\bibitem{gelfand2000calculus}
I.~Gelfand, S.~Fomin, and R.~Silverman, {\em Calculus of Variations}.
\newblock Dovers on Mathematics. Dover Publications, 2000.

\bibitem{Teh:2014xva}
R.~Teh, B.-L. Ng, and K.-M. Wong, {\it {Half-monopole in the
  Weinberg\textendash{}Salam model}},  {\sl Annals Phys.} {\bf 354} (2015)
  489--498, [\href{http://arxiv.org/abs/1406.0978}{{\sf arXiv:1406.0978}}],
  [\href{http://dx.doi.org/10.1016/j.aop.2015.01.018}{{\sf
  doi:10.1016/j.aop.2015.01.018}}].

\bibitem{GVin}
R.~Gervalle and M.~S. Volkov, {\it in preparation}.

\bibitem{Ridgway:1994sm}
S.~A. Ridgway and E.~J. Weinberg, {\it {Instabilities of magnetically charged
  black holes}},  {\sl Phys. Rev. D} {\bf 51} (1995) 638--646,
  [\href{http://arxiv.org/abs/hep-th/9409013}{{\sf arXiv:hep-th/9409013}}],
  [\href{http://dx.doi.org/10.1103/PhysRevD.51.638}{{\sf
  doi:10.1103/PhysRevD.51.638}}].

\bibitem{Ridgway:1995ke}
S.~A. Ridgway and E.~J. Weinberg, {\it {Static black hole solutions without
  rotational symmetry}},  {\sl Phys. Rev. D} {\bf 52} (1995) 3440--3456,
  [\href{http://arxiv.org/abs/gr-qc/9503035}{{\sf arXiv:gr-qc/9503035}}],
  [\href{http://dx.doi.org/10.1103/PhysRevD.52.3440}{{\sf
  doi:10.1103/PhysRevD.52.3440}}].

\bibitem{Ridgway:1995ac}
S.~A. Ridgway and E.~J. Weinberg, {\it {Are all static black hole solutions
  spherically symmetric?}},  {\sl Gen. Rel. Grav.} {\bf 27} (1995) 1017--1021,
  [\href{http://arxiv.org/abs/gr-qc/9504003}{{\sf arXiv:gr-qc/9504003}}],
  [\href{http://dx.doi.org/10.1007/BF02148644}{{\sf doi:10.1007/BF02148644}}].

\bibitem{Maldacena:2020skw}
J.~Maldacena, {\it {Comments on magnetic black holes}},  {\sl JHEP} {\bf 04}
  (2021) 079, [\href{http://arxiv.org/abs/2004.06084}{{\sf arXiv:2004.06084}}],
  [\href{http://dx.doi.org/10.1007/JHEP04(2021)079}{{\sf
  doi:10.1007/JHEP04(2021)079}}].

\bibitem{Forgacs:1980ym}
P.~Forgacs, Z.~Horvath, and L.~Palla, {\it {Exact multi - monopole solutions in
  the Bogomolny-Prasad-Sommerfield Limit}},  {\sl Phys. Lett. B} {\bf 99}
  (1981) 232, [\href{http://dx.doi.org/10.1016/0370-2693(81)91115-1}{{\sf
  doi:10.1016/0370-2693(81)91115-1}}]. [Erratum: Phys.Lett.B 101, 457 (1981)].

\bibitem{Arafune:1974uy}
J.~Arafune, P.~G.~O. Freund, and C.~J. Goebel, {\it {Topology of Higgs
  fields}},  {\sl J. Math. Phys.} {\bf 16} (1975) 433,
  [\href{http://dx.doi.org/10.1063/1.522518}{{\sf doi:10.1063/1.522518}}].

\end{thebibliography}

\providecommand{\href}[2]{#2}\begingroup\raggedright\endgroup

\end{document}